\documentclass{iucrjournals}

\usepackage{amsmath}
 \usepackage{amssymb}
 \usepackage{bm}
 \usepackage{cleveref}
 \usepackage{tikz}
 \usepackage{subcaption}
\usepackage{multirow}
\usepackage{makecell}
\usepackage{subcaption}  %
\graphicspath{{figures/}}
\nolinenumbers

\newcommand{\revised}[1]{#1}
\title{Learning Lattice Parameters from Powder X-Ray Diffraction Data Using Invariants}

\author[a]{Elyssa Hofgard\IUCrCemaillink{ehofgard@mit.edu}\IUCrOrcidlink{0000-0002-0745-9477}}%
\author[a]{Kyucheol Min\IUCrOrcidlink{0009-0002-3079-8845}}%
\author[a]{Nofit Segal\IUCrOrcidlink{0000-0002-8891-8590}}%
\author[b]{David W. Mittan-Moreau\IUCrOrcidlink{0000-0002-3193-571X}}%
\author[a]{Aria Mansouri Tehrani\IUCrOrcidlink{0000-0003-1968-0379}}%
\author[b]{Vanessa Oklejas\IUCrOrcidlink{0000-0001-5696-4720}}%
\author[a]{Rafael Gómez-Bombarelli\IUCrOrcidlink{0000-0002-9495-8599}}%
\author[a]{Jigyasa Nigam\IUCrOrcidlink{0000-0001-6857-4332}}%
\author[b]{Daniel W. Paley\IUCrOrcidlink{0000-0003-1161-5142}}%
\author[b]{Aaron S. Brewster\IUCrOrcidlink{0000-0002-0908-7822}}%
\author[a]{Tess Smidt\IUCrOrcidlink{0000-0001-5581-5344}}%

\affil[a]{Massachusetts Institute of Technology, Cambridge, MA}
\affil[b]{Molecular Biophysics and Integrated Bioimaging Division, Lawrence Berkeley National Lab, Berkeley, CA}

\begin{document}
\addtocontents{toc}{\protect\setcounter{tocdepth}{-1}} %
\maketitle
\begin{synopsis}
We introduce a smooth, invariant representation of the reciprocal lattice that avoids the discontinuities of unit cell parameterizations, and show it improves machine-learned lattice parameter prediction from powder X-ray diffraction data.
\end{synopsis}

\begin{abstract}
We present a machine learning (ML) method to determine unit cell parameters from powder X-Ray diffraction (XRD) data using a novel invariant lattice representation. In ML, the data representation used can have a substantial impact on the prediction quality. Previous approaches have directly predicted lattice parameters ($a,b,c,\alpha,\beta,\gamma$) from XRD inputs. However, these parameters depend strongly on the unit cell reduction or convention used. In this work, we construct an invariant representation of the reciprocal lattice that is independent of primitive cell convention, based on the bispectrum\textemdash a descriptor built from spherical harmonic projections of lattice points. The calculation of the lattice bispectrum is differentiable, and we demonstrate how to invert it using \revised{gradient-based optimization initialized from a nearest-neighbor lookup.} We show that with a fixed ML model architecture, using the lattice bispectrum as the ML target rather than the unit cell parameters leads to more accurate lattice parameter predictions. For example, using the MP-20 dataset, the bispectrum reduces length mean absolute percentage error (MAPE) from \revised{11.08\% to 2.43\% and angle MAPE from 12.15\% to 2.45\%} compared to direct prediction with the same model architecture. We additionally benchmark our approach against pre-existing XRD to crystal structure models such as Crystalyze and assess its performance on experimental data. Beyond unit cell representation, we anticipate this invariant lattice representation could serve more broadly as a geometry-aware target for other crystallographic machine learning tasks such as structure generation.

\end{abstract}


\section{Introduction}

Powder X-Ray diffraction (XRD) is a core tool for characterizing materials. In general, traditional methods extract peak positions and then assess trial unit cells in order to determine the correct lattice parameters. With the advent of machine learning (ML), interest has grown in using ML to aid in unit cell determination. ML models are able to learn patterns in large-scale datasets and thus could be used in conjunction with crystallographic expertise to narrow the lattice parameter search space \cite{chitturiAutomatedPredictionLattice2021, dong2021deep, habershon2004powder}. However, even setting aside challenges in signal quality such as peak overlap \cite{david2008structure}, ML approaches face a fundamental representation challenge, as the unit cell is not a unique representation for a given lattice. Small distortions can change the reduced cell convention discontinuously, creating an inherently non-smooth learning target. Previous ML approaches for learning lattice parameters from XRD data have usually directly predicted the lattice parameters ($a,b,c,\alpha,\beta,\gamma$) which depend on the unit cell convention used.

Throughout this work, we refer to cubic, hexagonal, and tetragonal systems as high-symmetry, as their lattice parameters are the most constrained. Low-symmetry triclinic and monoclinic systems are challenging due to having the greatest number of free lattice parameters. Previous work has struggled to obtain accurate predictions for low symmetry crystal systems, as the parameter space is larger and multiple reduced cell conventions can represent the same lattice \cite{chitturiAutomatedPredictionLattice2021,andrejevicAlphaDiffractAutomatedCrystallographic2026,shu2025,gomez-peraltaConvolutionalNeuralNetworks2023,gomezperalta2025insights}. We address this by introducing an invariant lattice representation in reciprocal space that is independent of primitive cell convention, based on the bispectrum \cite{bartokRepresentingChemicalEnvironments2013}. The lattice bispectrum is constructed from spherical harmonic projections of lattice points in reciprocal space and varies smoothly under continuous lattice deformations, providing a more learnable target for a neural network. We demonstrate that this representation can be inverted in an auto-differentiable framework.

While many previous ML works use convolutional neural networks (CNNs) \cite{chitturiAutomatedPredictionLattice2021,gomez-peraltaConvolutionalNeuralNetworks2023}, these models lose information in long-range XRD peak dependencies. By the construction of their architecture, CNNs are translation invariant. This means that while a CNN may detect individual peaks, it cannot directly relate peaks across the full diffraction pattern. This is a useful inductive bias for images, yet it does not align with our understanding of XRD data. Peak positions carry absolute physical meaning and cannot be treated as interchangeable regardless of their location. In contrast, self-attention mechanisms in transformers compute pairwise interactions between all positions simultaneously, allowing every peak to attend to every other peak across the full diffraction pattern in a single layer.

\begin{figure}[htbp]
    \centering
    \includegraphics[width=1.0\textwidth]{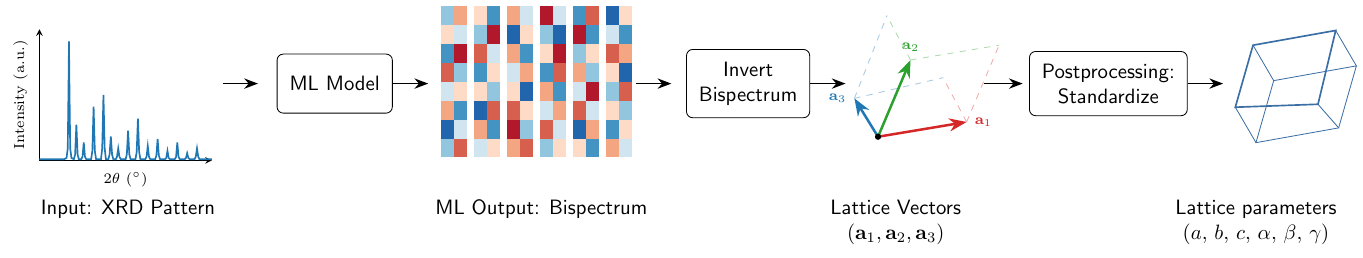}
    \caption{Workflow for predicting the bispectrum and then inverting to obtain lattice parameters. The model is given an input XRD pattern represented as an array of (2$\theta$, intensity) values and is trained to predict the lattice bispectrum. The prediction can then be inverted to obtain a set of spanning lattice vectors, see \Cref{sec:bispec_invert}. Post-processing using cctbx or pymatgen can then be done to obtain lattice parameters in a standardized or conventional setting.}
    \label{fig:proposed_workflow}
\end{figure}

We thus train a transformer to predict the lattice bispectrum from XRD data and compare performance with directly predicting the lattice parameters. The model takes as input a simulated or experimental XRD pattern and outputs the lattice bispectrum, from which lattice parameters are recovered via inversion, leading to more accurate predictions than directly predicting lattice parameters. See \Cref{fig:proposed_workflow} for our proposed workflow. We train our model using simulated XRD patterns from the Materials Project with physically-informed data augmentation. We benchmark model performance on experimental XRD patterns in the RRUFF database \cite{Lafuente2015RRUFF} and on unit cells with dominant zones. To summarize, our main contributions are:
\begin{itemize}
    \item We introduce a novel reciprocal-space representation of crystal lattices based on an extension of the bispectrum, yielding an invariant reciprocal space lattice representation that is independent of primitive cell choice. We demonstrate that the lattice bispectrum can be inverted to obtain the crystal lattice \revised{through an optimization through the differentiable computation of the bispectrum.}
    \item We train a transformer to predict the lattice bispectrum from XRD signals. We demonstrate that predicting the lattice bispectrum instead of directly predicting the lattice parameters aids model performance.
    \item We benchmark the two prediction approaches on generalization to the experimental RRUFF dataset and unit cells with dominant zones. Additionally, we explore the differences between training the model on MP-20 (commonly used for benchmarking) vs. the entire Materials Project dataset, finding that the bispectrum outperforms direct prediction consistently across both datasets.
\end{itemize}

\section{Background}

\subsection{Traditional Indexing Methods}

The goal of unit-cell determination is to identify the lattice parameters $a,b,c,\alpha,\beta,\gamma$ and assign the corresponding Miller indices $(h,k,l)$ to the observed peaks in the XRD data. Extensive research has focused on developing indexing algorithms, most of which operate by searching the lattice-parameter space to find the best match between predicted and observed diffraction patterns (e.g. see the dichotomy method implemented in DICVOL91 \cite{boultif1991} and the zone-finding algorithm \cite{de1957determination} implemented in the program ITO). More recent algorithms include SVD-Index implemented in TOPAS \cite{Coelho2003} that uses the singular value decomposition to iteratively solve linear equations relating $hkl$ to $d$-spacings. In contrast, LP-Search \cite{Coelho:to5164}, does not require peak positions and uses a Monte Carlo search of lattice parameter space with a Pawley refinement at the end of each step.

However, despite the vast array of algorithms available for use, indexing can be a non-trivial task. Current methods are quite effective for high-symmetry phases, yet struggle in cases of data with severe peak overlap or the presence of crystalline impurity phases in the powder sample. Even for high-quality single-phase samples, peak overlap becomes significant for low-symmetry crystals or materials with large unit cells (the “dominant zone” problem), where the density of reflections is high \cite{Harris2022}. For these structures, automatic peak finding can even fail on high-quality simulated data. Additionally, unit cells defining different lattices can yield an identical set of unique calculated $d$-spacings, leading to indexing algorithms to report different lattices with the same figure of merit \cite{Mighell_2000}. More broadly, indexing is often inherently ambiguous as lower symmetry cells possess greater degrees of freedom and can yield higher figures of merit than the true higher-symmetry solution. Discriminating between competing solutions can become an ``art form'' relying on the prior knowledge of the crystallographer. This highlights a deeper issue: the unit cell representation lacks smoothness under small structural distortions \cite{Andrewsniggli,andrewsSpaceLatticeRepresentation2019}. Two geometrically similar lattices may be assigned different reduced cells or symmetry classes, creating artificial discontinuities that ML models struggle to learn across. Our bispectrum descriptor addresses this directly by providing a representation that varies continuously under smooth lattice deformations.

\section{Related Work}
Previous work has applied ML to the unit cell prediction problem, primarily by classifying the space group/lattice system or predicting lattice parameters directly from XRD patterns. In general, the use of a benchmark dataset has not been standardized, making it difficult to compare models and accuracies directly. We provide a non-exhaustive literature review, focusing on the datasets used and the failure modes of each method.

\subsection{Classifying Lattice Symmetries from XRD}
Most previous studies have focused on classifying data into space groups or lattice systems from XRD patterns. We can glean insight from both model performance and the methods used to featurize XRD patterns. \cite{suzukiSymmetryPredictionKnowledge2020} trains an extremely randomized tree (exRT) based model on simulated XRD patterns from the ICSD for crystal system and space group classification. The authors use eleven features as input to the model: the first ten corresponding to the first ten peaks and the last corresponding to the total number of peaks in the 2$\theta$ range from 0 to 90.
Notably, they find that the difference in model performance when working in different units is negligible, with all crystal systems except triclinic attaining $\sim$90\% accuracy. The model is unable to classify triclinic crystal systems correctly, which the authors attribute to an imbalanced training set distribution.
\cite{corrieroCrystalMELANewCrystallographic2023} introduces an ML-based web platform (CrystalMELA) for crystal system classification using a 1D CNN model, trained using simulated XRD patterns from organic, inorganic, and metal compounds from the COD. Performance was found to vary among crystal systems, with triclinic, monoclinic, and orthorhombic being the most difficult to classify correctly.

Crucially, the above approaches did not augment simulated patterns to contain characteristics of experimental patterns (e.g., peak shifts/broadening). \cite{leeDeepLearningApproach2023} addresses these issues by incorporating data augmentation, and uses fully connected neural networks to classify crystal system and space group from XRD patterns, finding a concentration of misclassified examples in low symmetry lattice classes (triclinic, monoclinic, and orthorhombic, coined the ``Seattle Zone''). %
Through an analysis of the failure modes of the ML model, it was found that a proportion of misclassified entries corresponded to slight distortions in unit cell angles $\alpha, \beta, \gamma$ from higher symmetry structures. This highlights an issue with the unit cell representation that we aim to address in our work\textemdash small changes in unit cell parameters can lead to discontinuous changes in crystal symmetry.

\subsection{Predicting Lattice Parameters}
Previous work has also explored directly predicting lattice parameters from XRD patterns. \cite{chitturiAutomatedPredictionLattice2021} uses 1D CNNs to predict lattice parameters, training a separate model for each of the seven crystal systems, training with simulated patterns from CSD and ICSD data. This work provides key insight into the impacts of various data augmentation strategies. They test the impacts of including peaks from $0-30^{\circ}$ vs $0-90^{\circ}$ as training data and find that models perform similarly for both angle ranges. Additionally, they consider data augmentation due to peak broadening, baseline noise, random intensity modulation, detector zero shifting, and the presence of multiple unknown phases. However, their models were unable to accurately predict $\alpha, \beta, \gamma$ for lower symmetry systems. %
They further integrate the ML model into LP-search (an indexing algorithm), and find that providing $a,b,c$ predictions led to a speed-up in convergence of LP-search. \cite{liMlatticeabcGenericLattice2021} instead predicts $a,b,c$ from composition-based descriptors, reporting high accuracy for cubic systems yet significantly lower for triclinic.

\cite{shu2025} employs an extinction group classification model and then a convolution neural network unit cell parameter regression model with augmented XRD patterns from the ICSD in order to predict unit cell parameters. \cite{choudhary2025} uses a generative pretrained transformer model that incorporates chemical information to predict structures from XRD patterns. Most recently, \cite{andrejevicAlphaDiffractAutomatedCrystallographic2026} builds a large database of over 31 million simulated diffraction patterns through augmenting the ICSD and Materials Project datasets in order to train a model to predict crystal system, space group, and lattice parameters from XRD patterns. However, there has been limited exploration of the impact of the discontinuity of the unit cell representation on prediction and investigation of invariant unit cell representations. For example, \cite{segal2025loss} investigates gradient-based optimization of lattice parameters from powder XRD, and empirically finds that recovering ground-truth lattices from moderately distorted initial structures is challenging due to a rough optimization landscape.

\subsection{Structure Generation}
Recent works have also attempted to obtain more structural detail from XRD data. Even though these models focus on structure generation, they still have to determine unit cell parameters from XRD patterns and thus may also be subject to issues arising from unit cell representations. \cite{rieselCrystalStructureDetermination2024} builds on the crystal diffusion variational autoencoder (CDVAE, \cite{xieCrystalDiffusionVariational2021}) to develop Crystalyze, a generative model that predicts the crystal structure from PXRD. Crystalyze converts diffraction patterns from a 1D signal to a latent embedding and then conditions on the XRD pattern in order to propose candidate structures. For the MP-20 dataset (the Materials Project dataset filtered to less than 20 atoms), Crystalyze accurately predicts 66.6\% of crystal structures after 64 attempts, and about 30\% when given a single attempt. Moreover, after training on augmented data that is meant to model experimental artifacts, Crystalyze achieves a 41.8\% match rate on the experimental RRUFF dataset after 64 attempts, and about 8\% with a single attempt. \cite{liPowderDiffractionCrystal2025} also uses a generative model conditioned on XRD patterns with a refinement module to achieve a higher match rate on MP-20. Notably, however, these models are limited to predicting structures with 20 atoms or fewer in the unit cell. Similarly, \cite{guoInitioStructureSolutions2025} introduces a diffusion-based generative model trained on nanocrystalline PXRD data that predicts full candidate crystal structures, including unit cell and atomic positions, from the diffraction pattern and chemical composition. \cite{segal2026xrdiff} introduces XRDiff, a diffusion-based model that generates crystal structures from PXRD, using a peak-featurized representation to generalize zero-shot to experimental data without noise augmentation.

\section{Methodology}

Our overarching goal is to use machine learning to help determine the lattice parameters of unknown materials. We first make two key observations. (1) The choice of primitive unit cell convention is not geometrically unique. Moreover, standard unit cell conventions are defined piecewise across crystal systems. Thus, small continuous changes in a lattice can produce discontinuous jumps in lattice parameters near symmetry boundaries. This leads to an ambiguity in data representation of the lattice parameters and make lattice parameters a poorly-behaved regression target for a neural network. (2) When X-rays scatter off a crystal, they produce an image in reciprocal space, e.g. any incident wave vector will lead to a diffraction peak if and only if it intersects with a $k$-space Bragg plane. We are thus motivated to develop a lattice descriptor in reciprocal space. The descriptor is inverted to a set of lattice vectors that span the primitive lattice; see \Cref{fig:proposed_workflow} and \Cref{sec:bispec_invert}. Tools such as \texttt{cctbx} or \texttt{pymatgen} can then be used to obtain the conventional lattice parameters to enable consistent comparison with ground truth parameters. We hypothesize that a geometrically meaningful representation of the unit cell will allow models to better learn the relationship between XRD patterns and lattice parameters.

\subsection{Bispectrum Lattice Descriptor}
A useful lattice descriptor should be (1) independent of primitive cell choice, (2) continuous under smooth lattice deformation, and (3) naturally expressed in reciprocal space where diffraction operates. \revised{Thus, we develop a lattice representation in reciprocal space that is independent of primitive cell convention and invariant under $SO(3)$ (the group of 3D rotations).} Denoting the reciprocal space lattice vectors by $\{\mathbf{b}_1,\mathbf{b}_2,\mathbf{b}_3\}$, each lattice point in reciprocal space can be represented as $\mathbf{G} = h\mathbf{b}_1 + k\mathbf{b}_2 + l\mathbf{b}_3$ where  $h,k,l \in \mathbb{Z}$. We tile the lattice spherically up to a cutoff $k_{\max}$. $k_{\text{max}}$ has a natural physical interpretation as the limiting resolution of the detector, using the relation for the scattering vector $k_{\max} = \frac{2 \sin \theta_{\text{max}}}{\lambda}$, where $\lambda$ is the detector wavelength.

We next use spherical harmonics $Y_{lm}$ to represent the lattice points in reciprocal space. \revised{Spherical harmonics are used to construct rotationally-invariant descriptors as they form an orthonormal basis for functions on the sphere.} One can think of spherical harmonics as basis functions for an ``angular Fourier transform,'' see \Cref{supp_info:group} for a more rigorous background. For example, quantities such as the structure factor or the electron density around an atom can be expanded in spherical harmonics. The lattice in reciprocal space can be written as
\begin{equation}
    \rho(\mathbf{k})=\sum_{\mathbf{G} \in k_{\max}} \delta(\mathbf{k}-\mathbf{G})
\end{equation}
\revised{where $\delta$ is the Dirac delta function.} Although written as a density in reciprocal space, $\rho$ is completely described by the discrete set of reciprocal lattice vectors $\{\mathbf{G}\}$. Consequently, the spherical harmonic expansion is only evaluated at these lattice points.
The spherical harmonics describe the angular part of each $\mathbf{G}$, so they depend on the unit vector $\hat{\mathbf{G}}$. To include radial information, we use radial basis functions $g_n(|\mathbf{G}|)$. The full expansion is then
\begin{equation}
    \rho({\mathbf{G}}) = \sum_n \sum_{l=0}^{l_{\text{max}}} \sum_{m=-l}^l c_{nlm}Y_{lm}(\hat{\mathbf{G}})g_n(|\mathbf{G}|)
\end{equation}
where the coefficients are found through
\begin{equation}
    c_{nlm} = \sum_{\mathbf{G \in k_{\max}}}Y_{lm}(\hat{\mathbf{G}})g_n(|\mathbf{G}|)
\end{equation}
The expansion is exact in the limit of infinitely many angular and radial basis functions.

\begin{figure}[!htbp]
    \centering
    \includegraphics[width=0.75\textwidth]{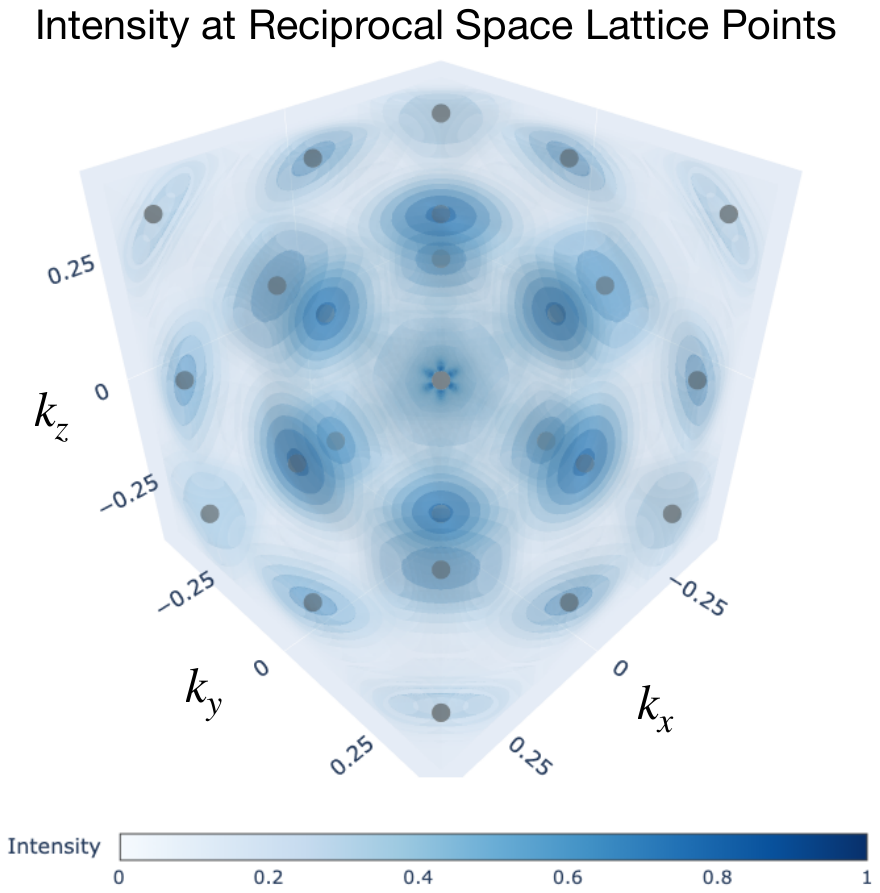}
    \caption{Plot of $\rho(\bf{k})$ for a cubic lattice with $a = 3$ \AA, $l_{\max} = 6$, 10 Bessel function radial basis functions. One can see that $\rho(\bf{k})$ attains its maximum values at reciprocal lattice points.%
    }
    \label{fig:vol_plot}
\end{figure}

As an example in Figure \ref{fig:vol_plot}, we plot $\rho(\bf{k})$ in reciprocal \revised{space} for a cubic lattice with lattice parameter $a = 3$ \AA. As expected, we observe $\rho(\bf{k})$ attains its maximum values at the lattice points in reciprocal space, with higher intensity at points near the $\Gamma$ point.
We can take successive tensor products of the expansion coefficients $c_{nlm}$ to obtain invariant descriptors, following standard literature such as \cite{bartokRepresentingChemicalEnvironments2013}. See \cite{nigamReconstructingLocalEnvironments2026} for the use of invariant descriptors to featurize local atomic environments in real space. The simplest invariant is known as the power spectrum, $\sum_m |c_{nlm}|^2$ for each allowed $l$. However, this loses information about angular correlations between different $l$ channels.

\begin{figure}[htbp]
    \centering
    \includegraphics[width=.9\textwidth, trim=0 70 0 70, clip]{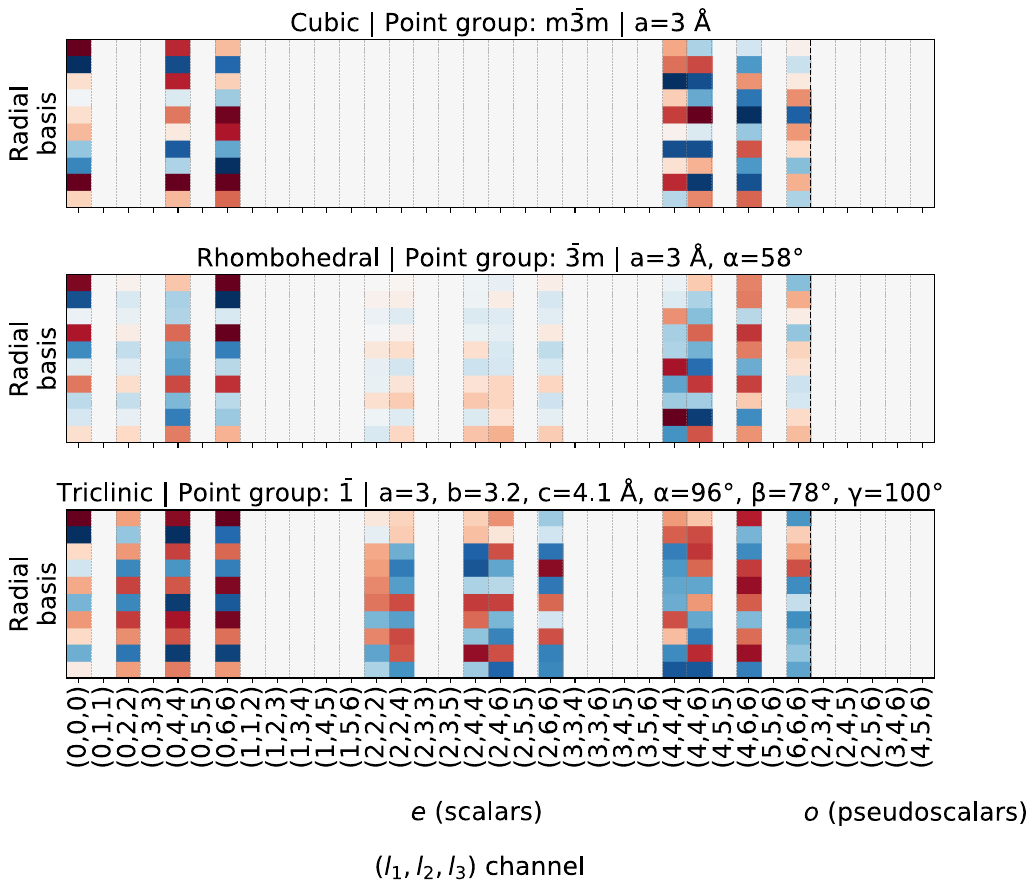}
    \caption{Sample bispectra for crystal systems with varying levels of symmetry. Each column corresponds to a bispectrum component $(l_1, l_2, l_3)$, grouped by angular channel. Scalar ($e$) and pseudoscalar ($o$) components are separated by the dashed line. All pseudoscalar components vanish, consistent with the centrosymmetry of Bravais lattices.}
    \label{fig:bispec_lat_viz}
\end{figure}

To retain more information, we combine three sets of coefficients. The bispectrum $b_{l_1l_2l_3}$ is constructed by coupling coefficients from three channels $(l_1,l_2,l_3)$ into a scalar quantity. To ensure invariance, the coupling is done through a symmetric Clebsch-Gordan tensor product (the same $|l_1-l_2| \leq l_3 \leq l_1+l_2$ that governs angular momentum selection rules in quantum mechanics, see \Cref{supp_info:group} and \Cref{supp:background_atom_rep}). The tensor product is also symmetrized over the indices $(l_1,l_2,l_3)$, guaranteeing that the result is permutation invariant. Note that to distinguish between structures related by inversion, one should include pseudoscalars along with the scalars from the coupling (corresponding to the $O(3)$ group rather than the $SO(3)$ group). This is not necessary for Bravais lattices as they are centrosymmetric but is included here for completeness. The tensor product can be written as
\begin{align}\label{eq:bispec}
     \mathbf{B} = ((\mathbf{c}_{l_3} \otimes (\mathbf{c}_{l_1} \otimes \mathbf{c}_{l_2}))^{(0_e \oplus 0_o)}
 \end{align}
where $(0_e \oplus 0_o)$ emphasizes we are retaining the scalars and pseudoscalars. For $N_r$ radial basis functions and
maximum $l_{\max}$, the final descriptor dimension is
\[
\text{dim}(\mathbf{B}) = \Big( N_r, \sum_{l_1=0}^{l_{\max}} \sum_{l_2=0}^{l_{\max}} \sum_{l_3=0}^{l_{\max}} N_{\text{allowed}}(l_1,l_2,l_3) \Big),
\]
where $N_{\text{allowed}}(l_1,l_2,l_3)$ counts the number of scalar/pseudoscalar contributions after the two tensor products. In this representation, each row corresponds to a radial basis function $n$, and each column corresponds to an allowed coupling. This makes the bispectrum naturally visualizable as a 2D heatmap, as in \Cref{fig:bispec_lat_viz}.

Because $\rho(\mathbf{k})$ transforms equivariantly under the crystal's point group, only specific $(l,m)$ channels can be non-zero. This is the same mathematical formalism that yields the first nontrivial charge density multipole of a given lattice. For example, for a cubic lattice with $O_h$ symmetry, only specific $l$ channels ($l=0,4,6,8, \ldots)$ contribute to the lattice bispectrum. This is confirmed in \Cref{fig:bispec_lat_viz}. The sparsity pattern of the bispectrum directly reflects the underlying symmetry of the unit cell. Consequently, higher symmetry lattices produce more zero entries. As unit cells retain the symmetry of a parallelepiped, the lattice bispectrum is constrained to have zeros in certain entries for all lattices (e.g. see the triclinic lattice in \Cref{fig:bispec_lat_viz}).

\revised{We enumerate the invariance properties of our construction of the reciprocal lattice bispectrum. Following the construction of the bispectrum in \cite{bartokRepresentingChemicalEnvironments2013}, we obtain the bispectrum as symmetrized products of spherical harmonics coefficients (implemented using Clebsch-Gordan coefficents) in \Cref{eq:bispec}. It is therefore invariant under $SO(3)$. If we include pseudoscalars, it then has components that change sign under inversion.  $(l_1,l_2,l_3)$ in \Cref{eq:bispec} are invariant under inversion or accumulate a negative sign depending on whether $l_1+l_2+l_3$ is even or odd. However, every reciprocal lattice is centrosymmetric, so all odd combinations of $l_1+l_2+l_3$ are zero, and the descriptor is thus invariant under the $O(3)$ group. We retain the pseudoscalar channels for generality. These invariances naturally align with the physics of powder diffraction. Rotation invariance means that the descriptor depends on relative geometry rather than absolute orientation. Inversion invariance is consistent with X-ray diffraction as we measure the intensity in Fourier space $|F(\bf{G})|^2$ at reciprocal lattice points $\bf{G}$, so we will have the same measurement if $\bf{G} \to -\bf{G}$.}

\revised{There are additional properties resulting from using the reciprocal lattice vectors to construct the lattice bispectrum. The reciprocal lattice depends only on the real-space lattice vectors, which are unchanged by a translation of the crystal, so the descriptor does not depend on the choice of real-space origin. Moreover, any set of primitive lattice vectors will generate the same set of lattice points. As $c_{nlm}$ is a sum over these points, the descriptor is independent of the choice of primitive cell, including permutations or re-indexing of the vectors $\{\mathbf{b_1},\mathbf{b_2},\mathbf{b_3}\}$. These properties ensure that the descriptor encodes the geometric information of relative distances and angles between reciprocal lattice points that determine the Bravais lattice. See \Cref{supp_info:theorysparse} for the theoretical justification of the sparsity pattern in the bispectrum.}

\begin{figure}[htbp]
    \centering
    \begin{subfigure}[b]{0.6\textwidth}
        \centering
        \includegraphics[width=\textwidth]{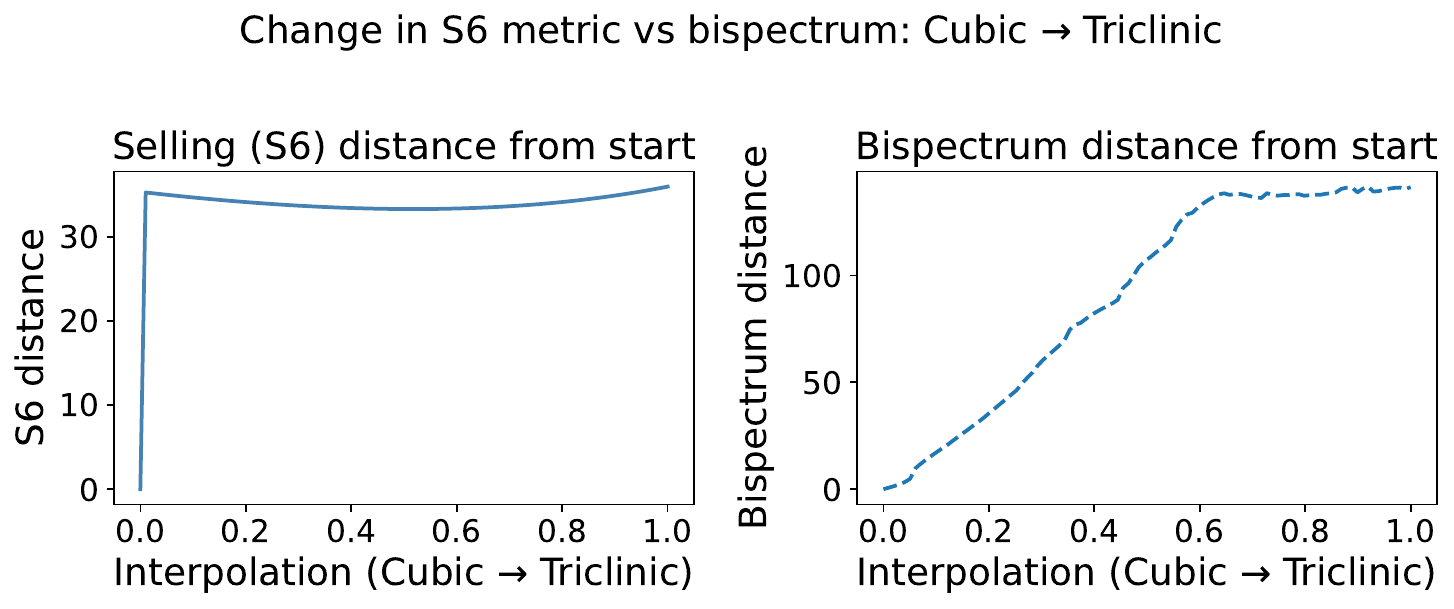}
        \label{fig:bispec_samples}
    \end{subfigure}
    \hfill
    \begin{subfigure}[b]{0.8\textwidth}
        \centering
        \includegraphics[width=\textwidth]{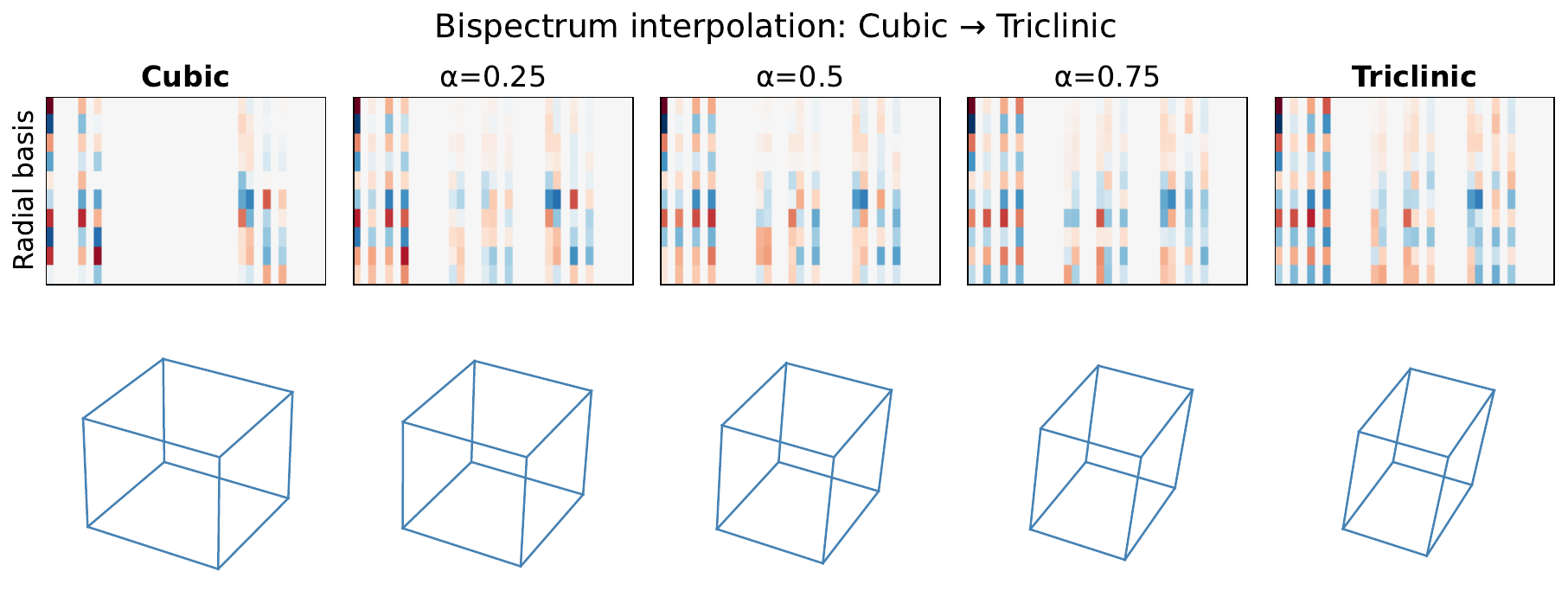}
        \label{fig:bispec_interp}
    \end{subfigure}
    \caption{Example interpolation between a cubic and triclinic lattice and the corresponding change in bispectrum and $S^6$ distance. The interpolation parameter $\alpha$ linearly deforms the lattice in real space. While the bispectrum varies smoothly in Euclidean ($L^2$) norm along this path, distances computed in reduced-cell representations (e.g. $S^6$) can exhibit non-smooth behavior due to changes in the chosen reduced basis during lattice reduction. Note that intermediate lattices are also triclinic, as the interpolation does not enforce any symmetry constraints.}
    \label{fig:bispec_interps}
\end{figure}

We note that this representation is motivated by the discontinuity that arises in lattice parameter representations under conventional space group settings. Small continuous distortions of a lattice can trigger a change in the reduced cell convention, producing artificial jumps even when the underlying geometry changes smoothly. This concept is illustrated in \Cref{fig:bispec_interps}, showing a cubic cell being uniformly distorted into a triclinic cell. Measured in $L^2$-norm from the starting bispectrum, the bispectrum is seen to change smoothly throughout the distortion.\footnote{modulo numerical artifacts}
This representation has the advantage of producing smooth variations in $L^2$ norm under continuous lattice deformations, instead of relying on reduced-cell based representations such as $S^6$, where distances between lattices are computed after a reduction step that can introduce non-uniqueness in the choice of basis \cite{andrewsMeasuringLattices2023}. As a result, such metrics can exhibit apparent non-smooth behavior along continuous deformation paths. See \Cref{fig:mape_vs_noise} for a sensitivity analysis of the inversion to Gaussian noise in the bispectrum.

\subsubsection{Practical Details}\label{sec:bispec_details}
In practice, this descriptor includes some hyperparameters that are user-defined. These include $k_{\max}$, $l_{\max}$, the type of radial basis function, and the number of radial basis functions. $k_{\max}$ depends on the limiting resolution of the detector. For our purposes, we use the standard CuK$\alpha$ wavelength of 1.5406 \AA{}.  Note here we use the crystallographic convention common in XRD, so there is no factor of $2\pi$. Assume that $2\theta_{\text{max}} = 60^{\circ}$, we then have $k_{\max} \approx \frac{2}{3}$ \AA{}.

\begin{figure}[htbp]
    \centering
    \includegraphics[width=0.8\textwidth, trim=0 80 0 80, clip]{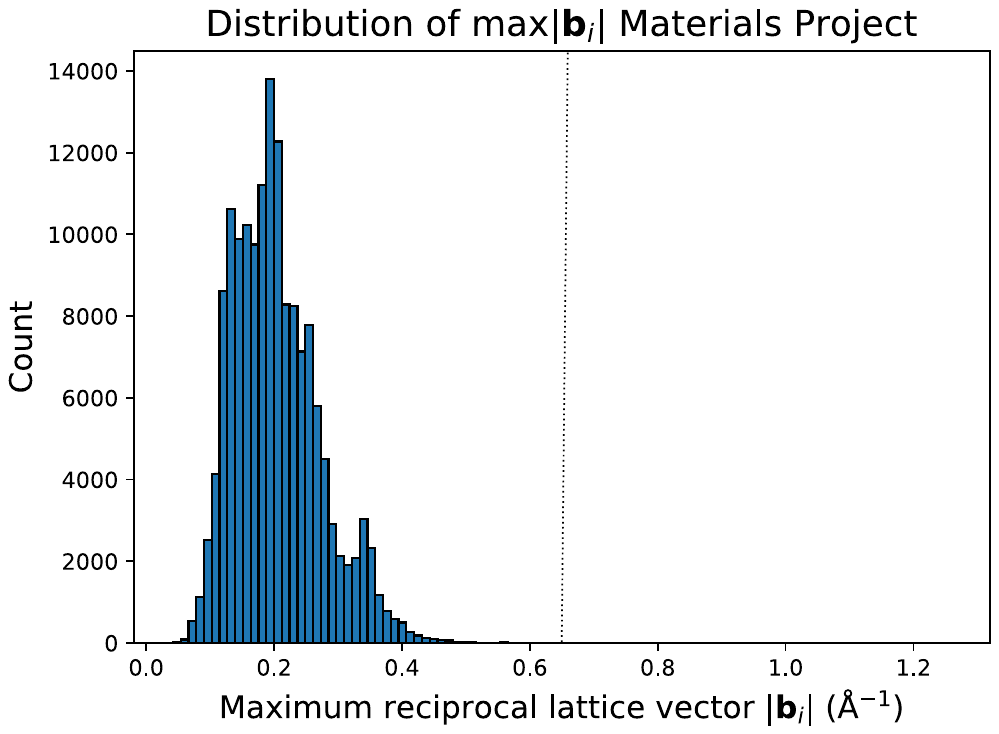}
    \caption{Distribution of the magnitude of the maximum reciprocal lattice vector (using the crystallographic convention) for the Materials Project dataset.}
    \label{fig:recip_lat_mag}
\end{figure}

As we are modeling a periodic structure, in practice, we also want $k_{\text{max}}$ large enough such that every reciprocal lattice point within a certain radius of the $\Gamma$ point is included in the descriptor at least once. For a unit cell, this is $k_{\text{max}} \gtrsim \frac{1}{a_{\text{min}}}$ where $a_{\text{min}}$ is the smallest lattice parameters or equivalently $k_{\text{max}} \gtrsim b_{\max}$ where $b_{\max}$ is the maximum reciprocal lattice vector. As seen in Figure \ref{fig:recip_lat_mag}, our cutoff does satisfy this condition for nearly all lattices in the Materials Project dataset. For different datasets, this could require further tuning.

We find $l_{\max} = 6$ is sufficient to resolve angular lattice information, following standard atomic descriptor literature \cite{bartokRepresentingChemicalEnvironments2013}. We choose to use Bessel functions as the radial basis function in order to weigh points near the $\Gamma$ point higher, but to also allow for variation outside of a narrow envelope (as opposed to Gaussian basis functions) and use 10 radial basis functions.

\subsection{Inversion of Lattice Bispectrum}\label{sec:bispec_invert}
\begin{figure}[htbp]
    \centering
    \includegraphics[width=1.0\textwidth]{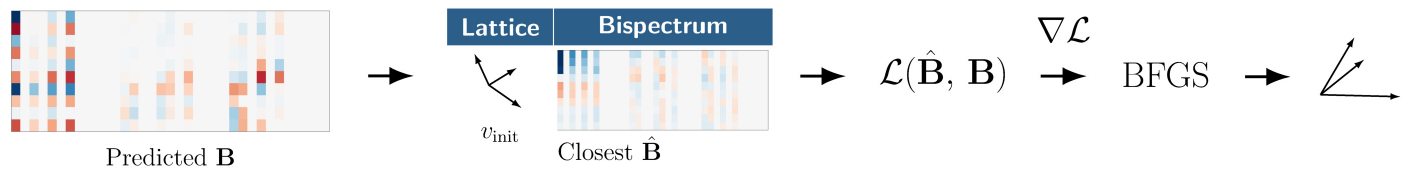}
    \caption{\revised{Algorithm to invert the lattice bispectrum. Given an initial bispectrum, the starting lattice parameters are initialized from the nearest entry in a precomputed database of bispectra. The residual between the true and predicted bispectra is then calculated and lattice parameters are updated with a L-BFGS step. After inversion, we obtain a set of vectors spanning the primitive lattice. Standard crystallographic tools such as pymatgen or cctbx can then be used for postprocessing.}
    }
    \label{fig:inversion_diagram}
\end{figure}

We demonstrate how to invert the lattice bispectrum to obtain the lattice parameters (the second step in \Cref{fig:proposed_workflow}.) While the bispectrum is not mathematically complete (i.e. it is not a one-to-one mapping\revised{; see \citeasnoun{bendoryOrbitRecoverySpherical2025} for generic uniqueness results}), it can be empirically inverted \cite{nigamReconstructingLocalEnvironments2026} \revised{in an auto-differentiable framework. We use a precomputed database of bispectra as a nearest-neighbor lookup to initialize the starting guess for the inversion.} We show the approach in \Cref{fig:inversion_diagram}. We aim to invert the mapping $f(a,b,c, \alpha, \beta, \gamma) = \mathbf{B}$ to obtain lattice parameters $(a,b,c, \alpha, \beta, \gamma)$ where $f$ represents the bispectrum calculation. Crucially, $f$ can be implemented in an \emph{auto-differentiable framework} such as \texttt{torch}. Thus, standard optimization tools can be used to invert $f$. In practice, we found that second-order optimization methods such as L-BFGS-B provide more accurate lattice parameter predictions. Additionally, the starting lattice guess can improve convergence. Thus, we initialize $v_{\text{guess}}=(a,b,c, \alpha, \beta, \gamma)_{\text{guess}}$ using precalculated bispectra from the Materials Project database (with distance between bispectra measured by the L2 norm).\footnote{One could also use other crystallographic databases. The calculated bispectra are available from \url{https://github.com/atomicarchitects/bispectrum-xrd-ml/tree/main}.} \revised{In practice, we use $n=15$ nearest entries by mean squared error are used as starting points. Each starting lattice is perturbed with Gaussian noise of standard deviation $0.05\,\|\mathbf{L}^*\|_F$ per element of the reciprocal lattice matrix $\mathbf{L}^*$ and whose rows are $\mathbf{b}_1, \mathbf{b}_2, \mathbf{b}_3$, and optimized with L-BFGS-B using SciPy \cite{2020SciPy-NMeth}. If a candidate reaches a bispectrum loss below $10^{-3}$, no further candidates are tried, and otherwise the lattice with the lowest loss is returned. Candidate selection uses only the predicted bispectrum.} As the calculation of $\mathbf{B}$ is differentiable, one can then calculate a loss between $\mathbf{B}$ and the calculated bispectra, and then update $v_{\text{guess}}$ based on this loss with an optimization step.

\subsection{Bispectrum Prediction from XRD Patterns}

\begin{figure}
    \centering
    \includegraphics[width=\linewidth]{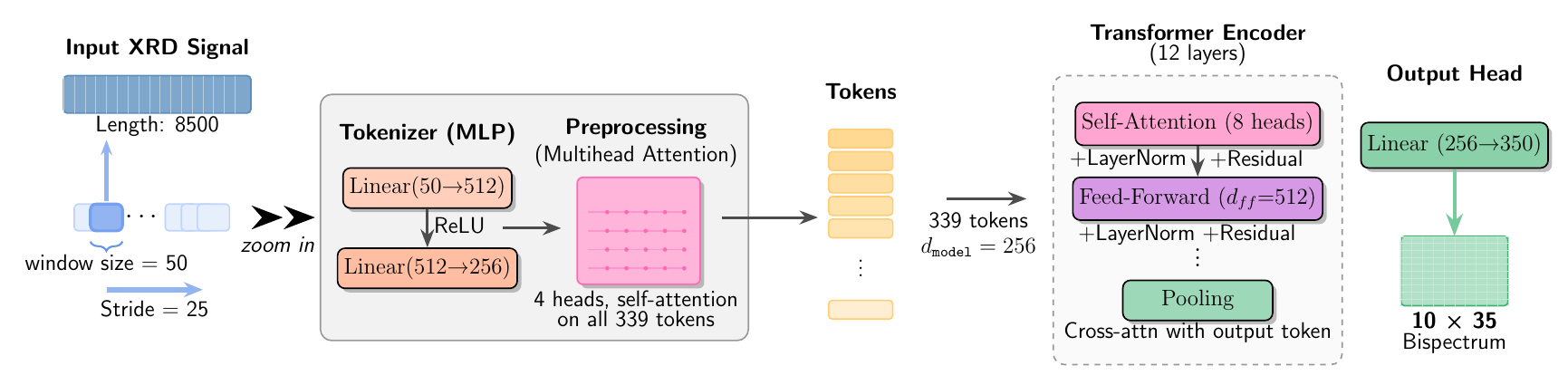}
    \caption{Transformer model architecture used for predicting the bispectrum from XRD input.}
    \label{fig:model}
\end{figure}

We use a transformer model to predict the lattice bispectrum from XRD patterns. We developed a hierarchical tokenization approach that segments the XRD pattern into overlapping local windows with a transformer-based architecture. This strategy succeeds at capturing relevant structural information across different scales while maintaining positional accuracy. After window extraction, each window is passed through a feedforward neural network. The resulting tokens are then refined with a multi-head self-attention mechanism. This methodology preserves positional precision that convolutional neural network (CNN)-based approaches lose through pooling, yet also captures long-range angular dependencies. Model hyperparameters and training details are in \Cref{tab:hyperparams} and the architecture is shown in \Cref{fig:model}.

\section{Results}

To test the utility of the lattice bispectrum as compared to directly predicting the lattice parameters, we first focus on the MP-20 dataset.
In general, there is no widely established benchmarking dataset for the unit cell prediction problem, as different works tend to use augmented versions of the Materials Project, ICSD, or CSD databases. However, recent crystal structure generation studies \cite{xieCrystalDiffusionVariational2021, rieselCrystalStructureDetermination2024, liPowderDiffractionCrystal2025} have used the MP-20 dataset. MP-20 consists of structures from the Materials Project with up to 20 atoms per unit cell, containing 45,231 inorganic materials. This dataset has the advantage of having predefined train, validation, and testing splits from \cite{xieCrystalDiffusionVariational2021}, making it suitable for benchmarking. We further explore the impact of data augmentation by incorporating physics-based XRD augmentations into the MP-20 dataset as described in \Cref{sec:mp20aug}. We then consider larger systems by using the full Materials Project dataset with the same physics-based augmentations (\Cref{sec:mpfull}). In each section, we assess the advantages of the bispectrum lattice representation vs. directly predicting lattice parameters by using the same model architecture.\footnote{with either an output dimension of 6 for lattice parameters or $\text{dim}(\mathbf{B})$ for the bispectrum.}. Results per dataset are shown in \Cref{tab:synth_results}.

Note that the bispectrum inversion procedure yields a set of vectors that span the lattice. Thus, we perform postprocessing steps using \texttt{pymatgen} and \texttt{cctbx} after the inversion in order to fairly compare our results to the true lattice. Both the true and predicted lattice are passed to \texttt{cctbx}, which finds the best-matching canonical representation. Lattice parameter errors are computed between the canonicalized true lattice for the direct prediction and bispectrum inversion. This ensures that errors are computed in a consistent setting.

We analyze direct vs. bispectrum model performance on structures with dominant zones in the Materials Project database in \Cref{sec:modelinterp}. The performance of each method is compared to benchmark predictions from Crystalyze and from a newer AlphaDiffract model \cite{andrejevicAlphaDiffractAutomatedCrystallographic2026} on RRUFF experimental data in \Cref{sec:rruff}.

\begin{table}[h]
\centering
\caption{MAE and MAPE per dataset for bispectrum + inversion and directly predicting lattice parameters. MP-20 and MP-20 aug. are evaluated on the MP-20 test set. MP-Full and MP-Full aug. are evaluated on the MP-Full test set (so this represents a more difficult learning problem with a larger set of diverse structures). Note that the evaluations in this table are done using synthetic patterns. The best method for each dataset is in bold.}
\resizebox{\columnwidth}{!}{
\begin{tabular}{llcccccc}
\toprule
Training data (num of pats.) & Method & Length MAE (\AA) & Length MAPE (\%) & Angle MAE ($^\circ$) & Angle MAPE (\%) & Volume MAPE (\%) \\
\midrule
MP-20 (27K) & Direct & 0.69 & 11.08 & 10.77 & 12.15 & 24.86 \\
MP-20 (27K) & Bispec + inv. & \textbf{0.18} & \textbf{2.43} & \textbf{2.19} & \textbf{2.45} & \textbf{5.15} \\
\midrule
MP-20 Aug (270K) & Direct & 0.80 & 13.20 & 12.36 & 13.33 & 27.96 \\
MP-20 Aug (270K) & Bispec + inv. & \textbf{0.23} & \textbf{3.19} & \textbf{2.58} & \textbf{2.90} & \textbf{6.61} \\
\midrule
MP-Full (93K) & Direct & 1.24 & 15.33 & 10.20 & 11.21 & 33.95 \\
MP-Full (93K) & Bispec + inv. & \textbf{0.44} & \textbf{4.46} & \textbf{3.92} & \textbf{4.41} & \textbf{10.09} \\
\midrule
MP-Full Aug (930K) & Direct & 1.27 & 15.75 & 9.87 & 10.91 & 33.67 \\
MP-Full Aug (930K) & Bispec + inv. & \textbf{0.62} & \textbf{6.35} & \textbf{5.27} & \textbf{5.94} & \textbf{14.25} \\
\bottomrule
\end{tabular}
}
\label{tab:synth_results}
\end{table}

\subsection{MP-20}\label{sec:mp20}
\begin{figure}[htbp]
    \centering
    \includegraphics[width=1.0\textwidth]{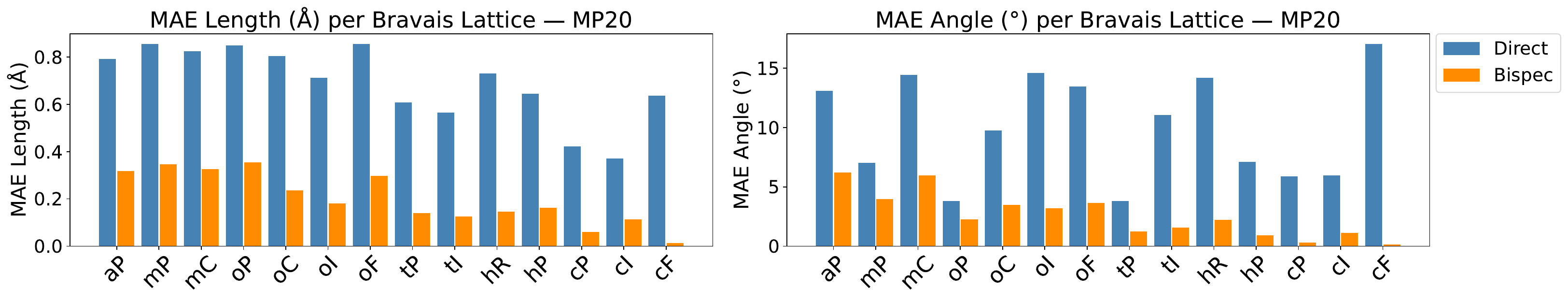}
    \caption{MAE for bispectrum + inversion compared to direct predictions per bravais lattice for MP-20.
    }
    \label{fig:mp20_res}
\end{figure}

In alignment with previous work \cite{chitturiAutomatedPredictionLattice2021, rieselCrystalStructureDetermination2024}, we report the mean absolute percentage error (MAPE) for angles, lengths, and unit cell volumes, defined as
\begin{equation}
    \text{MAPE}=\frac{|\text{true}-\text{pred}|}{\text{true}} \times 100
 \end{equation}
as well as the mean absolute error (MAE) per Bravais lattice.
We first consider the MP-20 dataset used in \cite{xieCrystalDiffusionVariational2021}, with the same train/test/validation splits used. The bispectrum prediction outperforms directly predicting the lattice parameters across Bravais lattices, see \Cref{fig:mp20_res}, mostly likely due to the fact that the bispectrum more strictly obeys symmetry constraints. For example, without training a model for separate crystal systems as in \cite{chitturiAutomatedPredictionLattice2021,shu2025}, the direct predictions may yield slightly different angle or length values that do not align with the correct Bravais lattice. This effect is not observed in the bispectrum prediction, where angle predictions are symmetry constrained and snap to the correct unit cell. See \Cref{supp_info:mp20_res} for additional results.

\subsubsection{Augmented MP-20}\label{sec:mp20aug}
In order to generalize to experimental XRD patterns, we augmented synthetic XRD patterns from MP-20 to mimic experimental data. We follow \cite{salgado2023automated, szymanskiProbabilisticDeepLearning2021,rieselCrystalStructureDetermination2024} which provide insight into physically-informed data augmentation of synthetic XRD patterns. The resulting augmented spectra sample possible changes in peak positions, intensities, and widths. (1) Shifts in peak positions ($2\theta$) were created using strain tensors for small distortions that preserve the space group of the parent structure. Modified unit cells were created with up to $\pm 4\%$ strain. (2) Peak intensities were varied through randomly selecting Miller indices ($hkl$) and then scaling peak intensities by $50\%$ of original values. (3) Peak widths were augmented through varying the Caglioti parameters $U,V,W$ (see \ref{supp_info:aug}), which parameterize the angular dependence of the full width at half maximum (FWHM) of diffraction peaks via $\text{FWHM}^2 = U\tan^2\theta + V\tan\theta + W$~\cite{caglioti1958choice}. Each structure was augmented 3 times per method, providing a diverse set of plausible XRD patterns. Following \cite{rieselCrystalStructureDetermination2024}, background noise sampled from a Gamma distribution is also added dynamically during training. The training set then has 271,360 examples instead of 27,136 as with MP-20. Note that data augmentation is applied to the training set, but the model is evaluated on the same testing set as MP-20 to ensure a fair comparison. Full details are in the supplementary material~\ref{supp_info:aug}. We again observe that the bispectrum outperforms the direct prediction method, yet there is a closer discrepancy.

\begin{figure}[htbp]
    \centering
    \includegraphics[width=1.0\textwidth]{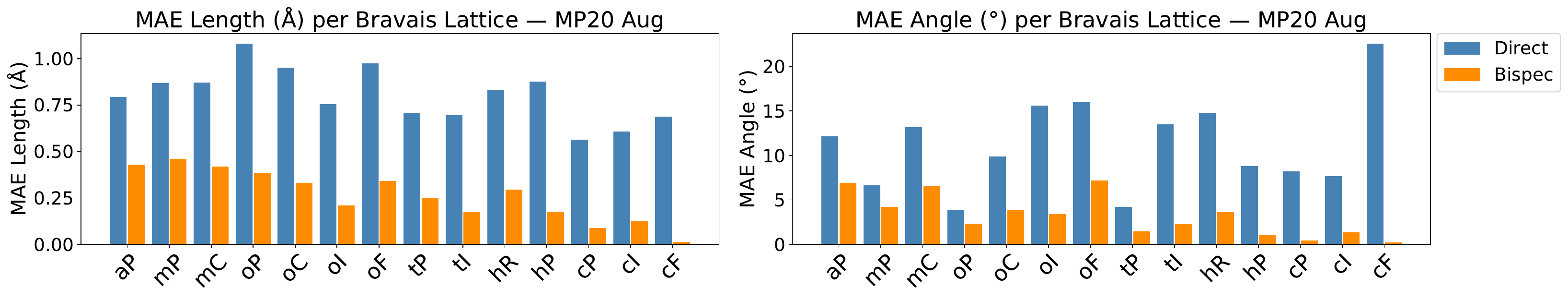}
    \caption{MAE for bispectrum + inversion compared to direct predictions per bravais lattice for MP-20 augmented.
    }
    \label{fig:mp20aug_res}
\end{figure}

\begin{figure}[htbp]
    \centering
    \begin{subfigure}[t]{0.8\textwidth}
        \centering
        \includegraphics[width=\textwidth]{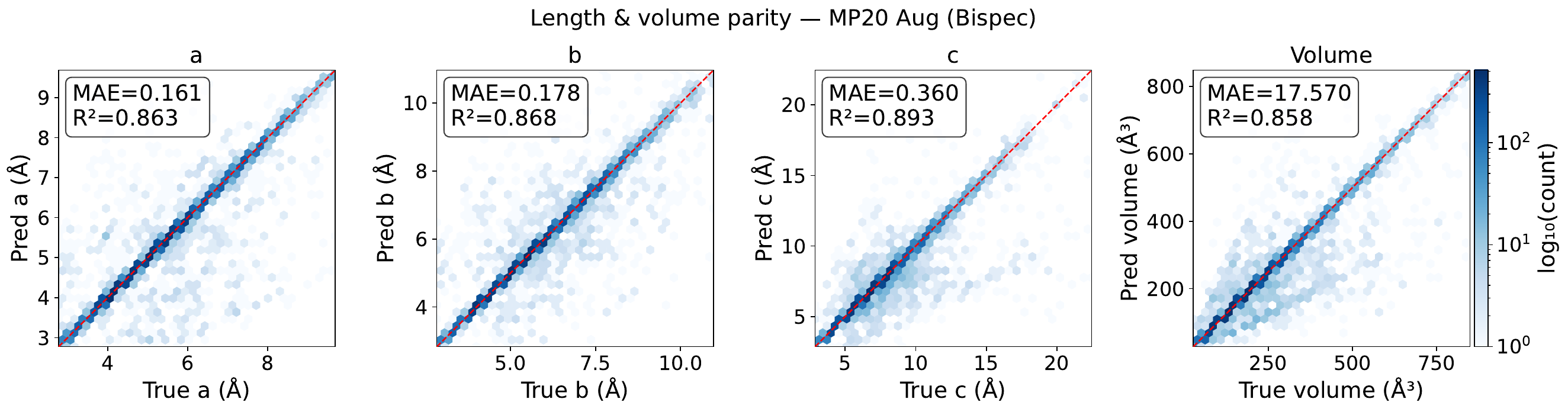}
    \end{subfigure}
    \hfill
    \begin{subfigure}[t]{0.8\textwidth}
        \centering
        \includegraphics[width=\textwidth]{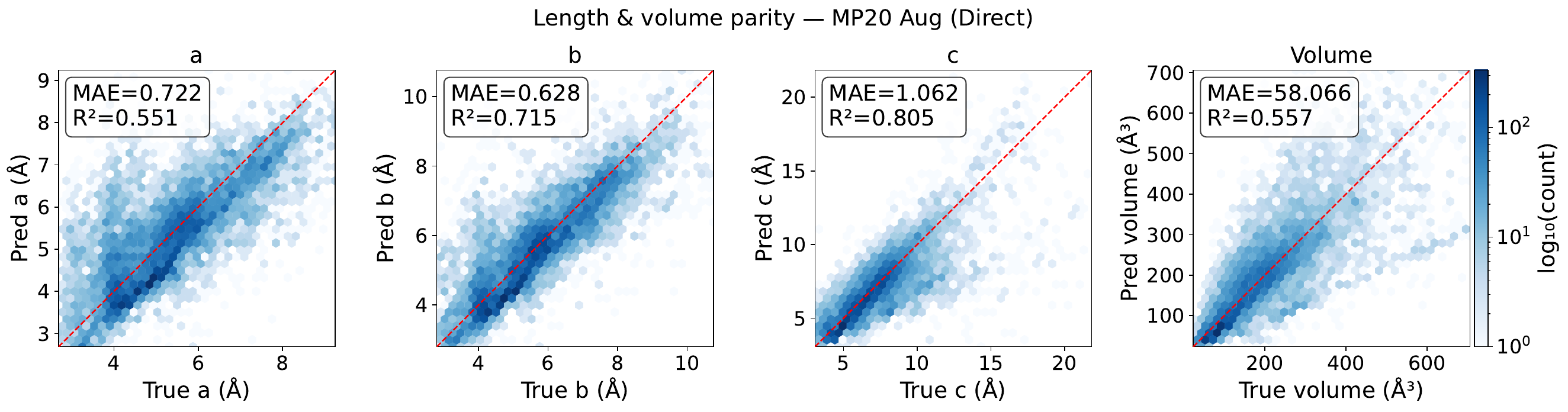}
    \end{subfigure}
    \caption{Predicted vs. true for $a,b,c$ and volume for bispec + inversion and directly predicting parameters using the MP20 augmented dataset.}
    \label{fig:pred_true_parity_mp20aug}
\end{figure}

\begin{figure}[htbp]
    \centering
    \begin{subfigure}[t]{0.7\textwidth}
        \centering
        \includegraphics[width=\textwidth]{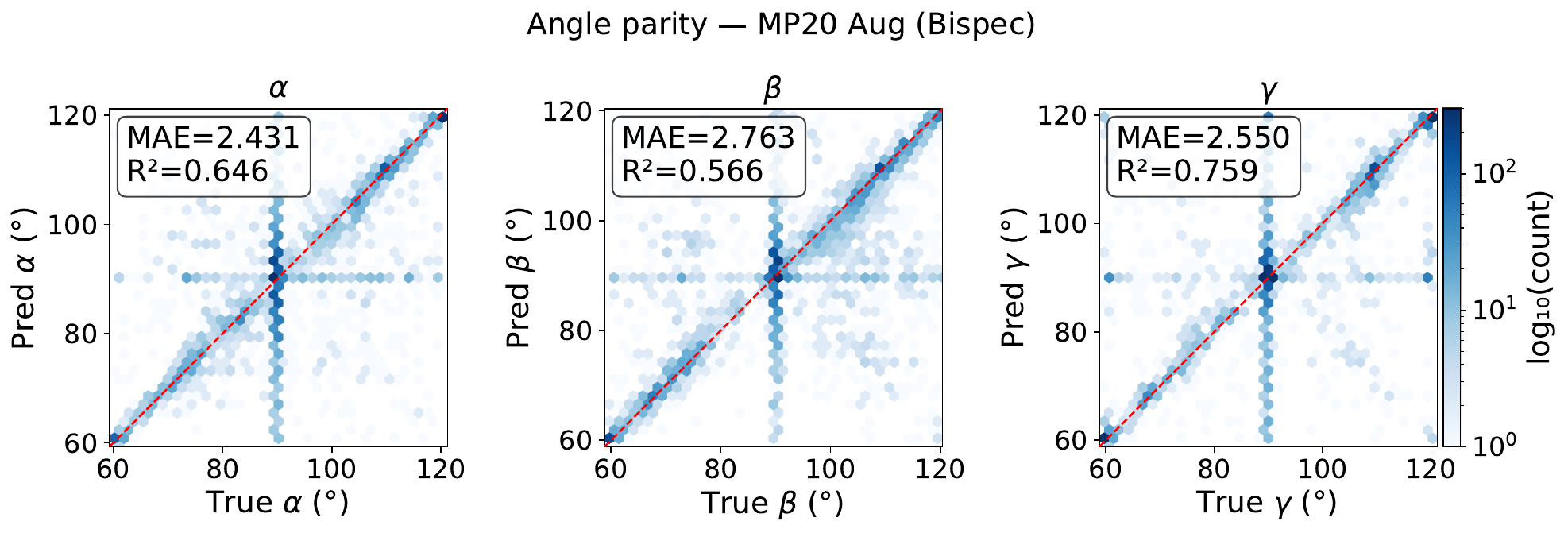}
    \end{subfigure}
    \hfill
    \begin{subfigure}[t]{0.7\textwidth}
        \centering
        \includegraphics[width=\textwidth]{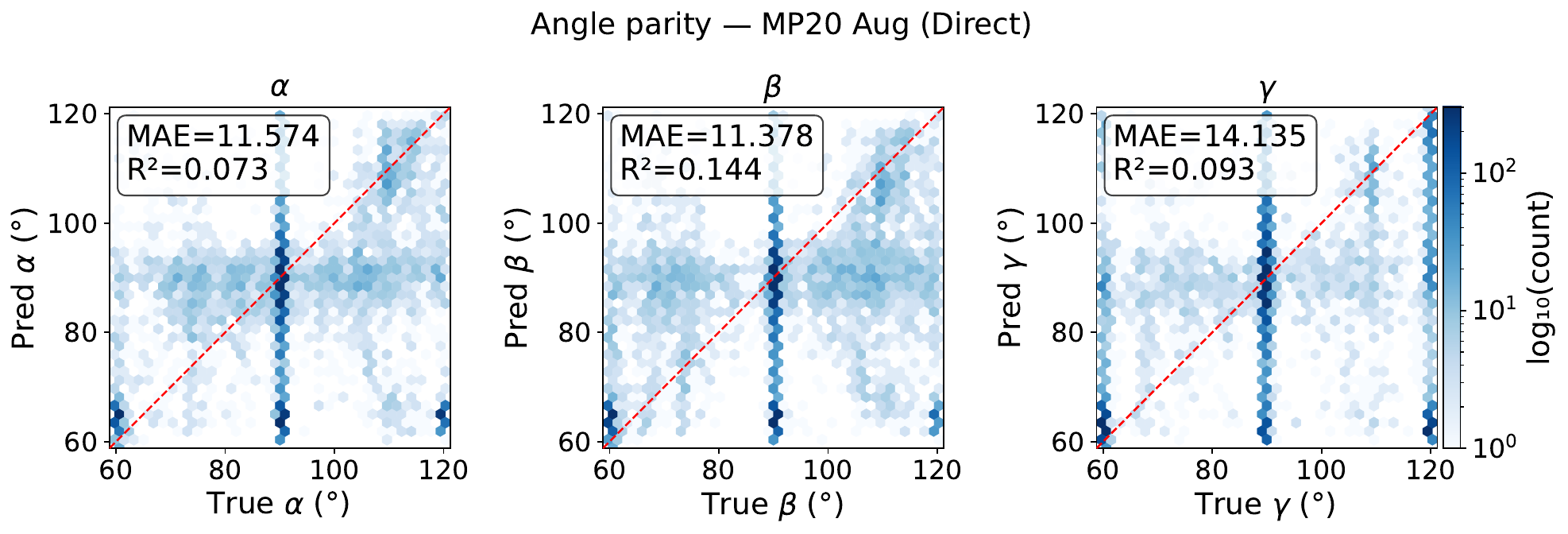}
    \end{subfigure}
    \caption{\revised{Predicted vs. true for $\alpha,\beta,\gamma$ for bispec + inversion and directly predicting parameters using the MP20 augmented dataset.}}
    \label{fig:pred_true_parity_angles_mp20aug}
\end{figure}

\subsection{Augmented Full Materials Project}\label{sec:mpfull}

Previous ML-XRD studies \cite{guoInitioStructureSolutions2025,rieselCrystalStructureDetermination2024} have rarely used the full Materials Project dataset. While the full Materials Project is invariably more complex than MP-20, we hypothesize that the greater diversity in crystal systems, unit cell sizes, and compositional complexity will lead to better generalization to structures with dominant zones. We split by reduced chemical formula into a 60\% training, 20\% validation, and 20\% test split. \revised{A data leakage audit based on XRD-pattern similarity is given in \Cref{supp_info:dataleak}.} We apply the same physics-based augmentations as in \Cref{sec:mp20aug} to the training set, resulting in a 930k pattern augmented dataset. This is a significantly more diverse dataset than MP-20 and thus represents a more difficult prediction problem. Nonetheless, we observe the same trend with the bispectrum generally outperforming the direct predictions. Angle predictions are more difficult than length predictions with more diversity in unit cells, although this trend is not uniform across Bravais lattices, see \Cref{fig:mpfullaug_threshold}. See \Cref{supp_info:mpfull_aug_res} for additional per-Bravais lattice MAE and parity plots comparing bispectrum and direct prediction. \Cref{supp_info:mpfull_res} additionally contains results training on the full Materials Project dataset without augmentation.

\begin{figure}[htbp]
    \centering
    \includegraphics[width=1.0\textwidth]{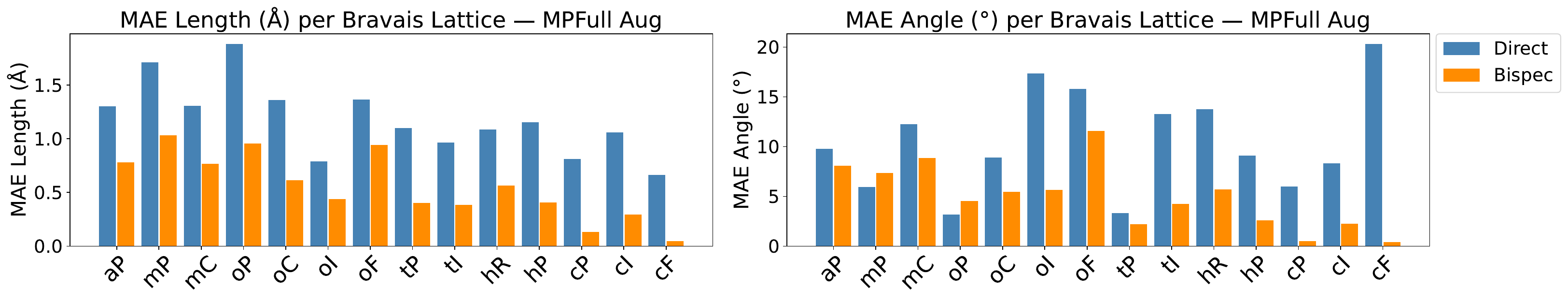}
    \caption{MAE for bispectrum + inversion compared to direct predictions per bravais lattice for MP full augmented.
    }
    \label{fig:mpfull_aug_res}
\end{figure}

\begin{figure}[htbp]
    \centering
    \begin{subfigure}[t]{0.8\textwidth}
        \centering
        \includegraphics[width=\textwidth]{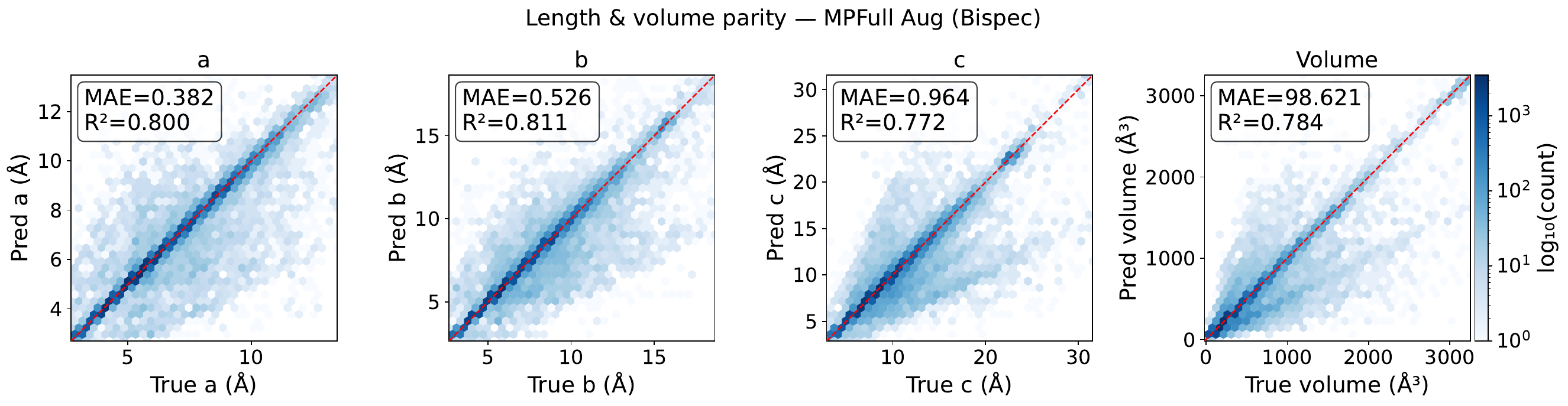}
    \end{subfigure}
    \hfill
    \begin{subfigure}[t]{0.8\textwidth}
        \centering
        \includegraphics[width=\textwidth]{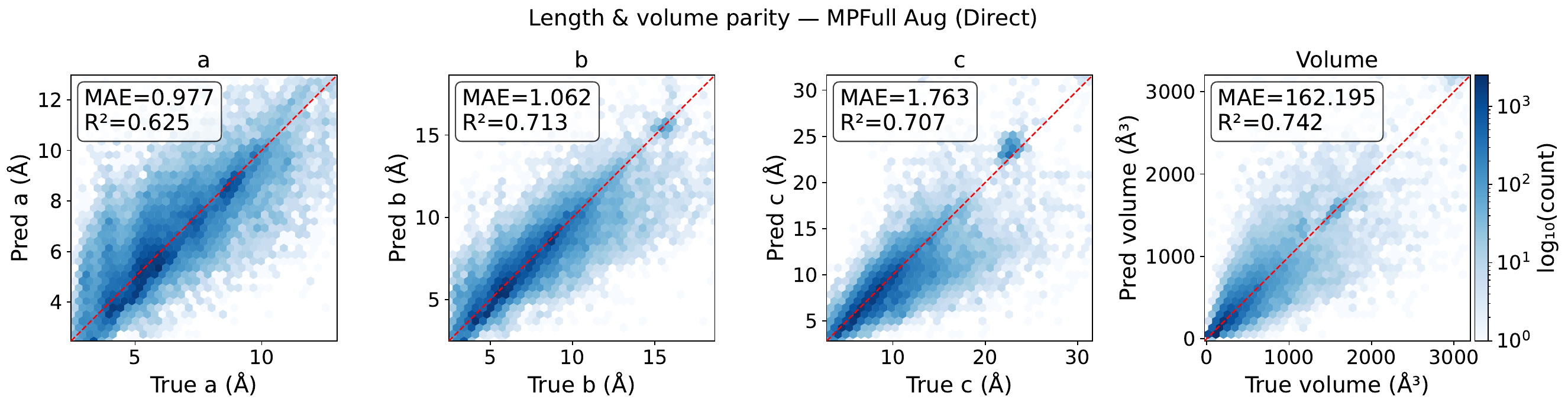}
    \end{subfigure}
    \caption{Predicted vs. true for $a,b,c$ and volume for bispec + inversion and directly predicting parameters using the MPFull augmented dataset.}
    \label{fig:pred_true_parity_mpfullaug}
\end{figure}

\begin{figure}[htbp]
    \centering
    \begin{subfigure}[t]{0.7\textwidth}
        \centering
        \includegraphics[width=\textwidth]{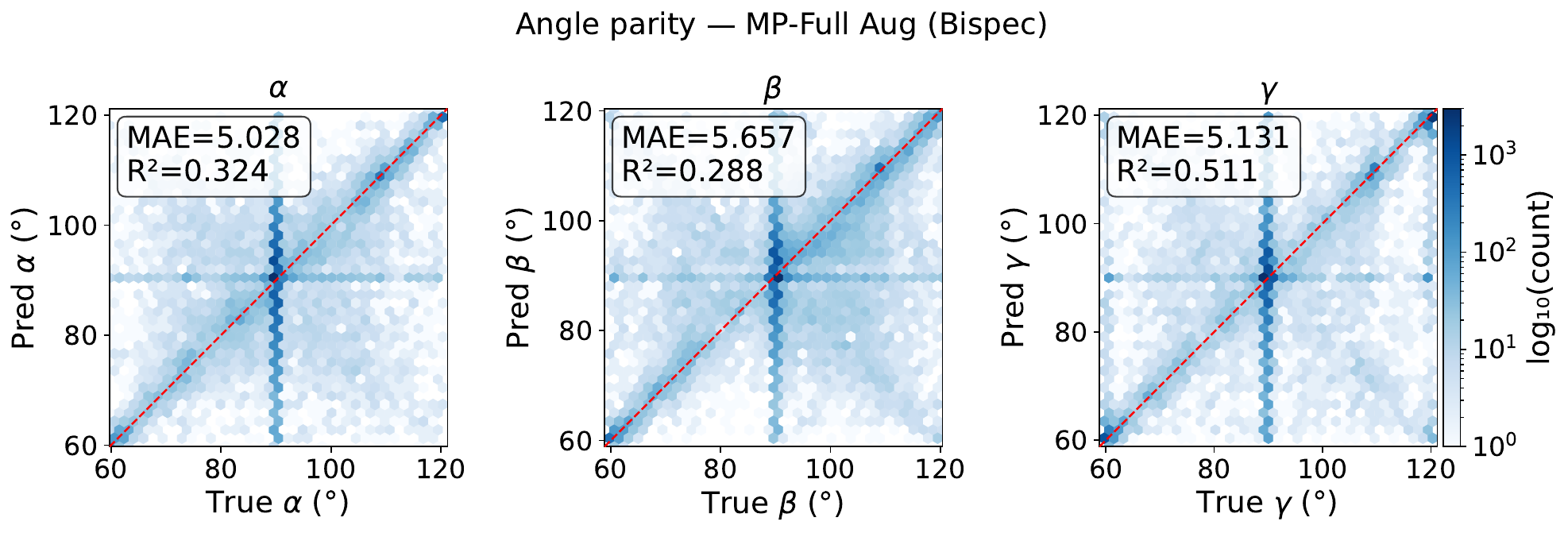}
    \end{subfigure}
    \hfill
    \begin{subfigure}[t]{0.7\textwidth}
        \centering
        \includegraphics[width=\textwidth]{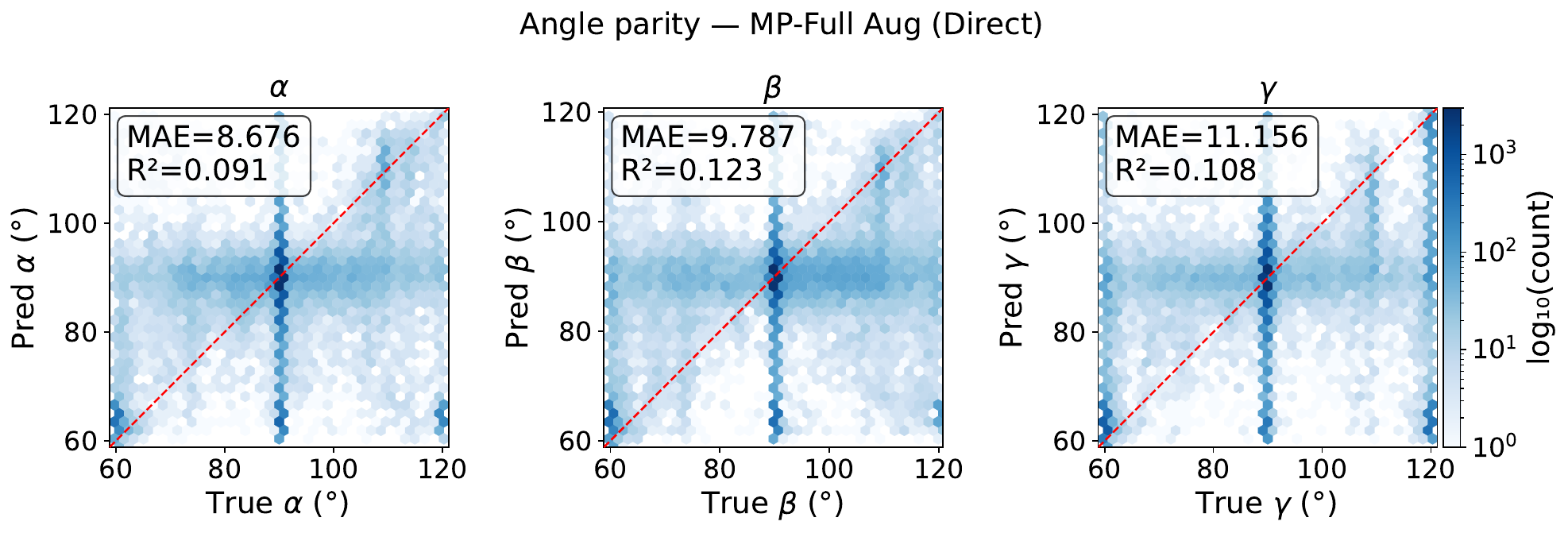}
    \end{subfigure}
    \caption{\revised{Predicted vs. true for $\alpha,\beta,\gamma$ for bispec + inversion and directly predicting parameters using the MPFull augmented dataset.}}
    \label{fig:pred_true_parity_angles_mpfullaug}
\end{figure}

\subsection{Dominant Zones}\label{sec:modelinterp}
A standard challenge in indexing pertains to dominant zone structures, in which most or all reflections can be indexed with two Miller indices (e.g. by $h0l$) \cite{Schlesinger2022}, usually where one axis in the unit cell is much larger than the others. We investigate the performance of the model trained on the full Materials Project data on dominant zone structures, comparing the bispectrum predictions with direct predictions of the lattice parameters. We define dominant-zone patterns as powder diffraction patterns in which a large fraction of total intensity ($\geq$ 50\%) is indexed by reflections belonging to a single zero-index family (i.e., $h=0$, $k=0$, or $l=0$). This provides a simple measure of indexing anisotropy in reciprocal space, capturing cases where diffraction intensity is strongly concentrated within coordinate-aligned Miller-index subspaces such as $00l$, $h0l$, or $hk0$. Note that we use a diffraction-based definition rather than considering real space lattice anisotropy as lattice recovery from XRD patterns depends on reciprocal-space indexing. Using this criteria, 1,100 out of 30,000 structures in the test set were identified as dominant zone structures.

\begin{figure}[htbp]
     \centering
     \includegraphics[width=0.5\textwidth]{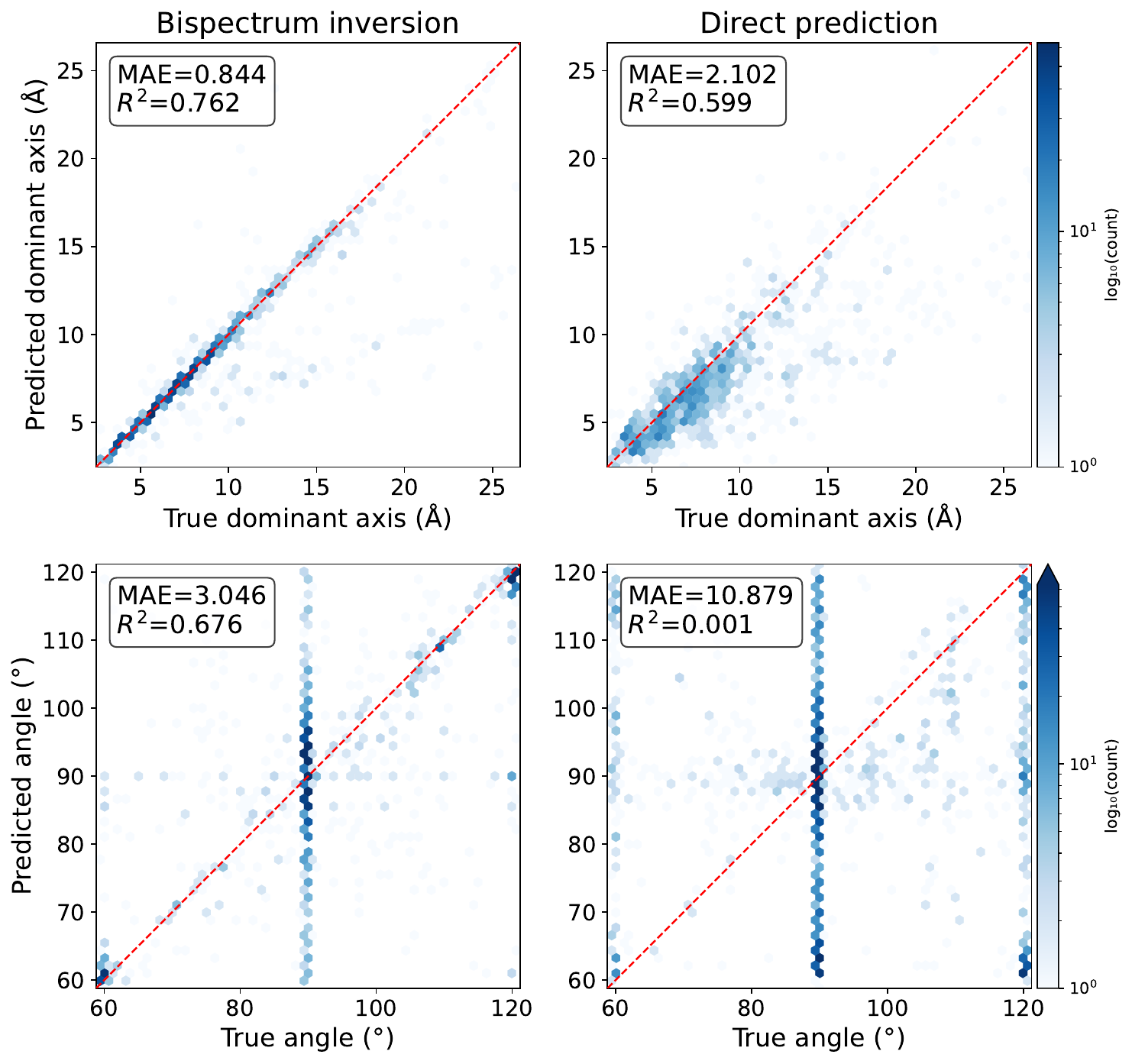}
     \caption{\revised{Parity plots for dominant zone structures (MP-Full test set, MP-Full Aug model), for bispectrum inversion (left) and direct lattice parameter prediction (right). Top: predicted vs.\ true dominant (longest) axis. Bottom: predicted vs.\ true lattice angles ($\alpha$, $\beta$, $\gamma$ pooled); the columns at 90$^\circ$ and 120$^\circ$ are angles fixed by symmetry, which bispectrum inversion largely preserves while direct prediction scatters them over 60--120$^\circ$.}}
     \label{fig:dominant_zones}
 \end{figure}

We find that the bispectrum prediction method performs much better on dominant zone structures, see \Cref{fig:dominant_zones}. For instance, 76.6\% of dominant zone structures have both length and angle MAPE less than 5\% when predicting the lattice bispectrum. However, this is only the case for 5.8\% of the direct parameter predictions. We hypothesize this advantage stems from the symmetry preserving properties of the lattice bispectrum. Dominant zone structures concentrate diffraction information into a reduced subspace of Miller indices, leaving the remaining lattice geometry underconstrained by the pattern. Direct prediction must resolve this ambiguity within an unconstrained six-parameter space, whereas bispectrum inversion benefits from a representation that already encodes the lattice's symmetry constraints.

\begin{table}[h]
\centering
\caption{Dominant zone success metrics for direct prediction and bispec + inv. Analysis was done using the MPFull test set and the MPFull-Aug trained model.}
\resizebox{\columnwidth}{!}{
\begin{tabular}{lccc}
\toprule
Method & Frac. $\leq$ 5\% Length MAPE & Frac. $\leq$ 5\% Angle MAPE & Frac. both $\leq$ 5\% \\
\midrule
Direct & 9.6\% & 47.5\% & 5.8\% \\
Bispec + inv & \textbf{82.0\%} & \textbf{81.7\%} & \textbf{76.6\%} \\
\bottomrule
\end{tabular}
}
\label{tab:dominant_zone_res}
\end{table}

\subsection{Experimental Data \& Comparisons to Benchmarks}\label{sec:rruff}
\begin{figure}[htbp]
    \centering
    \includegraphics[width=0.8\textwidth]{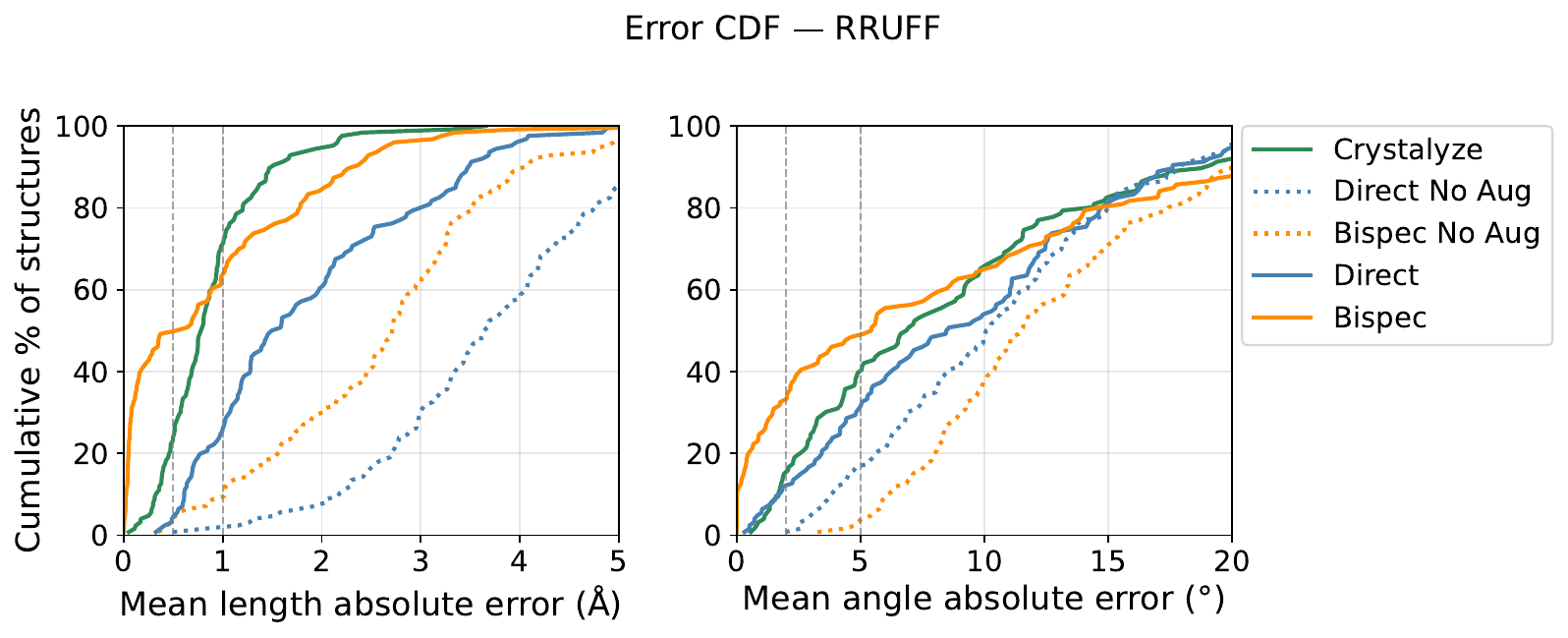}
    \caption{Cumulative distribution function for length and angle MAE for the RRUFF Crystalyze test set. The dashed lines correspond to the model trained on the MP20 dataset with no augmentation, and the solid lines correspond to the model trained on the MP20 dataset with augmentation as described in \Cref{sec:mp20aug}. Augmentation clearly improves performance.
    }
    \label{fig:crystalyze_ruff_cdf}
\end{figure}

Direct comparison across methods in this field is challenging due to inconsistent evaluation protocols, training datasets, and test splits; where possible, we have attempted to ensure fair comparison by rerunning baseline models with identical postprocessing. In line with prior works such as \cite{rieselCrystalStructureDetermination2024,choudhary2025,andrejevicAlphaDiffractAutomatedCrystallographic2026}, we consider the RRUFF dataset to test the generalization of our model to experimental data. First, we consider the same 134 RRUFF structures as used in Crystalyze \cite{rieselCrystalStructureDetermination2024} to have a consistent benchmark. We obtain the predicted lattice parameters from the MP-20 augmented model evaluated on the Crystalyze RRUFF dataset. Crystalyze is a structure generation model that predicts composition, lattice parameters, and number of atoms from an XRD pattern before denoising random atom positions to obtain a sampled structure. It is therefore a substantially more complex model than ours. We average output Crystalyze predictions over 5 sampled structures. As seen in \Cref{tab:rruff_results}, despite predicting only lattice parameters rather than full crystal structures, the bispectrum model achieves comparable performance to Crystalyze on experimental data. Additionally, our data augmentation strategy clearly aids in performance, as seen in \Cref{fig:crystalyze_ruff_cdf} and \Cref{supp_info:rruff_res}.

\begin{table}[h]
\centering
\caption{MAE and MAPE on RRUFF experimental data. We evaluate on the same 134-structure test set as Crystalyze, with all predictions postprocessed using cctbx to project onto the nearest valid Bravais lattice. Note these structures are filtered to be less than 20 atoms per unit cell.}
\resizebox{\columnwidth}{!}{
\begin{tabular}{llccccc}
\toprule
Training data & Method & Length MAE (\AA) & Length MAPE (\%) & Angle MAE ($^\circ$) & Angle MAPE (\%) & Volume MAPE (\%) \\
\midrule
MP-20 Aug & Crystalyze & \textbf{0.90} & 16.08 & 8.74 & 9.93 & 43.37 \\
MP-20 Aug (270K) & Direct (ours) & 1.86 & 30.92 & 9.46 & 10.29 & 94.82 \\
MP-20 Aug (270K) & Bispec + inv. (ours) & 0.91 & \textbf{14.41} & \textbf{8.37} & \textbf{8.94} & \textbf{35.72} \\
\bottomrule
\end{tabular}
}
\label{tab:rruff_results}
\end{table}

We further compare to AlphaDiffract performance on the AlphaDiffract RRUFF test set. This is a different set of 240 RRUFF structures than those used by Crystalyze, as AlphaDiffract does not limit training data to less than 20 atoms per unit cell \cite{andrejevicAlphaDiffractAutomatedCrystallographic2026}. As seen in \Cref{tab:rruff_alpha_results}, our model underperforms AlphaDiffract, which is expected given that AlphaDiffract trains on 31M patterns\textemdash 33x our training set size. Notably, training on 3\% of AlphaDiffract's data, our bispectrum model approached comparable length MAE (2.50 vs. 2.11 {}\AA) though the angle error gap remains larger.
\begin{table}[h]
\centering
\caption{MAE and MAPE on RRUFF AlphaDiffract experimental data. We evaluate on the same 240-structure test set as AlphaDiffract, with all predictions postprocessed using cctbx to project onto the nearest valid Bravais lattice.}
\resizebox{\columnwidth}{!}{
\begin{tabular}{llcccc}
\toprule
Training data & Method & Length MAE (\AA) & Length MAPE (\%) & Angle MAE ($^\circ$) & Angle MAPE (\%) \\
\midrule
AlphaDiffract MP-Full (31M) & AlphaDiffract & 2.11 & 23.50 & 2.72 & 2.91 \\
MP-Full Aug (930K) & Direct (ours) & 4.36 & 50.75 & 8.76 & 9.39 \\
MP-Full Aug (930K) & Bispec + inv. (ours) & 2.50 & 31.78 & 9.90 & 10.86 \\
\bottomrule
\end{tabular}
}
\label{tab:rruff_alpha_results}
\end{table}

We also explored scaling the model by 1.5x and 2x, finding small improvements on the RRUFF AlphaDiffract set ($\approx 1^\circ$ and $\approx 2^\circ$ in angle improvement). This suggests that model capacity is not the primary bottleneck, and that closing the sim-to-real gap requires larger or more diverse training data. On the RRUFF dataset, we observe that our data augmentation aids in performance, albeit not as much as Alpha Diffract, for example see \Cref{fig:crystalyze_ruff_cdf}.

\revised{We additionally evaluate both the bispectrum and direct prediction models on a curated subset of single-phase patterns from the CNRS/COD contribution to the opXRD database \cite{hollarekOpXRDOpenExperimental}. Both models show larger errors on this dataset than on RRUFF. Following standard practice \cite{rieselCrystalStructureDetermination2024}, our noise augmentation was fit to RRUFF. We hypothesize this differing performance is due to the background noise augmentation specifically being fit to the RRUFF dataset. Exploratory ablations support this: retraining with CNRS-informed background noise alone substantially reduces error, and peak-list inputs lower the average length error further. Direct prediction also rarely recovers the full cell: on every experimental set, fewer than $5\%$ of its predictions have both length and angle MAPE below $5\%$, whereas the bispectrum model trained with CNRS-informed noise reaches $25\%$ on these patterns. This is an important caution for practitioners that models trained with noise augmentation fit to one experimental dataset may not generalize to others. Identifying training procedures that generalize across experimental datasets is an important direction for future work (full results in \Cref{supp_info:cnrs_res}).}
\begin{table}[h]
\centering
\caption{\revised{MAE and MAPE on CNRS opXRD patterns with at most 20 atoms per primitive cell, for models trained on MP-20 Aug and postprocessed with cctbx. All rows use the same patterns. ``+ CNRS noise'': background noise informed by CNRS patterns during training; ``Peak list'': up to 30 extracted peaks as input. Best value per column in bold. Full results, including MP-Full Aug and RRUFF, are in \Cref{tab:cnrs_rruff}.}}
\resizebox{\columnwidth}{!}{
\begin{tabular}{llcccc}
\toprule
Training data & Method & Length MAE (\AA) & Length MAPE (\%) & Angle MAE ($^\circ$) & Angle MAPE (\%) \\
\midrule
MP-20 Aug Dense & Direct & 2.15 & 39.17 & \textbf{8.26} & \textbf{8.88} \\
MP-20 Aug Dense & Bispec + inv. & 2.54 & 43.36 & 13.85 & 14.95 \\
MP-20 Aug Dense + CNRS noise & Bispec + inv. & 1.62 & 26.60 & 10.87 & 11.67 \\
MP-20 Aug Peak list + CNRS noise & Direct & 1.59 & 26.37 & 10.66 & 11.40 \\
MP-20 Aug Peak list + CNRS noise & Bispec + inv. & \textbf{1.14} & \textbf{17.34} & 9.49 & 10.01 \\
\bottomrule
\end{tabular}
}
\label{tab:cnrs_small}
\end{table}

\section{Discussion}

We develop an invertible unit cell representation in reciprocal space. Crucially, this is invariant to choice of primitive unit cell. While previous ML approaches to predicting lattice parameters from XRD patterns have not taken the unit cell representation into account, we hypothesize that our geometry-informed representation will lead to more robust predictions. We test our hypothesis using a transformer-based architecture. We find that the bispectrum representation aids in predictions for both the MP-20 and the full Materials Project dataset for simulated XRD patterns.

The bispectrum acts to eliminate the ambiguity of reduced cell conventions. When predicting lattice parameters correctly, the model must implicitly learn a single consistent convention, whereas the training data may contain equivalent cells expressed differently. Second, the bispectrum naturally enforces symmetry constraints. As shown in \Cref{fig:bispec_lat_viz}, higher symmetry lattices produce sparser bispectra, and the inversion procedure recovers lattice parameters that respect these constraints without requiring separate per-crystal models as in prior work. In contrast, with direct prediction, angles or lengths are more likely to deviate slightly from their symmetry-constrained values (e.g. predicting cubic angles that are not exactly 90 degrees). Third, the bispectrum varies smoothly under continuous lattice deformations, whereas reduced cell parameters can change discontinuously at symmetry boundaries. This may produce a more learnable loss landscape for the neural network.

Several limitations should be noted. \revised{The inversion step adds computational overhead (a median of 88\,s per structure on a single CPU core with 15 starting candidates; see \Cref{supp:inversioncost}) but is parallelizable}. With more difficult experimental patterns that are out of distribution of the training data, bispectrum predictions are further from the ground truth. Thus, inversion becomes more difficult and can lead to degradation of the predicted lattice parameters (see \Cref{fig:bispec_lat_error}). Performance on experimental data remains below synthetic data, reflecting the well-documented simulation to real gap. Closing this gap will likely require a much larger dataset and augmentation strategies that better capture the full range of experimental artifacts, as in \cite{andrejevicAlphaDiffractAutomatedCrystallographic2026}, which is beyond the scope of this work. Additionally, our augmentation strategy, while physically motivated, is a simplification of real experimental variability (see \Cref{supp_info:augfuture}).

A general obstacle to progress in this area is the absence of a standardized augmentation and evaluation protocol. Throughout this work, we have tried to make comparisons fair by matching test splits and postprocessing pipelines to prior methods where feasible. We believe the community would benefit from a shared benchmark dataset analogous to MP-20 and collecting experimental data, much of which is scattered across personal repositories. This would ideally fix the training corpus, augmentation protocol, and evaluation split so that lattice-parameter prediction methods can be compared directly.

In terms of future work, the lattice bispectrum could serve as a useful intermediate representation for other crystallographic ML tasks. For example, structure generation models such as Crystalyze could potentially benefit from conditioning on the bispectrum rather than on raw lattice parameters. Additionally, we note that our model is predictive as it outputs lattice parameters given an XRD pattern. An interesting avenue for future work would be to extend our approach to be generative to sample candidate lattice parameters. This could represent a more faithful formulation of the XRD to lattice parameter problem, as it is fundamentally a one-to-many mapping. An interesting extension of the lattice bispectrum could be to incorporate the structure factor, yielding an atomic descriptor sensitive to both the lattice geometry and the atomic basis. This could aid in full structure determination beyond solely lattice parameters.

\begin{acknowledgements}
We thank Nina Andrejevic for helpful information about AlphaDiffract and Eric Riesel for discussions regarding Crystalyze. We also would like to thank Emily Oliphant and Ameya Daigavane for helpful discussions regarding our methodology.
\end{acknowledgements}

\begin{funding}
EH was supported by the U.S. Department
of Energy, Office of Science, Office of Advanced Scientific Computing Research, Department of Energy Computational Science Graduate Fellowship under Award Number DE-SC0024386. The National Institutes of Health, National Institute of General Medical Sciences supported DWMM and DWP via grant R24GM154040 to ASB. This research used resources of the National Energy Research Scientific Computing Center (NERSC), a Department of Energy User Facility using NERSC award ERCAP0033254 and ERCAP0036437. The authors also gratefully acknowledge the support by the Department of Energy Office of Science under the ICDI (Integrated Computational and Data Infrastructure) grant DE-SC0022215 and the Air Force Office of Scientific Research under Award No. FA9550-24-1-0067.

This report was prepared as an account of work sponsored by an agency of the
United States Government. Neither the United States Government nor any agency thereof, nor any of their employees, makes any warranty, express or implied, or assumes any legal liability or responsibility for the accuracy, completeness, or usefulness of any information, apparatus, product, or process disclosed, or represents that its use would not infringe privately owned rights. Reference herein to any specific commercial product, process, or service by trade name,
trademark, manufacturer, or otherwise does not necessarily constitute or imply its
endorsement, recommendation, or favoring by the United States Government or any agency thereof. The views and opinions of authors expressed herein do not necessarily state or reflect those of the United States Government or any agency thereof.
\end{funding}

\ConflictsOfInterest{There are no known conflicts of interests.
}

\DataAvailability{\revised{We have made the code available at \\ \url{https://github.com/atomicarchitects/bispectrum-xrd-ml} including the Materials Project identifiers for all dataset splits and the identifiers of the RRUFF/CNRS subsets.}
}

\bibliography{iucr} %

@incollection{Lafuente2015RRUFF,
  author    = {Lafuente, B. and Downs, R. T. and Yang, H. and Stone, N.},
  title     = {The power of databases: the {RRUFF} project},
  booktitle = {Highlights in Mineralogical Crystallography},
  editor    = {Armbruster, T. and Danisi, R. M.},
  year      = {2015},
  publisher = {W. De Gruyter},
  address   = {Berlin, Germany},
  pages     = {1--30}
}

@article{caglioti1958choice,
  title   = {Choice of collimators for a crystal spectrometer for neutron diffraction},
  author  = {Caglioti, G. and Paoletti, A. and Ricci, F. P.},
  journal = {Nuclear Instruments},
  volume  = {3},
  number  = {4},
  pages   = {223--228},
  year    = {1958},
  doi     = {10.1016/0369-643X(58)90029-X},
  publisher = {Elsevier}
}

@book{altmannPointGroupTheory1994,
  author    = {Altmann, Simon L. and Herzig, Peter},
  title     = {Point-Group Theory Tables},
  publisher = {Clarendon Press},
  address   = {Oxford},
  year      = {1994},
  isbn      = {9780198552260}
}

@ARTICLE{2020SciPy-NMeth,
  author  = {Virtanen, Pauli and Gommers, Ralf and Oliphant, Travis E. and
            Haberland, Matt and Reddy, Tyler and Cournapeau, David and
            Burovski, Evgeni and Peterson, Pearu and Weckesser, Warren and
            Bright, Jonathan and {van der Walt}, St{\'e}fan J. and
            Brett, Matthew and Wilson, Joshua and Millman, K. Jarrod and
            Mayorov, Nikolay and Nelson, Andrew R. J. and Jones, Eric and
            Kern, Robert and Larson, Eric and Carey, C J and
            Polat, {\.I}lhan and Feng, Yu and Moore, Eric W. and
            {VanderPlas}, Jake and Laxalde, Denis and Perktold, Josef and
            Cimrman, Robert and Henriksen, Ian and Quintero, E. A. and
            Harris, Charles R. and Archibald, Anne M. and
            Ribeiro, Ant{\^o}nio H. and Pedregosa, Fabian and
            {van Mulbregt}, Paul and {SciPy 1.0 Contributors}},
  title   = {{{SciPy} 1.0: Fundamental Algorithms for Scientific
            Computing in Python}},
  journal = {Nature Methods},
  year    = {2020},
  volume  = {17},
  pages   = {261--272},
  adsurl  = {https://rdcu.be/b08Wh},
  doi     = {10.1038/s41592-019-0686-2},
}

@online{nigamReconstructingLocalEnvironments2026,
  title = {Reconstructing Local Environments from Concise Atomistic Representations},
  author = {Nigam, Jigyasa and Phung, Tuong and Daigavane, Ameya and Tehrani, Aria Mansouri and Smidt, Tess},
  date = {2026-07-22},
  url = {https://arxiv.org/abs/2607.20338v1},
  urldate = {2026-07-23},
  year = {2026},
  langid = {english},
  organization = {arXiv.org}
}

@article{Schlesinger2022,
  author    = {Carina Schlesinger and Arnd Fitterer and Christian Buchsbaum and Stefan Habermehl and Michele R. Chierotti and Carlo Nervi and Martin U. Schmidt},
  title     = {Ambiguous structure determination from powder data: four different structural models of 4,11-difluoroquinacridone with similar X-ray powder patterns, fit to the PDF, SSNMR and DFT-D},
  journal   = {IUCrJ},
  year      = {2022},
  volume    = {9},
  number    = {4},
  pages     = {406--424},
  doi       = {10.1107/S2052252522004237},
  issn      = {2052-2525},
  url       = {https://doi.org/10.1107/S2052252522004237},
  affiliation = {aInstitute of Inorganic and Analytical Chemistry, Johann Wolfgang Goethe University, Frankfurt, Germany; bDepartment of Chemistry and NIS Centre, University of Torino, Italy}
}

@article{Harris2022,
  author    = {Kenneth D. M. Harris},
  title     = {Circumventing a challenging aspect of crystal structure determination from powder diffraction data},
  journal   = {Acta Crystallographica Section B: Structural Science, Crystal Engineering and Materials},
  year      = {2022},
  volume    = {78},
  number    = {2},
  pages     = {96--99},
  doi       = {10.1107/S2052520622003717},
  url       = {https://doi.org/10.1107/S2052520622003717}
}

@article{david2008structure,
  title={Structure determination from powder diffraction data},
  author={David, W. I. F. and Shankland, K.},
  journal={Acta Crystallographica Section A: Foundations and Advances},
  volume={64},
  number={1},
  pages={52--64},
  year={2008},
  publisher={International Union of Crystallography},
  doi={10.1107/S0108767307064252},
  url={https://doi.org/10.1107/S0108767307064252}
}

@article{chitturiAutomatedPredictionLattice2021,
	title = {Automated prediction of lattice parameters from {X}-ray powder diffraction patterns},
	volume = {54},
	copyright = {https://creativecommons.org/licenses/by/4.0/},
	issn = {1600-5767},
	url = {https://journals.iucr.org/j/issues/2021/06/00/vb5020/},
	doi = {10.1107/S1600576721010840},
	language = {en},
	number = {6},
	urldate = {2023-01-17},
	journal = {Journal of Applied Crystallography},
	author = {Chitturi, S. R. and Ratner, D. and Walroth, R. C. and Thampy, V. and Reed, E. J. and Dunne, M. and Tassone, C. J. and Stone, K. H.},
	month = dec,
	year = {2021},
	note = {Number: 6
Publisher: International Union of Crystallography},
	pages = {1799--1810},
}

@article{liMlatticeabcGenericLattice2021,
	title = {Mlatticeabc: {Generic} {Lattice} {Constant} {Prediction} of {Crystal} {Materials} {Using} {Machine} {Learning}},
	copyright = {© 2021 The Authors. Published by American Chemical Society},
	shorttitle = {Mlatticeabc},
	url = {https://pubs.acs.org/doi/full/10.1021/acsomega.1c00781},
	doi = {10.1021/acsomega.1c00781},
	language = {en},
	urldate = {2023-01-27},
	journal = {ACS Omega},
	author = {Li, Yuxin and Yang, Wenhui and Dong, Rongzhi and Hu, Jianjun},
	month = apr,
	year = {2021},
	note = {Publisher: American Chemical Society},
}

@article{andrewsSpaceLatticeRepresentation2019,
	title = {A space for lattice representation and clustering},
	volume = {75},
	copyright = {https://creativecommons.org/licenses/by/4.0/},
	issn = {2053-2733},
	url = {https://journals.iucr.org/a/issues/2019/03/00/ae5061/},
	doi = {10.1107/S2053273319002729},
	language = {en},
	number = {3},
	urldate = {2023-03-23},
	journal = {Acta Crystallographica Section A: Foundations and Advances},
	author = {Andrews, L. C. and Bernstein, H. J. and Sauter, N. K.},
	month = may,
	year = {2019},
	note = {Number: 3
Publisher: International Union of Crystallography},
	pages = {593--599},
}

@misc{andrewsMeasuringLattices2023,
	title = {Measuring {Lattices}},
	url = {https://arxiv.org/abs/2302.10240v1},
	language = {en},
	urldate = {2023-03-23},
	journal = {arXiv.org},
	author = {Andrews, Lawrence C. and Bernstein, Herbert J.},
	month = jan,
	year = {2023},
}

@article{corrieroCrystalMELANewCrystallographic2023,
	title = {\textit{{CrystalMELA}} : a new crystallographic machine learning platform for crystal system determination},
	volume = {56},
	issn = {1600-5767},
	shorttitle = {\textit{{CrystalMELA}}},
	url = {https://scripts.iucr.org/cgi-bin/paper?S1600576723000596},
	doi = {10.1107/S1600576723000596},
	language = {en},
	number = {2},
	urldate = {2023-06-08},
	journal = {Journal of Applied Crystallography},
	author = {Corriero, Nicola and Rizzi, Rosanna and Settembre, Gaetano and Del Buono, Nicoletta and Diacono, Domenico},
	month = apr,
	year = {2023},
	pages = {409--419},
}

@article{leeDeepLearningApproach2023,
	title = {A {Deep} {Learning} {Approach} to {Powder} {X}-{Ray} {Diffraction} {Pattern} {Analysis}: {Addressing} {Generalizability} and {Perturbation} {Issues} {Simultaneously}},
	issn = {2640-4567},
	shorttitle = {A {Deep} {Learning} {Approach} to {Powder} {X}-{Ray} {Diffraction} {Pattern} {Analysis}},
	url = {https://onlinelibrary.wiley.com/doi/10.1002/aisy.202300140},
	doi = {10.1002/aisy.202300140},
	language = {en},
	urldate = {2023-06-12},
	journal = {Advanced Intelligent Systems},
	author = {Lee, Byung Do and Lee, Jin-Woong and Ahn, Junuk and Kim, Seonghwan and Park, Woon Bae and Sohn, Kee-Sun},
	month = jun,
	year = {2023},
	note = {Publisher: John Wiley \& Sons, Ltd},
	pages = {2300140},
}

@misc{nigamCompletenessAtomicStructure2023,
	title = {Completeness of {Atomic} {Structure} {Representations}},
	url = {https://arxiv.org/abs/2302.14770v1},
	language = {en},
	urldate = {2023-07-12},
	journal = {arXiv.org},
	author = {Nigam, Jigyasa and Pozdnyakov, Sergey N. and Huguenin-Dumittan, Kevin K. and Ceriotti, Michele},
	month = feb,
	year = {2023},
}

@misc{pozdnyakovCompletenessAtomicStructure2020,
	title = {On the {Completeness} of {Atomic} {Structure} {Representations}},
	url = {https://arxiv.org/abs/2001.11696v2},
	language = {en},
	urldate = {2023-07-12},
	journal = {arXiv.org},
	author = {Pozdnyakov, Sergey N. and Willatt, Michael J. and Bartók, Albert P. and Ortner, Christoph and Csányi, Gábor and Ceriotti, Michele},
	month = jan,
	year = {2020},
	doi = {10.1103/PhysRevLett.125.166001},
}

@article{suzukiSymmetryPredictionKnowledge2020,
	title = {Symmetry prediction and knowledge discovery from {X}-ray diffraction patterns using an interpretable machine learning approach},
	volume = {10},
	copyright = {2020 The Author(s)},
	issn = {2045-2322},
	url = {https://www.nature.com/articles/s41598-020-77474-4},
	doi = {10.1038/s41598-020-77474-4},
	language = {en},
	number = {1},
	urldate = {2025-07-02},
	journal = {Scientific Reports},
	author = {Suzuki, Yuta and Hino, Hideitsu and Hawai, Takafumi and Saito, Kotaro and Kotsugi, Masato and Ono, Kanta},
	month = dec,
	year = {2020},
	note = {Number: 1
Publisher: Nature Publishing Group},
	pages = {1--11},
}

@article{Coelho:to5164,
author = "Coelho, Alan A.",
title = "{An indexing algorithm independent of peak position extraction for X-ray powder diffraction patterns}",
journal = "Journal of Applied Crystallography",
year = "2017",
volume = "50",
number = "5",
pages = "1323--1330",
month = "Oct",
doi = {10.1107/S1600576717011359},
url = {https://doi.org/10.1107/S1600576717011359},
}

@article{rieselCrystalStructureDetermination2024,
  title = {Crystal {{Structure Determination}} from {{Powder Diffraction Patterns}} with {{Generative Machine Learning}}},
  author = {Riesel, Eric A. and Mackey, Tsach and Nilforoshan, Hamed and Xu, Minkai and Badding, Catherine K. and Altman, Alison B. and Leskovec, Jure and Freedman, Danna E.},
  year = {2024},
  month = sep,
  journal = {Journal of the American Chemical Society},
  publisher = {American Chemical Society},
  doi = {10.1021/jacs.4c10244},
  urldate = {2025-07-12},
  copyright = {{\copyright} 2024 American Chemical Society},
  langid = {english}
}

@article{szymanskiProbabilisticDeepLearning2021,
  title = {Probabilistic {{Deep Learning Approach}} to {{Automate}} the {{Interpretation}} of {{Multi-phase Diffraction Spectra}}},
  author = {Szymanski, Nathan J. and Bartel, Christopher J. and Zeng, Yan and Tu, Qingsong and Ceder, Gerbrand},
  year = {2021},
  month = may,
  journal = {Chemistry of Materials},
  publisher = {American Chemical Society},
  doi = {10.1021/acs.chemmater.1c01071},
  urldate = {2025-07-12},
  copyright = {{\copyright} 2021 American Chemical Society},
  langid = {english}
}

@article{bartokRepresentingChemicalEnvironments2013,
  title = {On Representing Chemical Environments},
  author = {Bart{\'o}k, Albert P. and Kondor, Risi and Cs{\'a}nyi, G{\'a}bor},
  year = {2013},
  month = may,
  journal = {Physical Review B},
  volume = {87},
  number = {18},
  pages = {184115},
  publisher = {American Physical Society},
  doi = {10.1103/PhysRevB.87.184115},
  urldate = {2025-07-18},
  langid = {english}
}

@article{geiger2022e3nn,
  title={e3nn: Euclidean neural networks},
  author={Geiger, Mario and Smidt, Tess},
  journal={arXiv preprint arXiv:2207.09453},
  year={2022}
}

@article{cobelliLocalInversionChemical2022,
  title = {Local Inversion of the Chemical Environment Representations},
  author = {Cobelli, Matteo and Cahalane, Paddy and Sanvito, Stefano},
  year = {2022},
  month = jul,
  journal = {Physical Review B},
  volume = {106},
  number = {3},
  pages = {035402},
  publisher = {American Physical Society},
  doi = {10.1103/PhysRevB.106.035402}
}

@ARTICLE{bispecinversionMRA,
  author={Bendory, Tamir and Boumal, Nicolas and Ma, Chao and Zhao, Zhizhen and Singer, Amit},
  journal={IEEE Transactions on Signal Processing}, 
  title={Bispectrum Inversion With Application to Multireference Alignment}, 
  year={2018},
  volume={66},
  number={4},
  pages={1037-1050},
  doi={10.1109/TSP.2017.2775591}}

@article{kakaralaBispectrumSourcePhaseSensitive2012,
  title = {The {{Bispectrum}} as a {{Source}} of {{Phase-Sensitive Invariants}} for {{Fourier Descriptors}}: {{A Group-Theoretic Approach}}},
  author = {Kakarala, Ramakrishna},
  year = {2012},
  month = nov,
  journal = {Journal of Mathematical Imaging and Vision},
  volume = {44},
  number = {3},
  pages = {341--353},
  issn = {1573-7683},
  doi = {10.1007/s10851-012-0330-6}
}

@INPROCEEDINGS{uniquebisinvert,
  author={Pinilla, Samuel and Mishra, Kumar Vijay and Sadler, Brian M.},
  booktitle={ICASSP 2023 - 2023 IEEE International Conference on Acoustics, Speech and Signal Processing (ICASSP)}, 
  title={Unique Bispectrum Inversion for Signals with Finite Spectral/Temporal Support}, 
  year={2023},
  volume={},
  number={},
  pages={1-5},
  doi={10.1109/ICASSP49357.2023.10095922}}

@misc{xieCrystalDiffusionVariational2021,
  title = {Crystal {{Diffusion Variational Autoencoder}} for {{Periodic Material Generation}}},
  author = {Xie, Tian and Fu, Xiang and Ganea, Octavian-Eugen and Barzilay, Regina and Jaakkola, Tommi},
  year = 2021,
  month = oct,
  journal = {arXiv.org},
  urldate = {2025-12-04},
  howpublished = {https://arxiv.org/abs/2110.06197v3},
  langid = {english}
}

@article{guoInitioStructureSolutions2025,
  title = {Ab Initio Structure Solutions from Nanocrystalline Powder Diffraction Data via Diffusion Models},
  author = {Guo, Gabe and Saidi, Tristan Luca and Terban, Maxwell W. and Valsecchi, Michele and Billinge, Simon J. L. and Lipson, Hod},
  year = 2025,
  month = nov,
  journal = {Nature Materials},
  volume = {24},
  number = {11},
  pages = {1726--1734},
  publisher = {Nature Publishing Group},
  issn = {1476-4660},
  doi = {10.1038/s41563-025-02220-y},
  urldate = {2025-12-04},
  copyright = {2025 The Author(s), under exclusive licence to Springer Nature Limited},
  langid = {english}
}

@article{liPowderDiffractionCrystal2025,
  title = {Powder Diffraction Crystal Structure Determination Using Generative Models},
  author = {Li, Qi and Jiao, Rui and Wu, Liming and Zhu, Tiannian and Huang, Wenbing and Jin, Shifeng and Liu, Yang and Weng, Hongming and Chen, Xiaolong},
  year = 2025,
  month = aug,
  journal = {Nature Communications},
  volume = {16},
  number = {1},
  pages = {7428},
  publisher = {Nature Publishing Group},
  issn = {2041-1723},
  doi = {10.1038/s41467-025-62708-8},
  urldate = {2025-12-04},
  copyright = {2025 The Author(s)},
  langid = {english}
}

@article{boultif1991,
issn = {0021-8898},
author = {Boultif, A and Louër, D},
address = {Copenhagen,},
journal = {Journal of applied crystallography.},
lccn = {68007471},
number = {6},
publisher = {Munksgaard International Booksellers and Publishers.},
title = {Indexing of powder diffraction patterns for low-symmetry lattices by the successive dichotomy method},
volume = {24},
year = {1991-12-01},
}

@article{de1957determination,
  title={On the determination of unit-cell dimensions from powder diffraction patterns},
  author={De Wolff, PM},
  journal={Acta Crystallographica},
  volume={10},
  number={9},
  pages={590--595},
  year={1957},
  publisher={International Union of Crystallography}
}

@article{Coelho2003,
author = "Coelho, A. A.",
title = "{Indexing of powder diffraction patterns by iterative use of singular value decomposition}",
journal = "Journal of Applied Crystallography",
year = "2003",
volume = "36",
number = "1",
pages = "86--95",
month = "Feb",
doi = {10.1107/S0021889802019878},
url = {https://doi.org/10.1107/S0021889802019878},
}

@article{Mighell_2000, title={Lattice metric singularities and their impact on the indexing of powder patterns}, volume={15}, DOI={10.1017/S0885715600010873}, number={2}, journal={Powder Diffraction}, author={Mighell, Alan D.}, year={2000}, pages={82–85}}

@Book{Dresselhaus2008,
  author    = {Dresselhaus, Mildred S. and Dresselhaus, Gene and Jorio, Ado},
  title     = {Group Theory: Application to the Physics of Condensed Matter},
  publisher = {Springer Berlin Heidelberg},
  address   = {Berlin, Heidelberg},
  year      = {2008},
  doi       = {10.1007/978-3-540-32899-5},
  url       = {https://doi.org/10.1007/978-3-540-32899-5}
}

@article{dong2021deep,
  title={A deep convolutional neural network for real-time full profile analysis of big powder diffraction data},
  author={Dong, Hongyang and Butler, Keith T and Matras, Dorota and Price, Stephen WT and Odarchenko, Yaroslav and Khatry, Rahul and Thompson, Andrew and Middelkoop, Vesna and Jacques, Simon DM and Beale, Andrew M and others},
  journal={npj Computational Materials},
  volume={7},
  number={1},
  pages={74},
  year={2021},
  publisher={Nature Publishing Group UK London}
}

@article{habershon2004powder,
  title={Powder diffraction indexing as a pattern recognition problem: a new approach for unit cell determination based on an artificial neural network},
  author={Habershon, Scott and Cheung, Eugene Y and Harris, Kenneth DM and Johnston, Roy L},
  journal={The Journal of Physical Chemistry A},
  volume={108},
  number={5},
  pages={711--716},
  year={2004},
  publisher={ACS Publications}
}

@article{segal2025loss,
  title={The Loss Landscape of Powder X-Ray Diffraction-Based Structure Optimization Is Too Rough for Gradient Descent},
  author={Segal, Nofit and Subramanian, Akshay and Li, Mingda and Miller, Benjamin Kurt and Gomez-Bombarelli, Rafael},
  journal={arXiv preprint arXiv:2512.04036},
  year={2025}
}

@article{salgado2023automated,
  title={Automated classification of big X-ray diffraction data using deep learning models},
  author={Salgado, Jerardo E and Lerman, Samuel and Du, Zhaotong and Xu, Chenliang and Abdolrahim, Niaz},
  journal={npj Computational Materials},
  volume={9},
  number={1},
  pages={214},
  year={2023},
  publisher={Nature Publishing Group UK London}
}

@article{shu2025,
author = {Shu, Ke and Gui, Dong-Yun and Yan, Wei-Xin and Wang, Chun-Hai},
title = {Machine Learning Tackles the Challenge of Powder X-ray Diffraction Indexing for All Crystal Systems},
journal = {Journal of Chemical Information and Modeling},
volume = {65},
number = {19},
pages = {10025-10036},
year = {2025},
doi = {10.1021/acs.jcim.5c01506},
    note ={PMID: 40980829}
}

@misc{andrejevicAlphaDiffractAutomatedCrystallographic2026,
  title = {{{AlphaDiffract}}: {{Automated Crystallographic Analysis}} of {{Powder X-ray Diffraction Data}}},
  shorttitle = {{{AlphaDiffract}}},
  author = {Andrejevic, Nina and Du, Ming and Sharma, Hemant and Horwath, James P. and Luo, Aileen and Yin, Xiangyu and Prince, Michael and Toby, Brian H. and Cherukara, Mathew J.},
  year = 2026,
  month = mar,
  journal = {arXiv.org},
  urldate = {2026-04-13}
}

@article{choudhary2025,
author = {Choudhary, Kamal},
title = {DiffractGPT: Atomic Structure Determination from X-ray Diffraction Patterns Using a Generative Pretrained Transformer},
journal = {The Journal of Physical Chemistry Letters},
volume = {16},
number = {8},
pages = {2110-2119},
year = {2025},
doi = {10.1021/acs.jpclett.4c03137},
    note ={PMID: 39976483}
}

@article{gomez-peraltaConvolutionalNeuralNetworks2023,
  title = {Convolutional {{Neural Networks}} to {{Assist}} the {{Assessment}} of {{Lattice Parameters}} from {{X-ray Powder Diffraction}}},
  author = {{G{\'o}mez-Peralta}, Juan Iv{\'a}n and Bokhimi, Xim and Quintana, Patricia},
  year = 2023,
  month = sep,
  journal = {The Journal of Physical Chemistry A},
  volume = {127},
  number = {36},
  pages = {7655--7664},
  publisher = {American Chemical Society},
  issn = {1089-5639},
  doi = {10.1021/acs.jpca.3c03860}
}

@article{gomezperalta2025insights,
  title = {Insights from the reciprocal space revealed by a convolutional neural network and transfer learning},
  author = {G{\'o}mez-Peralta, J. I. and Bokhimi, X. and Quintana-Owen, P.},
  journal = {Scripta Materialia},
  volume = {263},
  pages = {116697},
  year = {2025},
  issn = {1359-6462},
  doi = {10.1016/j.scriptamat.2025.116697},
  url = {https://www.sciencedirect.com/science/article/pii/S1359646225001605}
}

@article{Andrewsniggli,
  title = {The Geometry of {{Niggli}} Reduction: {{{\emph{BGAOL}}}} --Embedding {{Niggli}} Reduction and Analysis of Boundaries},
  author = {Andrews, Lawrence C. and Bernstein, Herbert J.},
  year = 2014,
  month = feb,
  journal = {Journal of Applied Crystallography},
  volume = {47},
  number = {1},
  pages = {346--359},
  doi = {10.1107/S1600576713031002}
}

@article{hollarekOpXRDOpenExperimental,
  title = {op{XRD}: {Open} {Experimental} {Powder} {X}-{Ray} {Diffraction} {Database}},
  shorttitle = {op{XRD}},
  author = {Hollarek, Daniel and Schopmans, Henrik and {\"O}streicher, Jona and Teufel, Jonas and Cao, Bin and Alwen, Adie and Schweidler, Simon and Singh, Mriganka and Kodalle, Tim and Hu, Hanlin and Heymans, Gregoire and Abdelsamie, Maged and Hardiagon, Arthur and Wieczorek, Alexander and Zhuk, Siarhei and Schwaiger, Ruth and Siol, Sebastian and Coudert, Fran{\c{c}}ois-Xavier and Wolf, Moritz and {Sutter-Fella}, Carolin M. and Breitung, Ben and Hodge, Andrea M. and Zhang, Tong-yi and Friederich, Pascal},
  doi = {10.1002/aidi.202500044},
  journal = {Advanced Intelligent Discovery},
  year = {2025},
  langid = {english}
}

@article{segal2026xrdiff,
  title={XRDiff: Crystal Structure Prediction from Powder X-Ray Diffraction Data Using Diffusion Models},
  author={Segal, Nofit and Li, Mingda and Miller, Benjamin Kurt and G{\'o}mez-Bombarelli, Rafael},
  journal={arXiv preprint arXiv:2606.14003},
  year={2026}
}

@misc{bendoryOrbitRecoverySpherical2025,
	title = {Orbit Recovery for Spherical Functions},
	url = {https://arxiv.org/abs/2508.02674v2},
	journal = {arXiv.org},
	author = {Bendory, Tamir and Edidin, Dan and Katz, Josh and Kreymer, Shay},
	month = jun,
	year = {2025},
}

\clearpage
\section*{Supplementary Information}
\renewcommand{\thesubsection}{S\arabic{subsection}}
\setcounter{subsection}{0}
\renewcommand{\thefigure}{S\arabic{figure}}
\setcounter{figure}{0}
\renewcommand{\thetable}{S\arabic{table}}
\setcounter{table}{0}
\renewcommand{\theequation}{S\arabic{equation}}
\setcounter{equation}{0}

\setcounter{topnumber}{4}
\setcounter{bottomnumber}{4}
\setcounter{totalnumber}{6}
\renewcommand{\topfraction}{0.9}
\renewcommand{\bottomfraction}{0.9}
\renewcommand{\textfraction}{0.07}
\renewcommand{\floatpagefraction}{0.7}
\makeatletter
\setlength{\@fptop}{0pt}
\setlength{\@fpsep}{12pt plus 1fil}
\setlength{\@fpbot}{0pt plus 1fil}
\makeatother

\addtocontents{toc}{\protect\setcounter{tocdepth}{3}}
\tableofcontents

\subsection{Model and Training Details}\label{supp_info:model}

Our tokenization module processes the XRD pattern through three stages. The first step is the local window extraction. The input XRD pattern is segmented into overlapping windows of size 50 with a stride of 25, producing 339 windows. The window size of 50 guarantees a peak will be fully captured in at least one of the windows. The overlap ensures that peak features near window boundaries are not lost. Subsequently, each window goes through a two-layer feedforward network that maps the 50-dimensional local pattern into a 256-dimensional token representation. This local processor learns  characteristic diffraction features such as peak widths and intensities. Finally, the local tokens are then refined through a multi-head self-attention mechanism of 4 heads that enables information exchange between tokens from different angular regions. This global processing stage allows the model to capture long-range dependencies in the XRD pattern.

The tokens are augmented with sinusoidal positional encodings to preserve positional information. This combination of local feature extraction and global attention maps the one-dimensional XRD input of length 8500 to a 339×256 token sequence, which is then processed by the downstream transformer encoder to predict the bispectrum coefficients.

Each model is trained with a learning rate of 2e-4 and the Adam optimizer. If the validation loss does not decrease for more than 50 epochs, training is stopped. The dimension of the bispectrum with 10 radial basis functions and $l_{\max}=6$ is (10,35). Thus, the final linear layer has output dimension 350, as shown in \Cref{fig:model}. For direct prediction, the final linear layer has output dimension 6 in order to predict $(a,b,c,\alpha,\beta,\gamma)$. The bispectrum prediction model is trained with the L1 loss. \revised{Because bispectrum coefficients empirically span many orders of magnitude}, we apply a signed cube-root normalization to stabilize optimization:
\begin{align}
    \mathbf{B}_{\text{norm}}= \mathrm{sign}(\mathbf{B})\,|\mathbf{B}|^{1/3}.
\end{align}
We mask bispectrum coefficients that are symmetry-forbidden (guaranteed to be zero for any unit cell, e.g. see \Cref{fig:bispec_lat_viz}), ensuring that the loss is computed over non-zero coefficients. For predicting the lattice parameters, we scale by the mean and standard deviation of the corresponding training data and train using the MSE loss.

\begin{table}[htbp]
\centering
\caption{Model and tokenization hyperparameters.}
\label{tab:hyperparams}
\begin{tabular}{lll}
\hline
\textbf{Component} & \textbf{Hyperparameter} & \textbf{Value} \\
\hline
\multicolumn{3}{l}{\textit{Input}} \\
 & XRD pattern length & 8500 \\
\hline
\multicolumn{3}{l}{\textit{Local Window Extraction}} \\
 & Window size & 50 \\
 & Stride & 25 \\
 & Number of windows & 339 \\
 & Overlap & 25 (50\%) \\
\hline
\multicolumn{3}{l}{\textit{Local Feedforward Network}} \\
 & Number of layers & 2 \\
 & Input dimension & 50 \\
 & Output (token) dimension & 256 \\
\hline
\multicolumn{3}{l}{\textit{Global Self-Attention (Tokenizer)}} \\
 & Number of attention heads & 4 \\
 & Positional encoding & Sinusoidal \\
\hline
\multicolumn{3}{l}{\textit{Transformer Encoder}} \\
 & Number of layers & 12 \\
 & Hidden dimension & 512 \\
 & Number of attention heads & 8 \\
 & Normalization & Layer normalization \\
 & Residual connections & Yes \\
\hline
\multicolumn{3}{l}{\textit{Output}} \\
 & Token sequence shape & $339 \times 256$ \\
\hline
\end{tabular}
\end{table}

We also found model performance to be relatively robust to architecture choices above a minimum capacity threshold. As shown in Table~\ref{tab:ablation}, validation loss drops substantially from 4 to 8 layers, but varies only modestly across larger configurations.
\begin{table}[htbp]
\centering
\caption{Ablation study over transformer encoder hyperparameters on the MP-20 validation set. All other hyperparameters are fixed as in Table~\ref{tab:hyperparams}. The selected configuration (\textbf{bold}) is used throughout this work.}
\label{tab:ablation}
\begin{tabular}{ccccc}
\toprule
\textbf{Num. Layers} & \textbf{Num. Heads} & \textbf{Model Dim.} &
\textbf{Feedforward Dim.} & \textbf{Val. Bispectrum Loss} \\
\midrule
4  & 4  & 128 & 256 & 5.79 \\
8  & 8  & 256 & 512 & 5.05 \\
\textbf{12} & \textbf{8} & \textbf{256} & \textbf{512} & \textbf{4.82} \\
12 & 12 & 384 & 756 & 4.88 \\
16 & 12 & 384 & 756 & 5.06 \\
\bottomrule
\end{tabular}
\end{table}

\subsection{Dataset Processing}\label{supp_info:datasets}

\subsubsection{Dataset Creation}
Unaugmented XRD patterns are generated from cif files using $\texttt{pymatgen}$ with Caglioti parameters $(U,V,W) = (0.05,-0.06,0.07)$, aligning with those used in Crystalyze \cite{rieselCrystalStructureDetermination2024}. As in Crystalyze, we use a uniform $2\theta$ grid ($5^{\circ}-90^{\circ}$ at 0.01 resolution) and normalize to unit maximum intensity. The primitive lattice for each structure was found using $\texttt{SpacegroupAnalyzer}$ with a symmetry tolerance of $10^{-2}$. This lattice was then used to calculate the lattice bispectrum for each structure with parameters in \Cref{sec:bispec_details}.

\subsubsection{Postprocessing Steps}

After prediction, lattice outputs are compared through a multi-stage postprocessing pipeline. First, lattices are reduced using the Niggli reduction. Second, we apply a CCTBX-based metric summary search over nearby unit cells using \texttt{cctbx.lattice\_symmetry.metric\_subgroups}. This step searches over metric subgroups and can recover more symmetry-consistent cells that lie within a small metric tolerance (we use 0.1). Both true and predicted lattices are compared across all candidate metric subgroups, and the best pair is selected by minimizing the distance between the respective metric tensors. This search is essential as the Niggli reduction does not guarantee consistent axis alignment between predicted and reference cells. \revised{Third, because low-symmetry lattices admit several nearly equivalent reduced cells, we check whether each predicted cell is related to the true cell by a change of setting, using cctbx \texttt{similarity\_transformations} with a tolerance of 5\% in lengths and 5$^\circ$ in angles. If so, we transform the prediction into the setting of the true cell before computing errors. A stricter tolerance of 2\%/2$^\circ$ leaves all comparisons unchanged. This step primarily reduces angle errors, and more so for the bispectrum model, whose inversion returns lattice vectors in an arbitrary basis, than for direct prediction.} We perform this post-processing procedure to ensure that errors reflect geometric deviations between the true and predicted lattice rather than differences in cell choice.

\subsubsection{\revised{Data Leakage Audit}}\label{supp_info:dataleak}
\revised{We performed a more detailed audit of possible data leakage. For the MP-20 dataset, we use pre-defined train/validation/test splits that are used by Crystalyze and prior work such as \cite{xieCrystalDiffusionVariational2021} to enable fair comparison. For MP-Full, we split by reduced chemical formula as stated. Both splitting strategies could potentially allow for similar XRD patterns in the train/validation/test set. As the input to the model is the XRD pattern, we performed a quantitative leakage audit based on the cosine-similarity of pairs of XRD patterns for MP-20 and MP-Full. For each material in the validation/test set for MP-20 and MP-Full, we compute the cosine-similarity between its XRD pattern and every pattern in the training set and take the maximum.}

\revised{We used cosine similarity $>0.999$ to flag candidates as near-duplicate pairs for further checks via lattice-parameter matching. For the MP-Full, 696 structures in the validation+test set (out of 62,171, 1.12\%) were flagged. 534 of these structures were confirmed as nearly identical lattices after Niggli reduction (mean absolute length difference $7\times10^{-4}$ \AA, angle difference $1.6\times10^{-3}$ degrees). The remaining structures either had substantially different structures despite high XRD-pattern similarity or were related superstructures (e.g. showing a lattice-length ratio of $\approx \sqrt{2}$).}

\begin{figure}[htbp]
    \centering
    \includegraphics[width=1.0\textwidth]{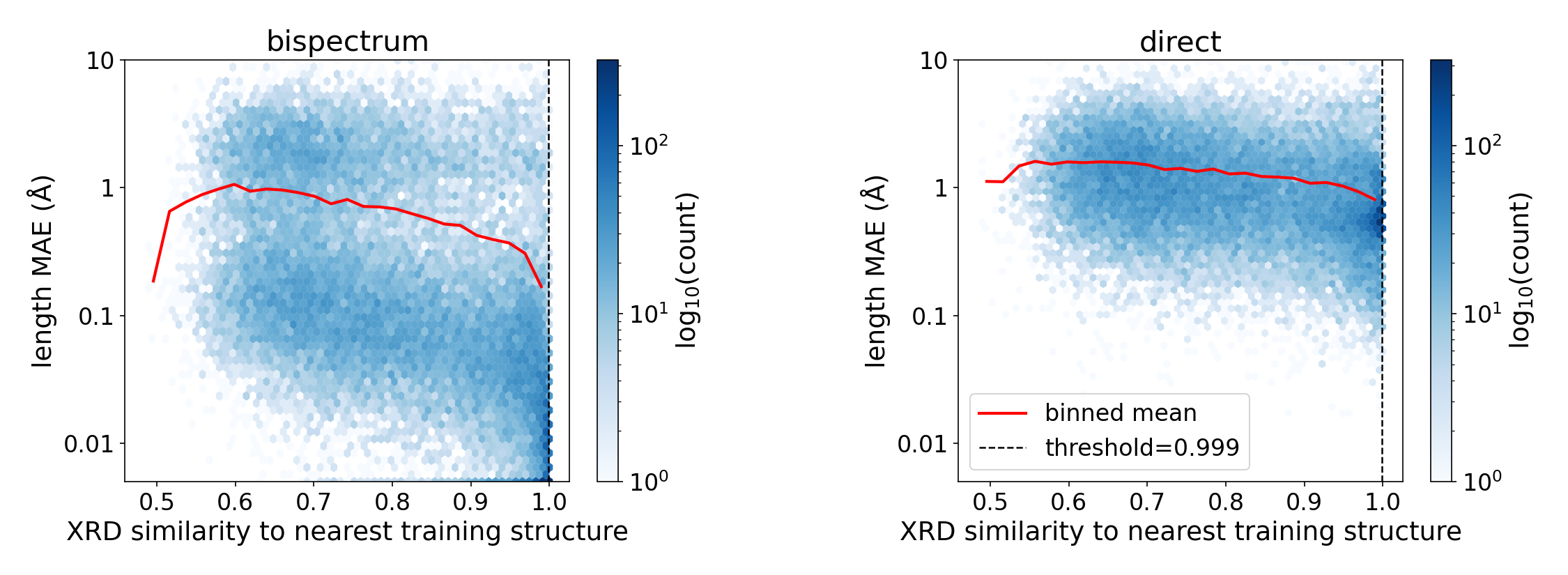}
    \caption{Length MAE vs. XRD-cosine similarity for patterns in the validation+test set for the full Materials Project.}
    \label{fig:length_mae}
\end{figure}

\begin{figure}[htbp]
    \centering
    \includegraphics[width=1.0\textwidth]{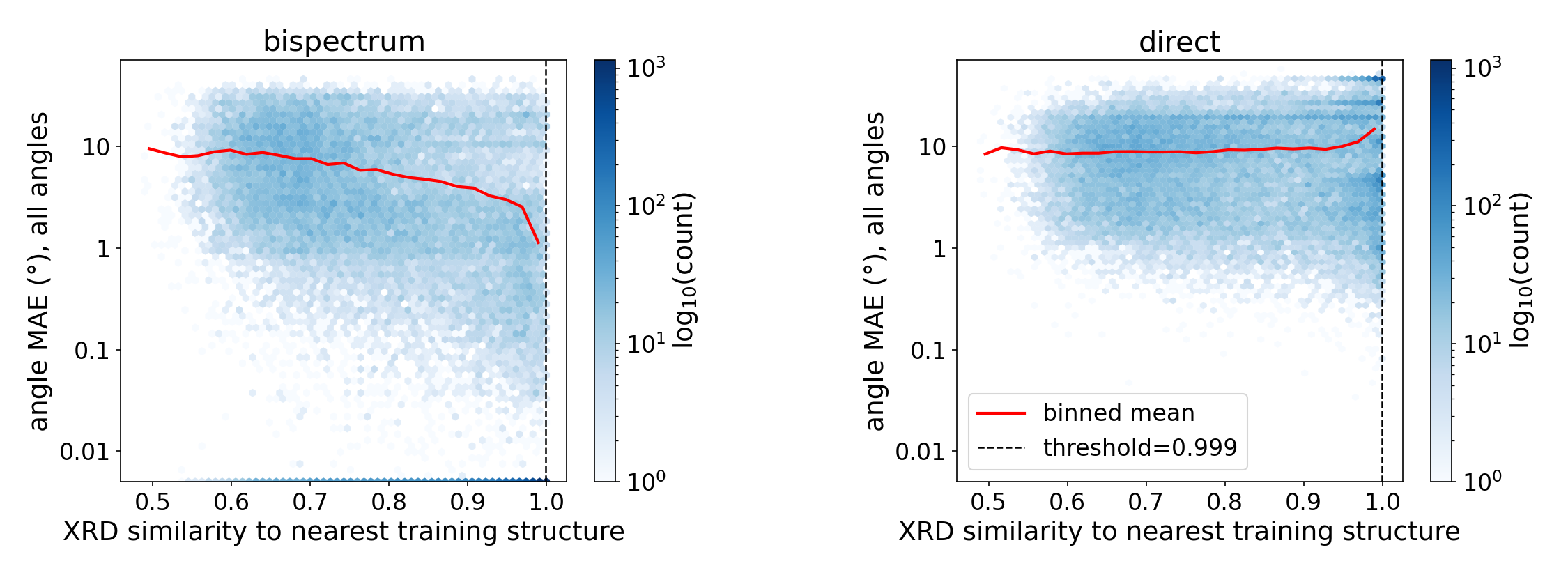}
    \caption{Angle MAE vs. XRD-cosine similarity for patterns in the validation+test set for the full Materials Project.}
    \label{fig:angle_mae}
\end{figure}

\revised{Restricting solely to the test set, $2.37\%$ of the MP-20 test set and $1.08\%$ of the MP-Full test set exceed this threshold. Excluding this entire flagged subset changes our reported metrics negligibly, as seen in \Cref{tab:leakage_exclusion}. To test robustness to the choice of threshold, we also exclude test structures above looser thresholds of 0.99 and 0.95, removing up to 27\% of the test set. A modest increase in error is expected when excluding test structures similar to the training set, as this makes the test set more difficult regardless of leakage. As shown in \Cref{fig:length_mae} and \Cref{fig:angle_mae}, the error of the direct model shows little dependence on cosine similarity, while the error of the bispectrum model is lower and has a small decrease with similarity.}
\begin{table}[h]
\centering
\caption{\revised{Effect of excluding progressively larger subsets of test structures with high XRD-pattern similarity (cosine similarity above the threshold) to their nearest training structure, on mean MAE for MP-Full Aug and MP-20 Aug. ``All'' is the full test set; percentages give the fraction of test structures excluded. Bispectrum errors increase modestly as more similar structures are excluded, and the angle advantage over direct prediction narrows, but the bispectrum continues to clearly outperform direct prediction at every threshold, including when over a fifth of the test set is excluded.}}
\footnotesize
\begin{tabular}{llcc}
\toprule
Dataset & Method (subset) & Length MAE (\AA) & Angle MAE ($^\circ$) \\
\midrule
MP-Full Aug & Direct, All & 1.27 & 9.87 \\
MP-Full Aug & Direct, Excl.\ $>0.999$ (1.1\%) & 1.27 & 9.76 \\
MP-Full Aug & Direct, Excl.\ $>0.99$ (7.7\%) & 1.31 & 9.33 \\
MP-Full Aug & Direct, Excl.\ $>0.95$ (21.1\%) & 1.37 & 9.01 \\
MP-Full Aug & Bispec + inv., All & 0.62 & 5.27 \\
MP-Full Aug & Bispec + inv., Excl.\ $>0.999$ (1.1\%) & 0.63 & 5.33 \\
MP-Full Aug & Bispec + inv., Excl.\ $>0.99$ (7.7\%) & 0.66 & 5.65 \\
MP-Full Aug & Bispec + inv., Excl.\ $>0.95$ (21.1\%) & 0.73 & 6.21 \\
\midrule
MP-20 Aug & Direct, All & 0.80 & 12.36 \\
MP-20 Aug & Direct, Excl.\ $>0.999$ (2.4\%) & 0.81 & 12.13 \\
MP-20 Aug & Direct, Excl.\ $>0.99$ (12.1\%) & 0.82 & 11.23 \\
MP-20 Aug & Direct, Excl.\ $>0.95$ (26.6\%) & 0.85 & 10.55 \\
MP-20 Aug & Bispec + inv., All & 0.23 & 2.58 \\
MP-20 Aug & Bispec + inv., Excl.\ $>0.999$ (2.4\%) & 0.24 & 2.63 \\
MP-20 Aug & Bispec + inv., Excl.\ $>0.99$ (12.1\%) & 0.26 & 2.86 \\
MP-20 Aug & Bispec + inv., Excl.\ $>0.95$ (26.6\%) & 0.29 & 3.20 \\
\bottomrule
\end{tabular}
\label{tab:leakage_exclusion}
\end{table}

\subsection{Bispectrum Numerical Experiments}\label{supp_info:bispec_num_exp}
\subsubsection{Bispectrum Sensitivity Analysis}\label{supp_info:sensitivity}
\begin{figure}[htbp]
    \centering
    \begin{subfigure}[t]{0.48\textwidth}
        \centering
        \includegraphics[width=\textwidth]{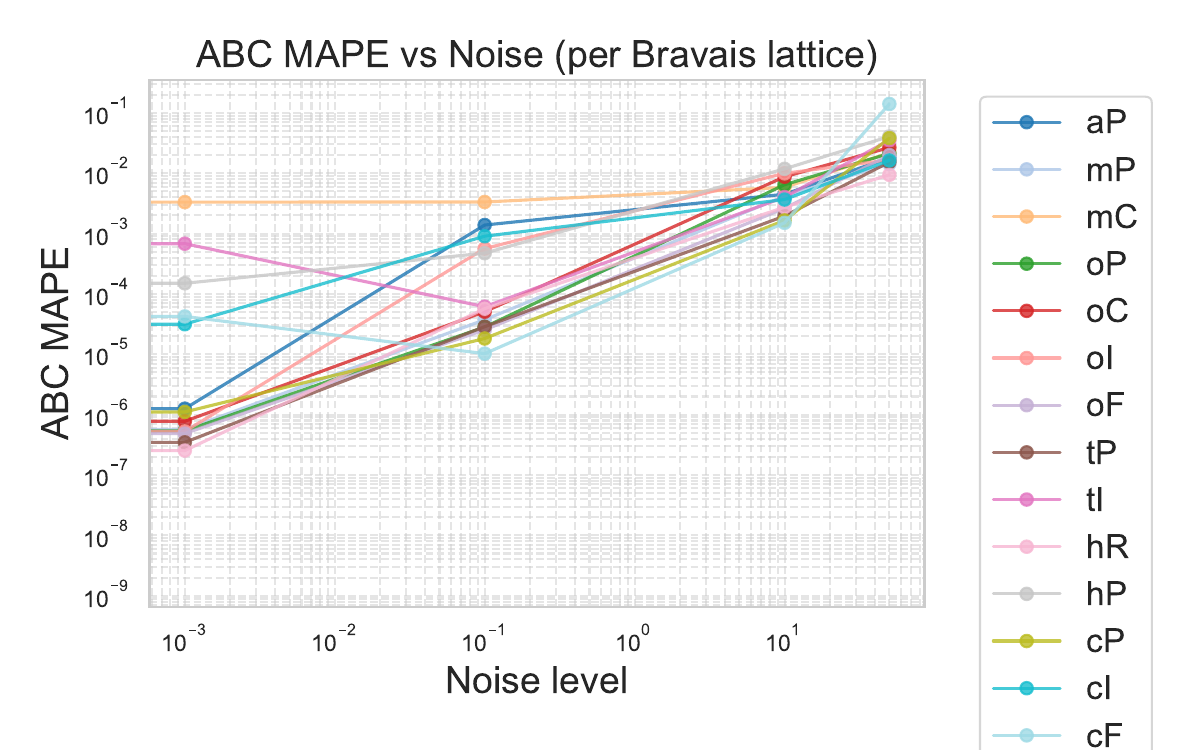}
        \caption{ABC MAPE vs noise.}
        \label{fig:perturb_abc_mape}
    \end{subfigure}
    \hfill
    \begin{subfigure}[t]{0.48\textwidth}
        \centering
        \includegraphics[width=\textwidth]{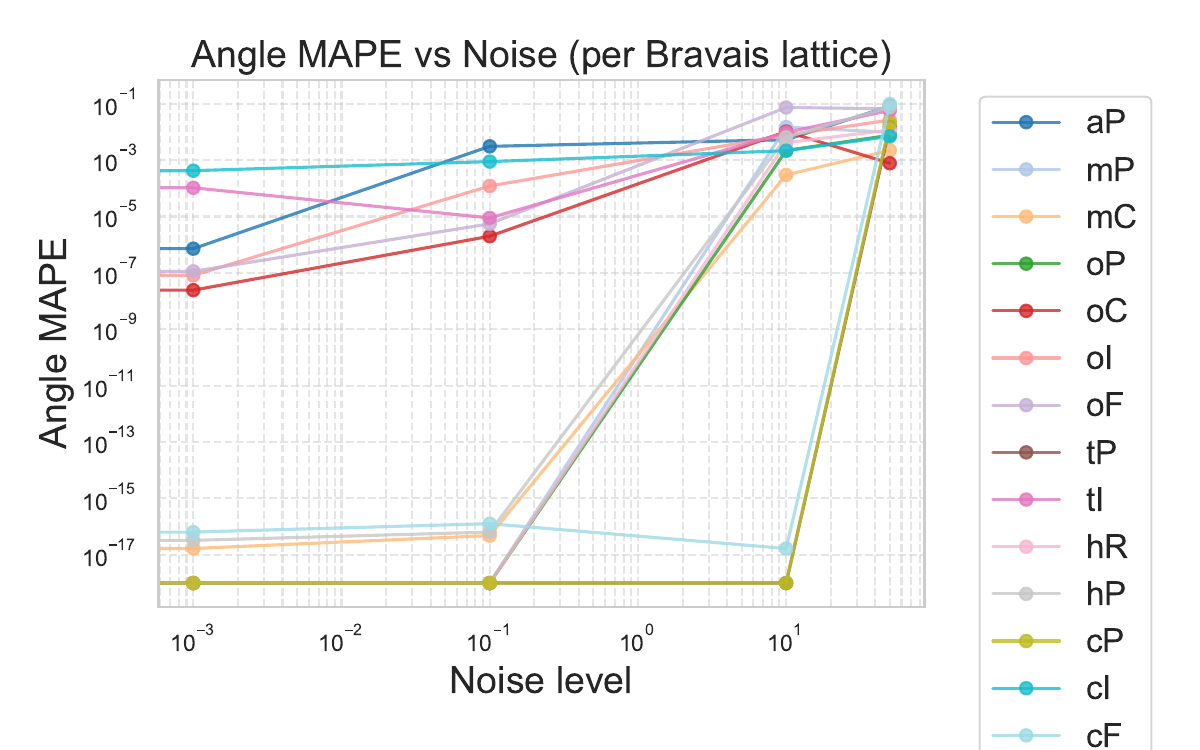}
        \caption{Angle MAPE vs noise.}
        \label{fig:perturb_angle_mape}
    \end{subfigure}
    \caption{Mean absolute percentage error as a function of noise level in the bispectrum $\mathbf{B}$ for different Bravais lattices.}
    \label{fig:mape_vs_noise}
\end{figure}

We also experiment with the sensitivity of the bispectrum representation to Gaussian noise. In \Cref{fig:perturb_abc_mape} and \Cref{fig:perturb_angle_mape}, we add different levels of Gaussian noise to the true bispectrum. We consider a sample lattice from the Materials Project for each Bravais lattice and average over 5 perturbed bispectra per lattice per noise level. Even at higher noise levels, cell parameters are recovered accurately across Bravais lattices, illustrating the robust nature of this representation.

\subsubsection{Bispectrum Error Analysis}\label{supp_info:error_analysis}
\begin{figure}[htbp]
    \centering
    \includegraphics[width=0.9\textwidth]{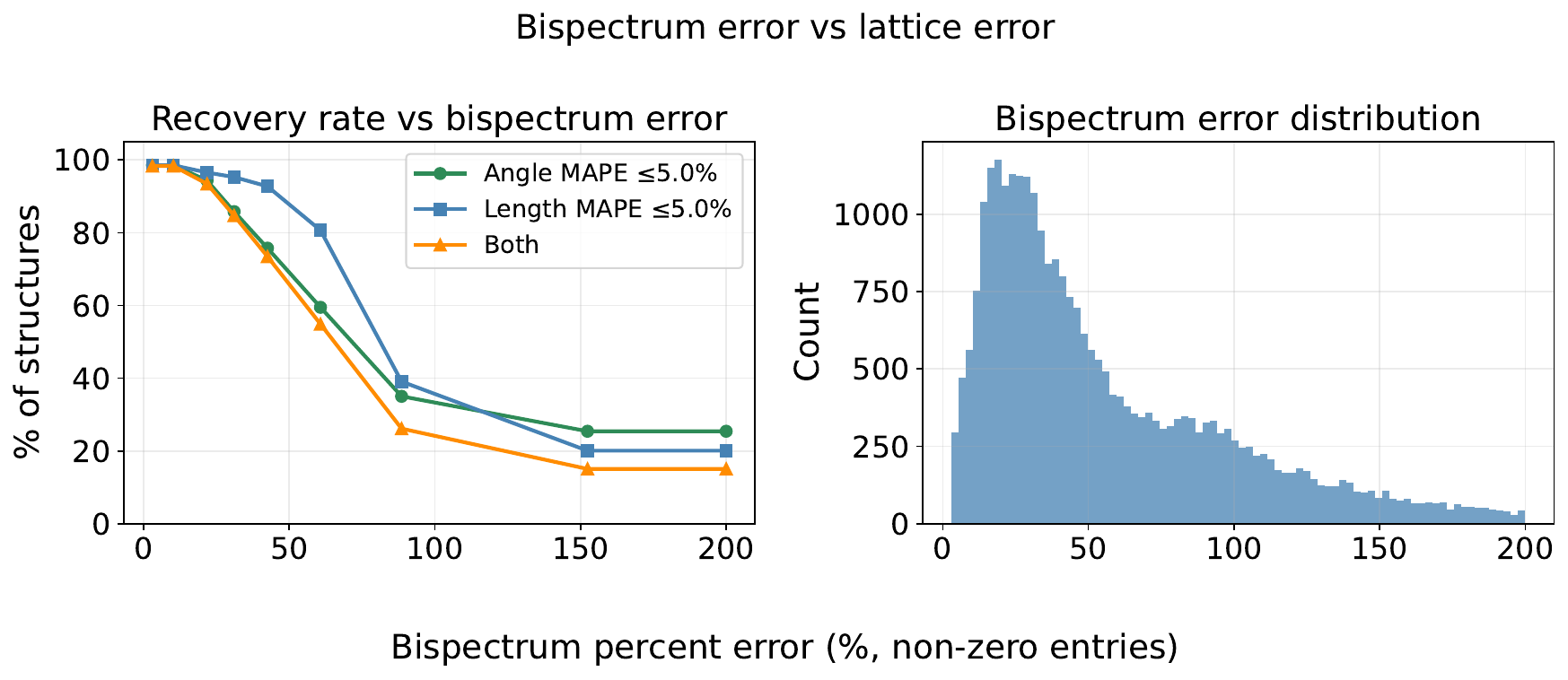}
    \caption{Recovery rate (\% of structures meeting length, angle, or
joint MAPE $\leq 5\%$ threshold) as a function of bispectrum MAPE computed over non-zero
entries (left), and distribution of per-structure bispectrum MAPE across the test set (right),
evaluated on the MPFull test set (after cctbx postprocessing). Both length and angle recovery are
high ($>90\%$) when bispectrum MAPE is low, and decline together as bispectrum error increases,
with length recovery consistently slightly above angle recovery across all bins.}
    \label{fig:bispec_lat_error}
\end{figure}

We analyze how bispectrum prediction error propagates to downstream lattice parameter error, motivated by the length/angle asymmetry observed in \Cref{sec:rruff}.
Total bispectrum MAPE correlates moderately with downstream length error ($R^2=0.388$,
$\rho=0.705$) but only weakly with angle error ($R^2=0.119$, $\rho=0.535$), consistent with a nonlinear relationship between bispectrum prediction quality and angle recovery. Despite the weak linear correlation, Figure~\ref{fig:bispec_lat_error} shows that both length and angle recovery are high when bispectrum error is low, suggesting the inversion succeeds when the upstream prediction is accurate. Additionally, the distribution of bispectrum MAPE is itself heavy-tailed (\Cref{fig:bispec_lat_error}, right), with a minority of structures showing very high prediction error; these correspond to the low-recovery regime in the left panel.

\revised{\subsubsection{Bispectrum $l_{\max}$ Truncation Error}}\label{supp_info:bispec_lmax_trunc}
\revised{
We computed the bispectrum for $l_{\max} \in \{2,4,6,8,10\}$ for a set of $n=5$ lattices per Bravais lattice drawn from the Materials Project dataset. We invert the bispectrum back to the lattice parameters using the same L-BFGS-B optimizer used in this work, starting from the true lattice with a small random perturbation, and measure the resulting mean absolute error in lattice length and angle. Results are shown below in Figure \ref{fig:lmax_exp}. Recovery error remains near zero at $l_{\max}=6$ and does not improve for higher truncation degrees. This indicates that the current choice of $l_{\max}=6$ is not limiting recovery accuracy. We also hypothesize that the lack of improvement at higher $l_{\max}$ reflects that the inversion landscape is harder to optimize with a larger number of bispectrum components, and may need a more sophisticated optimization approach.
    }
    \begin{figure}[htbp]
    \centering
    \includegraphics[width=1.0\textwidth]{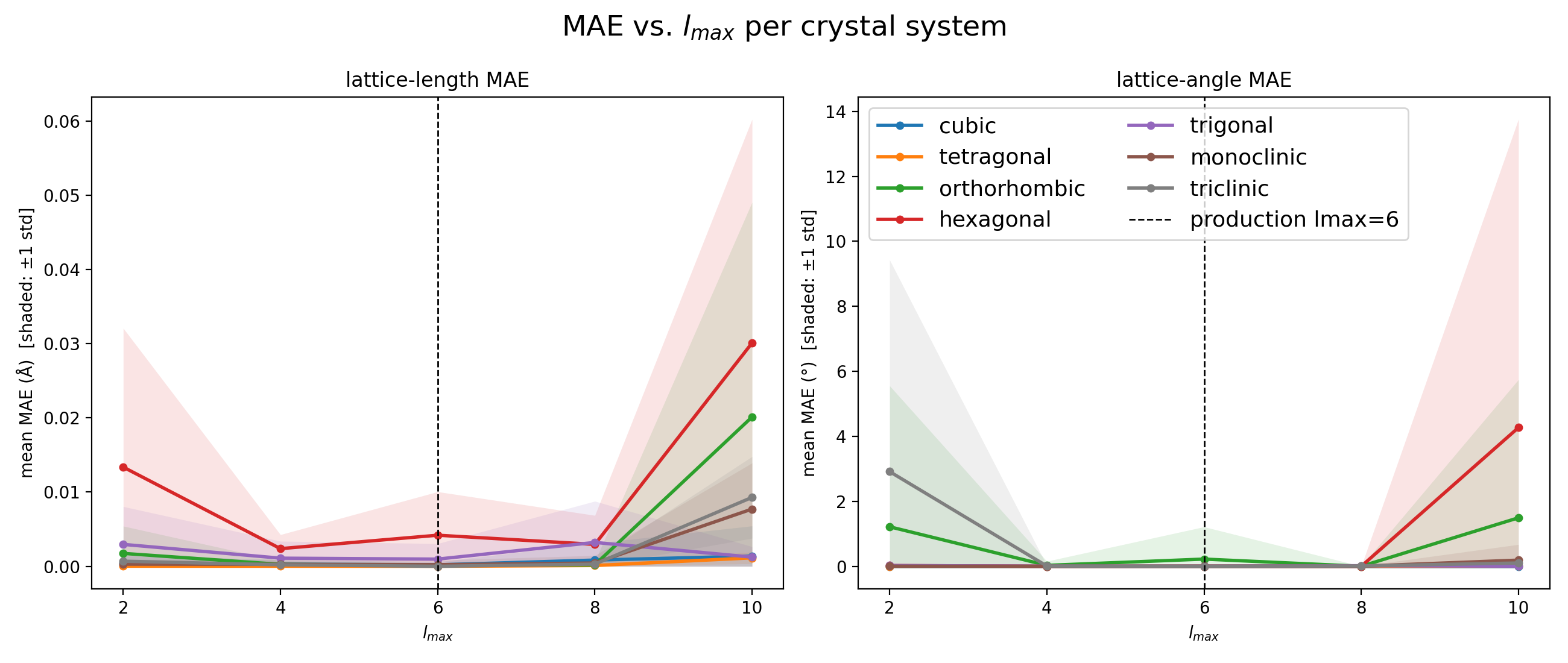}
    \caption{\revised{Mean absolute error (MAE) in recovered lattice length (\AA) and angle (\textdegree), per crystal system, after inverting the bispectrum computed at $l_{\max} \in \{2,4,6,8,10\}$ back to lattice parameters via L-BFGS-B, for $n=5$ real
structures per Bravais lattice drawn from the Materials Project. Shaded regions show $\pm 1$
standard deviation across the sampled structures in each system. The dashed line marks $l_{\max}=6$.}}
    \label{fig:lmax_exp}
\end{figure}

\subsubsection{\revised{Inversion and Database Computational Cost}}\label{supp:inversioncost}

\revised{We provide further details about the lookup dataset. Crucially, the bispectrum calculation per material is parallelizable. The database currently has $156{,}801$ entries. Each entry stores the bispectrum of the structure's primitive reciprocal lattice, the reciprocal lattice matrix ($3\times3$), and the material identifier. The whole
    database is 452\,MB on disk stored as an HDF5 file (439\,MB bispectra, 11\,MB lattices, 1.6\,MB identifiers). Each bispectrum takes a mean of 22\,ms per entry on a single CPU core. Building the full database therefore takes about one hour on one core, and is much faster when parallelized depending on the number of CPU cores available. The database is built once and is independent of the trained network. It can thus be extended with additional structures from other sources (e.g. ICSD or COD) without retraining.}

    \revised{At inference time, we compute the mean squared error between the predicted bispectrum and every database entry (after the same cube-root scaling used in training), excluding the query's own entry, and select the $n=15$ nearest entries. This search takes an average of 0.92 sec on one thread and scales linearly with database size. For much larger databases, approximate nearest-neighbor searches could be used instead. Each of the 15 candidate lattices is perturbed with Gaussian noise ($\sigma = 0.05 ||\mathbf{L}^*||_F$) scaling the Frobenius norm of the reciprocal lattice matrix and optimized with L-BFGS-B. The inversion stops once the bispectrum loss falls below $10^{-3}$, no further candidates are tried. Otherwise, the inverted lattice with the lowest loss across all candidates is returned. Candidate selection uses only the predicted bispectrum, not the true lattice.}

\begin{figure}[htbp]
    \centering
    \includegraphics[width=0.5\textwidth]{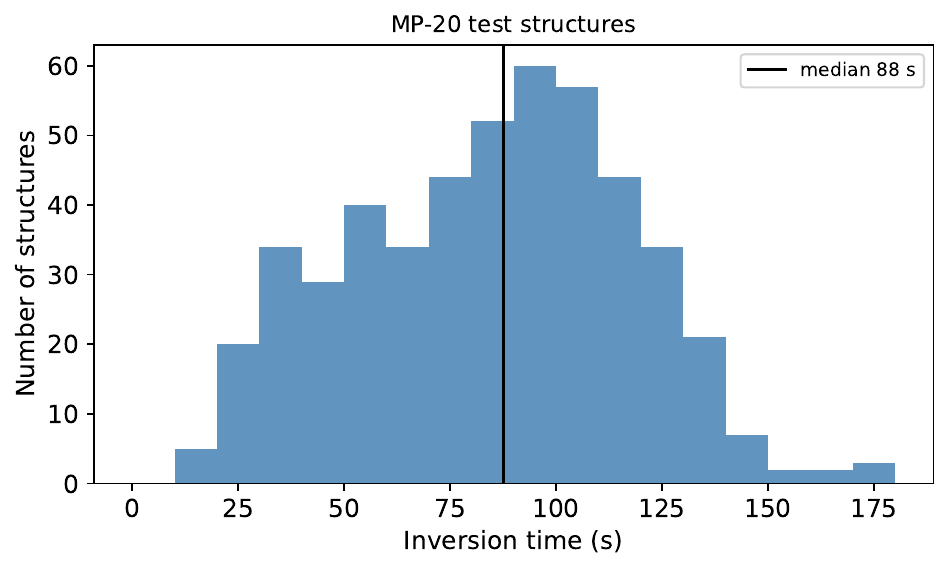}
    \caption{\revised{Inversion time for 500 test structures from the MP20 test set using predicted bispectra}.}
    \label{fig:timing_exp}
    \end{figure}

    \revised{We timed the inversion using $n=15$ candidates for 500 structures randomly sampled from the MP-20 test set, using bispectra predicted by the model trained on MP-20 Aug, on a node with 128 CPU cores. The median time was 88~s, (90th percentile 126\,s, maximum 175\,s); the distribution is shown in \Cref{fig:timing_exp}. Thus, inversion is the dominant cost at inference, but structures are inverted independently, so it is parallelized across across CPU cores. For this experiment, we used 32 parallel worker processes, each single-threaded. Processing 500 structures took about 34 minutes. Improving the efficiency of the inversion is a useful direction for future work.}

\subsection{Physically Informed XRD Data Augmentation}
\label{supp_info:aug}
To improve robustness to experimental variability, we applied a suite of physically informed augmentations to synthetic powder X-ray diffraction (XRD) patterns derived from the MP-20 and MP-Full dataset. The three augmentation strategies target complementary aspects of experimental variability: lattice strain (peak positions), preferred orientation (peak intensities), and instrumental broadening (peak widths).

\paragraph{Symmetry-Preserving Strain Augmentation}
To simulate peak-position shifts arising from elastic lattice distortions, we apply strain tensors directly to crystal lattices prior to diffraction simulation. Each structure is first represented in its conventional cell, and strain is applied multiplicatively to the lattice matrix.
For a given structure with lattice matrix $\mathbf{L}$, a strained lattice $\mathbf{L}'$ is generated as
\begin{equation}
\mathbf{L}' = \mathbf{S}\,\mathbf{L},
\end{equation}

where $\mathbf{S}$ is a $3\times3$ strain tensor. The entries of $\mathbf{S}$ are sampled randomly, with the tensor structure conditioned on the crystal’s space-group symmetry to preserve the original space group.

Structures are grouped into symmetry classes based on their space-group number: cubic (195–230), orthorhombic (16–74), monoclinic (3–15), triclinic (1–2), and hexagonal/tetragonal (75–194), with further distinction between high- and low-symmetry hexagonal/tetragonal settings. For each symmetry class, only symmetry-allowed entries of the strain tensor are activated. For example, cubic systems are restricted to isotropic strain (main diagonal entries), while lower-symmetry systems permit selected shear components (off-diagonal entries).

Diagonal components of the strain tensor, corresponding to normal strain along the lattice vectors, are sampled independently from a uniform grid spanning
\[
[1-\epsilon_{\max},\,1+\epsilon_{\max}],
\]
with $\epsilon_{\max}=0.04$. Off-diagonal components, corresponding to shear strain, are sampled (when symmetry-allowed) from a uniform grid spanning
\[
[-\epsilon_{\max},\,\epsilon_{\max}].
\]
All other tensor entries are set to zero according to the symmetry constraints of the crystal class.

For each structure, three independently sampled strain tensors are generated, producing three strained lattices. Atomic fractional coordinates are kept fixed, and only the lattice vectors are modified. Each strained structure is then used to generate a synthetic powder XRD pattern, resulting in augmented diffraction patterns that primarily exhibit shifts in peak positions ($2\theta$) while preserving relative peak intensities and peak shapes. \revised{As the lattice vectors are modified, the bispectrum for each strained lattice is also re-calculated. The other augmentations do not modify the bispectra.}

\paragraph{Intensity Augmentation via Preferred Orientation (Texture)}
To simulate variations in relative peak intensities arising from preferred orientation (texture) effects in powder diffraction experiments, we apply a random, directionally dependent intensity scaling to the synthetic diffraction patterns. This augmentation modifies peak intensities while leaving peak positions unchanged.

For a given powder XRD pattern, each diffraction peak is associated with a Miller index vector $\mathbf{h}$ (either three-index $(hkl)$ or four-index $(hkil)$ notation, depending on crystal symmetry).

A random non-zero preferred orientation direction $\mathbf{p}$ is then sampled in the same index space, with each component drawn from $\{0,1\}$ and resampled until $\|\mathbf{p}\| \neq 0$. This vector defines the crystallographic direction along which texture is imposed.
For each diffraction peak, an orientation-dependent texture factor is computed as the normalized absolute dot product
\begin{equation}
t(\mathbf{h}) = \left| \frac{\mathbf{h} \cdot \mathbf{p}}{\|\mathbf{h}\| \, \|\mathbf{p}\|} \right|,
\end{equation}
which lies in the interval $[0,1]$. This value is then linearly mapped to a bounded scaling interval
\begin{equation}
f(\mathbf{h}) \in [1 - \alpha,\, 1],
\end{equation}
where $\alpha$ is the maximum texture strength. In this work, $\alpha = 0.5$, allowing peak intensities to be reduced by up to 50\% relative to their original values. The scaled peak intensity is given by
\begin{equation}
I'(\mathbf{h}) = f(\mathbf{h})\, I(\mathbf{h}).
\end{equation}
The resulting set of scaled peak intensities is then used to reconstruct a continuous diffraction pattern using the pseudo-Voigt profile. Fixed Caglioti parameters $(U,V,W)=(0.1,0.1,0.1)$ are used during this step, such that the texture augmentation affects only relative peak intensities and not peak widths or positions.

\paragraph{Peak Broadening via Caglioti Parameter Augmentation}
To simulate variations in diffraction peak widths and shapes arising from microstructural effects such as finite crystallite size and instrumental resolution, we augment synthetic diffraction patterns by modifying the parameters in the Caglioti formulation. In this model, the full width at half maximum (FWHM) of a diffraction peak depends on the Bragg angle according to
\begin{equation}
\mathrm{FWHM}^2 = U \tan^2\theta + V \tan\theta + W,
\end{equation}
where $U$, $V$, and $W$ are instrumental broadening parameters and $\theta$ is the Bragg angle.

We generate broadened diffraction patterns using four parameter configurations designed to capture a range of realistic peak shapes. Three configurations use fixed parameter sets: $(U,V,W)=(0.05, 0.06, 0.07)$, $(0.05,-0.01,0.01)$, and $(0,0,0.01)$. These combinations represent different angular dependencies of peak broadening commonly observed in powder diffraction instruments. A fourth configuration samples the constant term $W$ uniformly from the interval $[0.001,\,0.1]$ while keeping $U=0$ and $V=0$, producing a spectrum of angle-independent broadening strengths.

This augmentation modifies peak widths and shapes while leaving peak positions and relative intensities unchanged.

\subsubsection{Limitations and Future Directions in Augmentation}\label{supp_info:augfuture}
The augmentation strategies above are hand-designed and parametric, and their realism is
ultimately bounded by how well a chosen analytic form (e.g. Gamma-distributed background,
symmetry-constrained strain) matches a given experimental dataset's actual acquisition
characteristics. This introduces a degree of circularity: strategies designed by inspecting
typical lab-source diffraction data will generalize well to data of that kind, but this is a
narrower claim than generalization to experimental data in general. A more rigorous test would evaluate on experimental data whose acquisition characteristics differ meaningfully from those assumed during augmentation design, for example synchrotron rather than lab-source patterns, or instruments with substantially different background or zero-shift behavior. One direction to address this is to learn augmentation distributions directly from experimental data rather than specifying them analytically, for example using a diverse multi-instrument corpus such as opXRD \cite{hollarekOpXRDOpenExperimental} or RRUFF \cite{Lafuente2015RRUFF} to fit empirical distributions over peak shift, broadening, and background. Additionally, it could be useful to include experimental patterns in the training set itself; this would require collating available experimental datasets across instruments, and would ideally lead to robustness across a wider range of experimental conditions rather than the specific conditions any one augmentation scheme was designed around.

We also note that the augmentations in \Cref{supp_info:aug} are not uniform in their effect across unit cell size. Strain augmentation applies a fixed relative magnitude regardless of absolute lattice parameters. However, larger unit cells have proportionally more dense reflections in $2\theta$. Thus, the same relative strain could alter the ordering or resolvability of closely spaced peaks for large cells more than small ones. Future work could mitigate this by augmenting with perturbations that scale with the unit cell rather than at fixed strain.

\subsection{Group and Representation Theory Background}\label{supp_info:group}

We provide a brief background of the group and representation theory needed to understand the invariant lattice descriptor (for more information, see \cite{Dresselhaus2008}). Symmetries form the basis of condensed matter physics and crystallography. They are described abstractly by algebraic groups. For example, $SO(3)$ is the group of three dimensional rotations, and $O(3)$ also includes inversion. All crystallographic point groups are subgroups of $O(3)$. Thus, for computational materials/machine learning applications, it can be advantageous to enforce $O(3)$ symmetries (e.g. through software frameworks such as \texttt{e3nn}, \cite{geiger2022e3nn}).

\textbf{Invariance/Equivariance}
A function $f: X \to Y$ is invariant under a group $G$ if $f(D_X(g)x) = f(x) \forall g \in G, x \in X$ where $D_X(g)$ is the representation of the group element $g$ on the vector space $X$ (for example, a rotation matrix). $f$ is equivariant if $f(D_X(g)x) = D_Y(g)f(x) \forall g \in G, x \in X$. These concepts are implicit in the field of crystallography. For example, X-ray diffraction patterns are invariant under a given crystal's point group, as rotating the crystal by a symmetry operation doesn't change the pattern. Higher-order tensor properties would be equivariant under these operations.

\textbf{Spherical Harmonics}
Spherical harmonics are an orthonormal basis for the class of square integrable functions on the sphere, organized by angular frequency $l$. One can thus think of them as basis functions for an ``angular Fourier transform.'' For example, quantities such as the structure factor or the electron density around an atom can be expanded in spherical harmonics. \revised{Spherical harmonics are essential to enable the construction of the bispectrum due to their transformation properties under $SO(3)$.}

They are equivariant under $SO(3)/O(3)$ (as they have a definite parity). The $2l+1$ spherical harmonics for a given $l$ are denoted by $Y_{lm}$ where $m=-l,\dots,0,\dots, l$. The spherical harmonics have a specific parity with even $l$ spherical harmonics having even parity and odd $l$ having odd parity. A rotation $R$ acting on spherical harmonic $Y_{lm}$ transforms it as
\begin{align}
    RY_{lm} = \sum_{m'=-l}^lD^l_{mm'}(R)Y_{lm'}
\end{align}
with the Wigner-D matrices. For any given $l$, the spherical harmonics form a basis for the irreducible representation $D^l$, as the elements of $D^l$ are given by
\begin{align}
    D^l_{mm'} = \langle Y_{lm}|\hat{R}|Y_{lm'} \rangle
\end{align}
Due to these properties, spherical harmonics are often used to construct rotationally-invariant descriptors.

\subsubsection{\revised{Theoretical Justification of Bispectrum Sparsity}}\label{supp_info:theorysparse}
\revised{For completeness, we provide a theoretical justification for the sparsity of the bispectrum as in \Cref{fig:bispec_lat_viz}. The spherical harmonics permitted under a given point group are used extensively in materials science. These are tabulated in for example \cite{altmannPointGroupTheory1994}. We sketch the derivation here for completeness. The sparsity pattern is fixed by the point group of the Bravais lattice. First, note that the point group of the lattice in real space is the same as that in reciprocal space. Let $H$ be the point group of the lattice and apply some element $R \in H$ to the lattice expanded in terms of spherical harmonics (Equation 3 in the paper). The coefficients $c_{nlm}$ are invariant under the action of $R \in H$.
    \begin{align}
        c_{nlm} = \sum_{\mathbf{G \in k_{\max}}}Y_{lm}(R\hat{\mathbf{G}})g_n(|\mathbf{G}|)
    \end{align}
    Using the transformation properties of spherical harmonics we have
    \begin{align}
        c_{nlm} = \sum_{\mathbf{G \in k_{\max}}}\sum_{m'=-l}^lD^{l}_{mm'}(R)Y_{lm'}(\hat{\mathbf{G}})g_n(|\mathbf{G}|) = \sum_{m'=-l}^l D^l_{mm'}(R)c_{nlm'} \quad \forall R \in H
    \end{align}
    This condition states that $\mathbf{c}_{nl}$ is left invariant by $D^{l}$ for $R \in H$ where $D^{l}$ is the Wigner-D matrix of degree $l$. While $D^l$ is an irreducible representation of $O(3)$, its restriction to $H$ is reducible and may contain the identity representation of $H$ multiple times. Thus, $\mathbf{c}_{nl}$ is zero for a given $l$ if $D^l$ does not contain the identity representation of $H$. The sparsity then extends to the bispectrum, as if $\mathbf{c}_{nl}=0$ for a given $l$, every bispectrum component whose ($l_1,l_2,l_3$) triplet includes that $l$ also vanishes.}

    \revised{We include the lattices in \Cref{fig:bispec_lat_viz} to provide intuition for this theoretical observation. Inversion symmetry is present in every Bravais lattice, so any term with odd $l$ is zero. For the rhombohedral and triclinic lattices, all terms with even $l_1,l_2,l_3$ are allowed. Additionally, for the cubic lattice, any term containing $l=2$ is zero, as a cubic lattice cannot host an anisotropic rank-2 tensor.}

\subsection{Powder X-Ray Diffraction}\label{supp:xrdbackground}
We assume that most of the readers are familiar with the principles of powder XRD. However, we provide a brief overview for completeness. Crystals are composed of a lattice and a basis. The lattice is an array of points $\{\bm{R}\}$ in space which satisfy
\begin{equation}
\bm{R} = n_1\bm{a}_1 + n_2\bm{a}_2 + n_3\bm{a}_3, \quad n_i \in \mathbb{Z}.
\end{equation}
The basis is the physical unit that is repeated. The vectors $\{\bm{a}_i\}$ are known as the primitive lattice vectors. Lattices are classified according to their symmetry properties under rotation and reflection, and the distinct crystal lattices are known as Bravais lattices. For example, in 3D, there are 14 Bravais lattices. The Fourier transform of the lattice is known as the reciprocal lattice $\{\bm{G}\}$, which satisfies $\mathbf{a}_i\cdot\mathbf{b}_j = 2\pi\delta_{ij}$, and is given by
\begin{equation}
    \bm{G} = h\bm{b}_1+k\bm{b}_2+l\bm{b}_3, \quad h,k,l \in \mathbb{Z},
\end{equation}
where $(h,k,l)$ are known as Miller indices. In powder X-ray diffraction, the sample is illuminated with a monochromatic incident beam of X-rays. The intensity is measured as a function of the angle between the incident beam and the detector. The locations of Bragg peaks are determined by the equation
\begin{align}
    n\lambda=2d\sin\theta
\end{align}
where $n$ is the diffraction order, $\lambda$ is the X-ray wavelength, $d$ is the crystal plane spacing, and the diffraction angle is $2\theta$. The Bragg condition can equivalently be written $\mathbf{k'}-\mathbf{k}=\mathbf{G}$ or $2\mathbf{k} \cdot \mathbf{G} =\mathbf{G}^2$ where $\mathbf{k}$ and $\mathbf{k}'$ are wavevectors of the incident and scattered beams. In PXRD, the lattice parameters and crystal symmetry determine the positions of the Bragg peaks, while the atomic positions in the basis determine the peak intensities. The observed peak positions provide direct information about the unit cell geometry, forming a natural target for regression or generative models.

\subsection{Representing Atomic Environments Using Spectra}\label{supp:background_atom_rep}
Atomic representations typically are known as descriptors, or a list of real values that correspond to physical parameters of the atomic system. An ideal representation is invariant with respect to permutation, rotation, reflection, and translation symmetries. For example, if atomic positions are rotated in space, we would like the representation to remain the same. Invariant descriptors $q_1,q_2,\dots,q_M$ are said to be complete if they uniquely determine the atomic environment up to symmetry \cite{bartokRepresentingChemicalEnvironments2013}. Additionally, it is advantageous if descriptors are continuous and differentiable. The density function associated with a given atom type can be written as
\begin{align}
    \rho(\mathbf{r})=\sum_{i \in \text{cutoff}}\delta(\mathbf{r}-\mathbf{r}_i)
\end{align}
where the sum contains neighbors of the atom within some cutoff, and $\mathbf{r}_i$ is the vector from the given atom to neighbor $i$. We can expand the density in terms of spherical harmonics as
\begin{align}\label{eq:sphproj}
    \rho(\hat{\mathbf{r}}) = \sum_{l=0}^{\infty}\sum_{m=-l}^l c_{lm}Y_{lm}({\hat{\mathbf{r}}})
\end{align}
In practice, one would not sum to $l=\infty$, rather the expansion would be truncated at some $l_{\max}$. The coefficients $c_{lm}$ are
\begin{align}
    c_{lm} = \langle \rho|Y_{lm}\rangle = \sum_i Y_{lm}(\hat{\mathbf{r}}_i)
\end{align}
For each $l$, we denote $\mathbf{c}_l$ as the $(2l+1)$ vector of expansion coefficients $c_{lm}$. Under rotations, $\mathbf{c}_l$ transforms according to the irreducible representation $D^{(l)}$ of $\mathrm{SO}(3)$.

The rotationally invariant power spectrum is defined as $\mathbf{c}_l^{\dagger}\mathbf{c}_l$. Higher-order invariants such as the bispectrum are constructed by taking successive tensor products of $\mathbf{c}_l$, coupling $l$ channels through the Clebsch-Gordan coefficients. The tensor product representation $\mathbf{c}_{l_1} \otimes \mathbf{c}_{l_2}$ can be decomposed as
\begin{align}
    \left(\mathbf{c}_{l_1} \otimes \mathbf{c}_{l_2}\right)_{lm}
    =
    \sum_{m_1,m_2}
    C^{\,l m}_{\,l_1 m_1\, l_2 m_2}
    \, c_{l_1 m_1} c_{l_2 m_2},
\end{align}
where $C^{\,l m}_{\,l_1 m_1\, l_2 m_2}$ are Clebsch--Gordan coefficients. The bispectrum components are labeled by triples $(l_1,l_2,l)$. For each such triple, the coupled tensor product is contracted with $\mathbf{c}_l^\dagger$, yielding a rotational scalar/pseudoscalar through the invariant product $l \otimes l^* \to 0$.
\begin{align}\label{eq:bispec_supp}
    b_{l_1 l_2 l}
    &=
    \mathbf{c}_l^\dagger
    \left(\mathbf{c}_{l_1} \otimes \mathbf{c}_{l_2}\right)_l \\
    &=
    \sum_{m,m_1,m_2}
    c_{l m}^*
    C^{\,l m}_{\,l_1 m_1\, l_2 m_2}
    c_{l_1 m_1}
    c_{l_2 m_2}.
\end{align}
See \cite{bartokRepresentingChemicalEnvironments2013} for a complete description and the proof that the bispectrum is rotationally invariant. \cite{bartokRepresentingChemicalEnvironments2013} also notes that one can introduce radial information with radial basis functions $g_n$, such that Equation \ref{eq:sphproj} becomes
\begin{align}\label{eq:sphprojrad}
    \rho(\hat{\mathbf{r}}) = \sum_n\sum_{l=0}^{\infty}\sum_{m=-l}^l c_{nlm}Y_{lm}g_n(|\mathbf{r}|)({\hat{\mathbf{r}}})
\end{align}
Assuming the radial basis functions are orthonormal, the coefficients are then
\begin{align}
    c_{nlm} = \langle g_n Y_{lm} | \rho \rangle
\end{align}
The construction of the bispectrum is the same as outlined above, with an additional index corresponding to the $n$ radial basis functions. A potential issue arises in that the bispectrum is not a theoretically complete representation \cite{pozdnyakovCompletenessAtomicStructure2020,nigamCompletenessAtomicStructure2023}\textemdash meaning that counterexamples can be constructed such that two different atomic environments yield the same bispectrum. A potential solution could be to use the union of all $n$-body correlations for a given dataset until environments from that dataset are distinguishable. However, in practice, the bispectrum can often be inverted to obtain the signal \cite{bispecinversionMRA,cobelliLocalInversionChemical2022} or constructive algorithms can be developed (with additional theoretical assumptions \cite{kakaralaBispectrumSourcePhaseSensitive2012,uniquebisinvert}.

In conventional SOAP-like descriptors, the density is defined in real space around a chosen atomic center. In contrast, experimental crystallographic techniques such as powder diffraction probe reciprocal space directly. Thus, in our work, we extend the bispectrum to \textbf{represent crystal lattices in reciprocal space}. Consequently, the origin is uniquely fixed at the $\Gamma$ point, and the representation encodes the lattice geometry rather than a local atomic environment. The resulting spectra thus reflect the symmetry of the Brillouin zone (a parallelepiped) rather than that of an atom-centered neighborhood.

\subsection{Additional Results}\label{supp_info:moreplots}

Here, we include parity plots for each dataset \revised{and additionally report angle MAE per $\alpha, \beta, \gamma$ per crystal system in \Cref{tab:angle_by_system}}. We also include recovery rates per Bravais lattice for each dataset. A predicted cell is considered correctly recovered if the mean absolute percentage error (MAPE) on the three lattice lengths is $\leq$ 5\% and the MAPE on the three lattice angles is $\leq$ 5\%, computed after postprocessing to a canonical form using cctbx. We report the fraction of test structures satisfying each condition individually and jointly. The 5\% threshold is chosen as a representative, dataset-comparable tolerance.

\begin{table}[htbp]
\centering
\caption{\revised{Angle MAE ($^\circ$) by true crystal system, separately for $\alpha$, $\beta$, $\gamma$, after cctbx post-processing, on the MP-20 and MP-Full test sets. Both methods are evaluated on the same test structures; $n$ is the number of them in each system. In monoclinic cells only $\beta$ is free; in triclinic cells all three angles are free. The lower error of the two methods for each angle is in bold.}}
\footnotesize
{
\begin{tabular}{lccccccc}
\toprule
& & \multicolumn{3}{c}{Bispec + inv.} & \multicolumn{3}{c}{Direct} \\
System & $n$ & $\alpha$ & $\beta$ & $\gamma$ & $\alpha$ & $\beta$ & $\gamma$ \\
\midrule
\multicolumn{8}{l}{\emph{MP-20 (27K)}} \\
Triclinic & 432 & \textbf{6.11} & \textbf{5.39} & \textbf{7.16} & 13.12 & 10.82 & 15.36 \\
Monoclinic & 1280 & \textbf{4.55} & \textbf{6.92} & \textbf{4.31} & 9.77 & 16.46 & 9.21 \\
Orthorhombic & 1715 & \textbf{3.31} & \textbf{2.86} & \textbf{2.69} & 8.25 & 7.57 & 9.64 \\
Tetragonal & 1547 & \textbf{1.47} & \textbf{1.71} & \textbf{1.12} & 8.50 & 8.30 & 7.38 \\
Trigonal & 627 & \textbf{1.71} & \textbf{2.04} & \textbf{2.89} & 11.02 & 11.46 & 20.05 \\
Hexagonal & 1296 & \textbf{0.61} & \textbf{0.63} & \textbf{1.51} & 4.20 & 2.82 & 14.25 \\
Cubic & 2149 & \textbf{0.14} & \textbf{0.17} & \textbf{0.36} & 15.33 & 13.53 & 15.22 \\
\midrule
\multicolumn{8}{l}{\emph{MP-20 Aug (270K)}} \\
Triclinic & 432 & \textbf{7.00} & \textbf{6.10} & \textbf{7.71} & 11.57 & 10.60 & 14.32 \\
Monoclinic & 1280 & \textbf{4.97} & \textbf{7.80} & \textbf{4.54} & 8.37 & 16.16 & 8.04 \\
Orthorhombic & 1715 & \textbf{3.67} & \textbf{3.14} & \textbf{3.09} & 8.23 & 7.81 & 10.52 \\
Tetragonal & 1547 & \textbf{2.16} & \textbf{2.42} & \textbf{1.23} & 10.41 & 10.70 & 7.82 \\
Trigonal & 627 & \textbf{2.47} & \textbf{3.10} & \textbf{5.39} & 12.07 & 12.10 & 20.17 \\
Hexagonal & 1296 & \textbf{0.77} & \textbf{0.68} & \textbf{1.68} & 4.67 & 4.45 & 17.31 \\
Cubic & 2149 & \textbf{0.20} & \textbf{0.20} & \textbf{0.54} & 21.01 & 15.99 & 21.48 \\
\midrule
\multicolumn{8}{l}{\emph{MP-Full (93K)}} \\
Triclinic & 5164 & \textbf{6.69} & \textbf{5.58} & \textbf{6.68} & 9.90 & 8.23 & 13.30 \\
Monoclinic & 6985 & \textbf{5.35} & \textbf{9.04} & \textbf{5.13} & 6.85 & 15.06 & 7.53 \\
Orthorhombic & 6419 & \textbf{4.11} & \textbf{3.21} & \textbf{3.69} & 7.34 & 6.23 & 9.61 \\
Tetragonal & 3424 & \textbf{1.99} & \textbf{2.25} & \textbf{1.46} & 8.27 & 8.14 & 7.21 \\
Trigonal & 1490 & \textbf{2.71} & \textbf{3.15} & \textbf{4.82} & 11.91 & 11.33 & 19.66 \\
Hexagonal & 2892 & \textbf{1.38} & \textbf{1.19} & \textbf{2.26} & 4.61 & 3.73 & 17.52 \\
Cubic & 4069 & \textbf{0.28} & \textbf{0.29} & \textbf{0.39} & 16.66 & 13.14 & 18.10 \\
\midrule
\multicolumn{8}{l}{\emph{MP-Full Aug (930K)}} \\
Triclinic & 5164 & \textbf{8.36} & \textbf{7.08} & \textbf{8.75} & 9.48 & 7.80 & 12.08 \\
Monoclinic & 6985 & 6.76 & \textbf{11.06} & 6.42 & \textbf{6.25} & 14.71 & \textbf{6.09} \\
Orthorhombic & 6419 & \textbf{5.66} & \textbf{4.74} & \textbf{5.25} & 7.35 & 6.67 & 8.94 \\
Tetragonal & 3424 & \textbf{3.42} & \textbf{3.75} & \textbf{2.25} & 8.65 & 8.70 & 6.23 \\
Trigonal & 1490 & \textbf{4.45} & \textbf{5.08} & \textbf{7.58} & 11.13 & 11.46 & 18.68 \\
Hexagonal & 2892 & \textbf{2.09} & \textbf{1.88} & \textbf{3.74} & 4.06 & 3.39 & 19.79 \\
Cubic & 4069 & \textbf{0.47} & \textbf{0.52} & \textbf{0.68} & 16.34 & 13.62 & 17.43 \\
\bottomrule
\end{tabular}
}
\label{tab:angle_by_system}
\end{table}

\subsubsection{MP20}\label{supp_info:mp20_res}
We include parity plots for MP20 without augmentation.
\begin{figure}[htbp]
    \centering
    \begin{subfigure}[t]{0.7\textwidth}
        \centering
        \includegraphics[width=\textwidth]{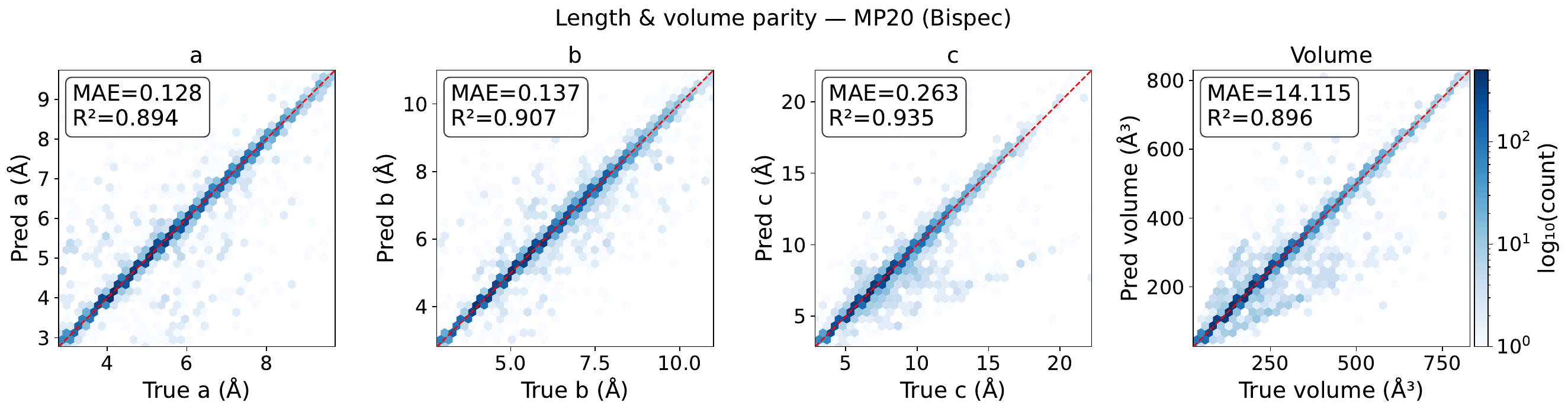}
    \end{subfigure}
    \hfill
    \begin{subfigure}[t]{0.7\textwidth}
        \centering
        \includegraphics[width=\textwidth]{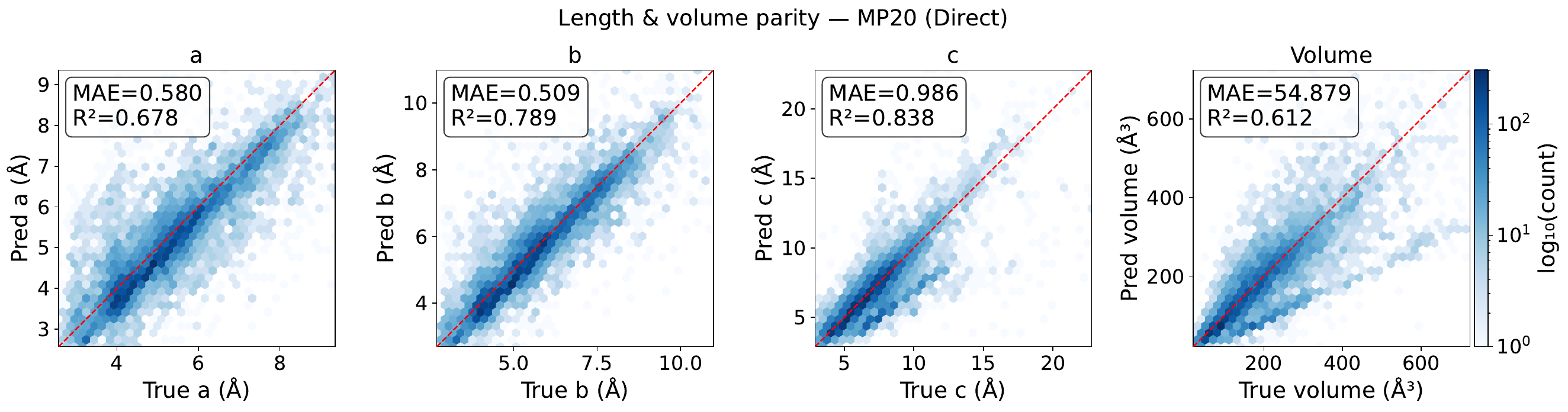}
    \end{subfigure}
    \caption{Predicted vs. true for $a,b,c$ and volume for bispec + inversion and directly predicting parameters using the MP20 dataset (no augmentations).}
    \label{fig:pred_true_parity_mp20}
\end{figure}

\begin{figure}[htbp]
    \centering
    \begin{subfigure}[t]{0.7\textwidth}
        \centering
        \includegraphics[width=\textwidth]{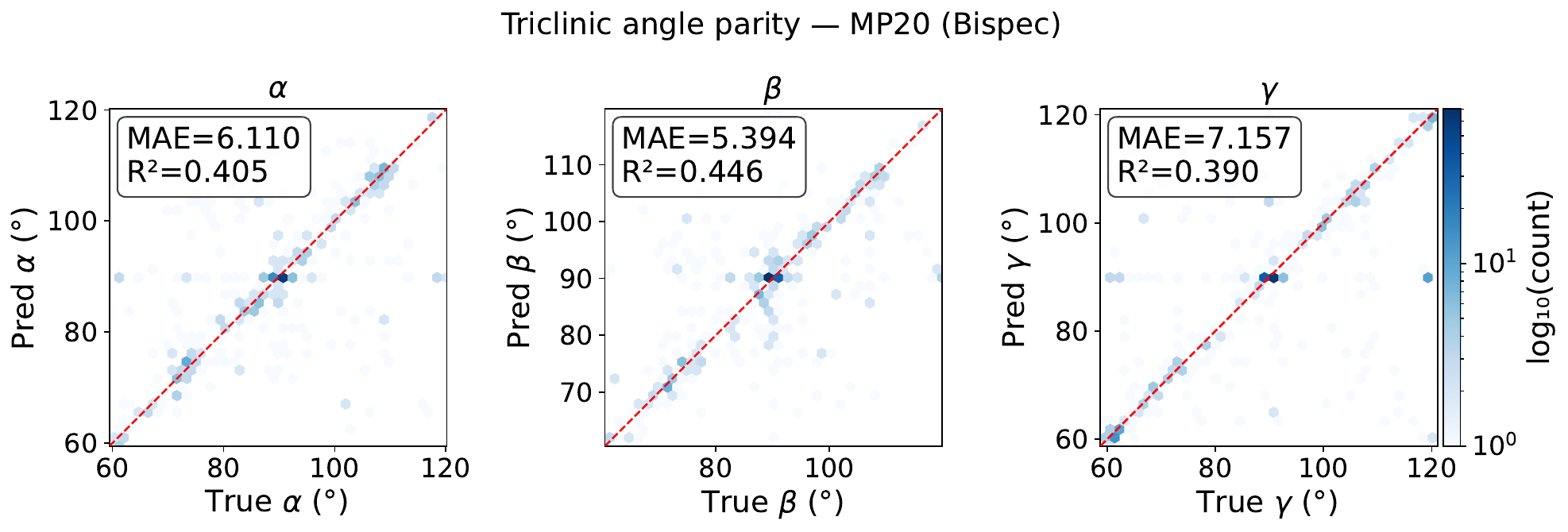}
    \end{subfigure}
    \hfill
    \begin{subfigure}[t]{0.7\textwidth}
        \centering
        \includegraphics[width=\textwidth]{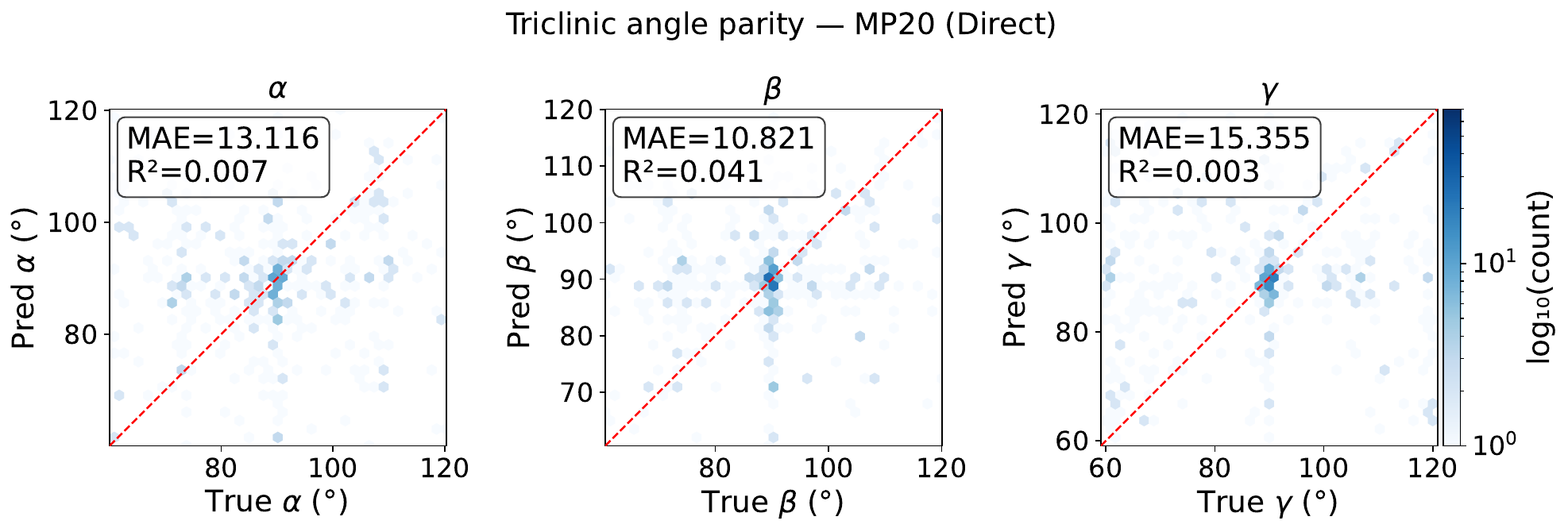}
    \end{subfigure}
    \caption{Parity plots for triclinic systems for the MP20 test set trained on MP20 (no augmentations).}
    \label{fig:mp20_triclinicparity}
\end{figure}

\begin{figure}[htbp]
    \centering
    \includegraphics[width=0.7\textwidth]{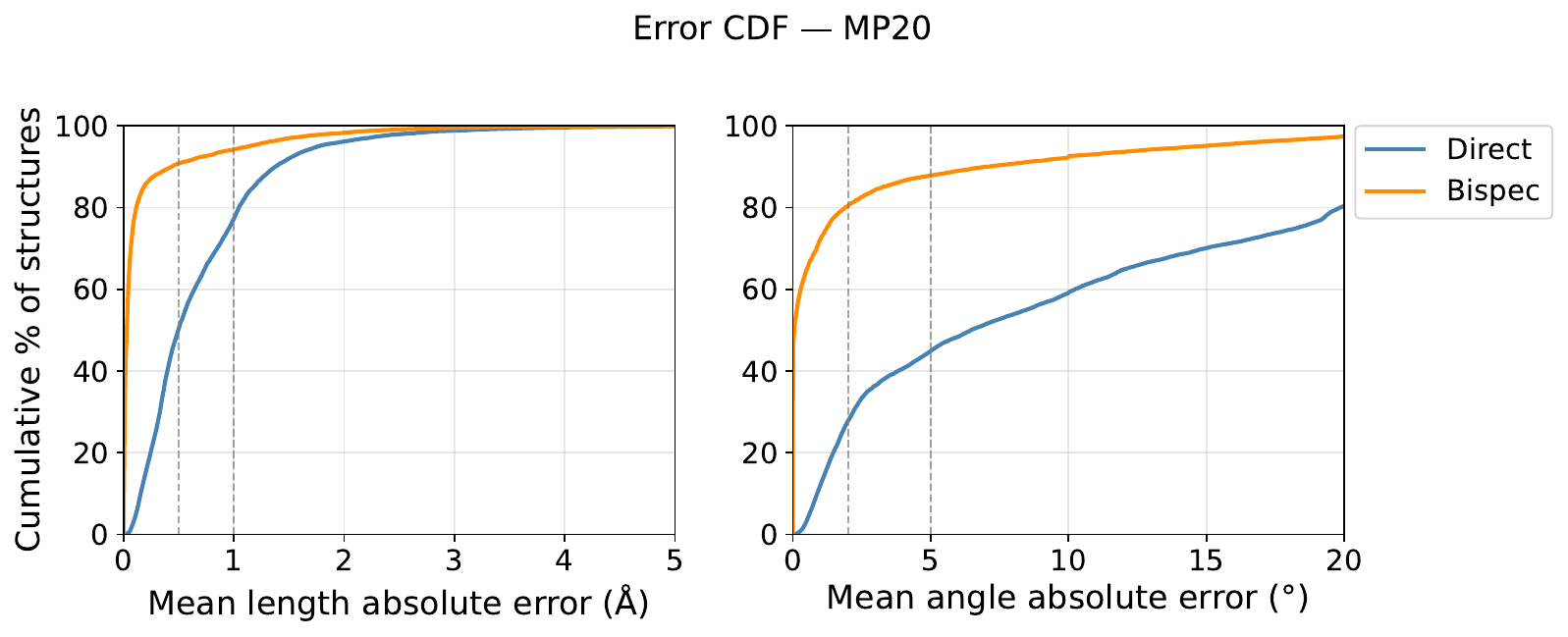}
    \caption{Cumulative distribution function for length and angle MAE MP20.
    }
    \label{fig:error_cdf_mp20}
\end{figure}

\begin{figure}[htbp]
    \centering
    \begin{subfigure}[t]{0.7\textwidth}
        \centering
        \includegraphics[width=\textwidth]{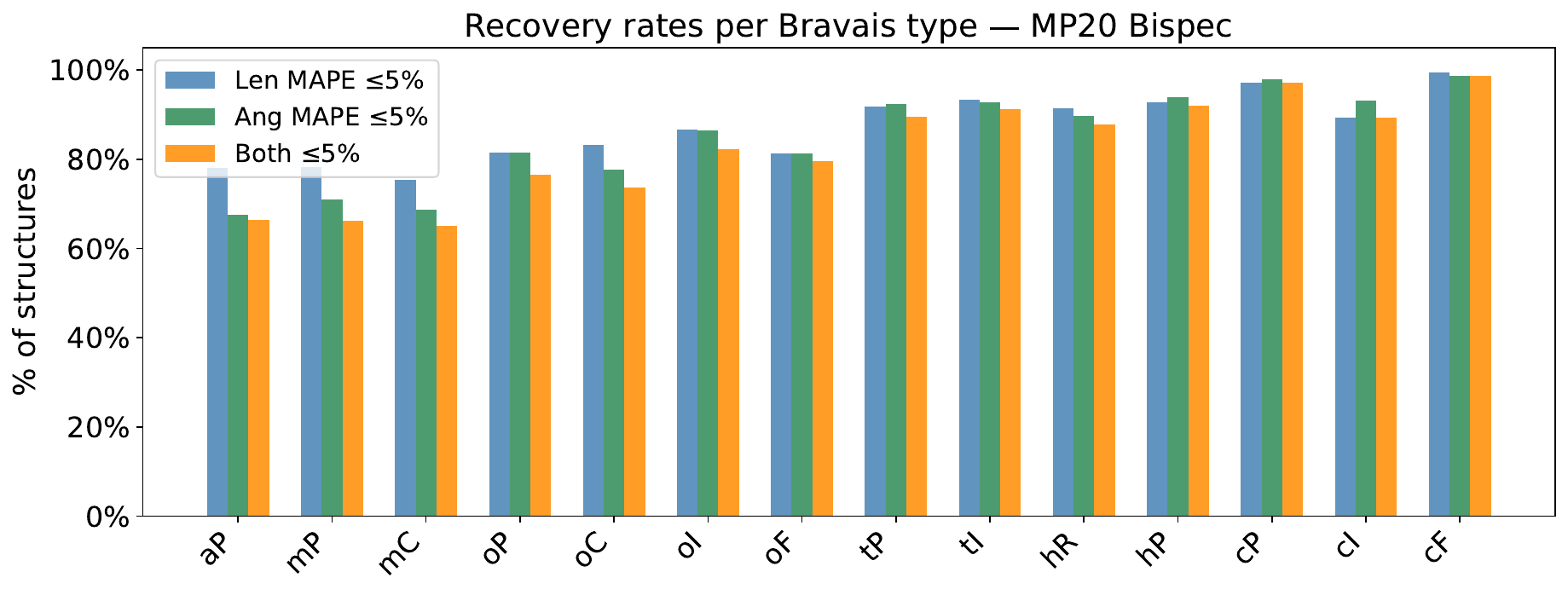}
    \end{subfigure}
    \hfill
    \begin{subfigure}[t]{0.7\textwidth}
        \centering
        \includegraphics[width=\textwidth]{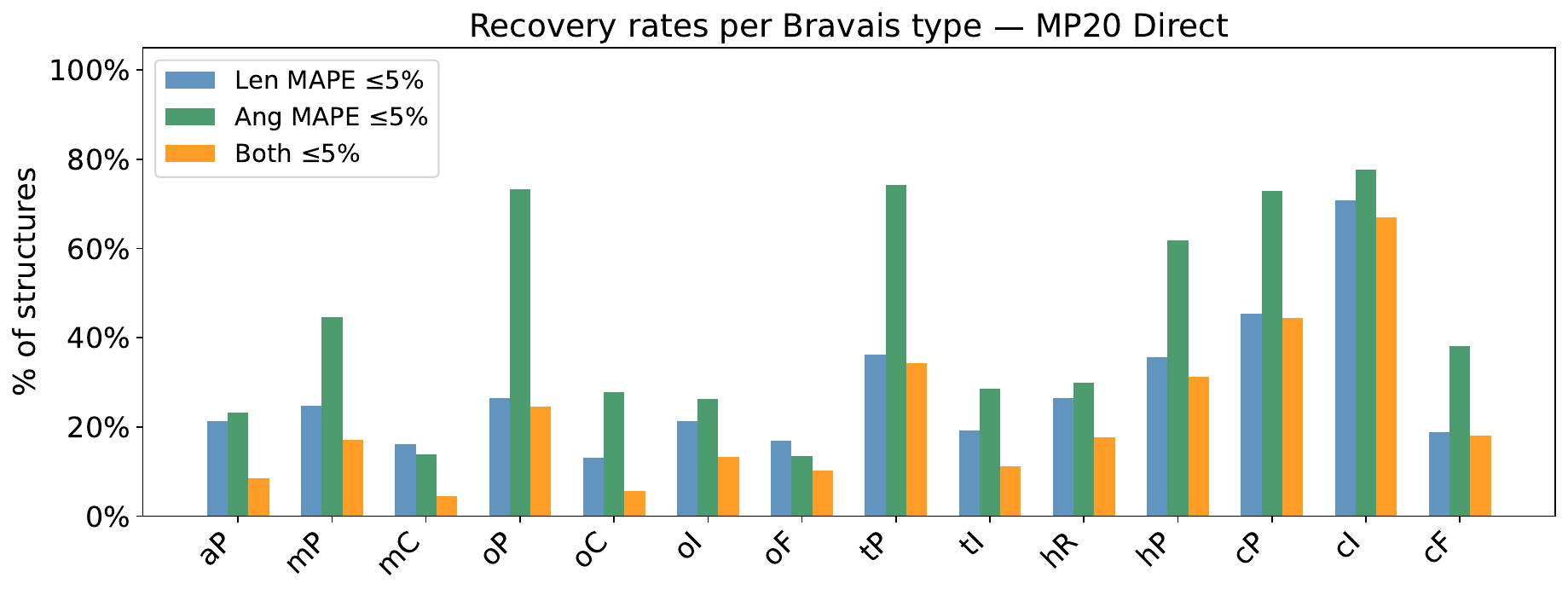}
    \end{subfigure}
    \caption{Recovery rates per Bravais lattice for the MP20 dataset (no augmentations).}
    \label{fig:mp20_threshold}
\end{figure}

\clearpage
\subsubsection{MP20 Augmented}\label{supp_info:mp20aug_res}

\begin{figure}[htbp]
    \centering
    \includegraphics[width=0.7\textwidth]{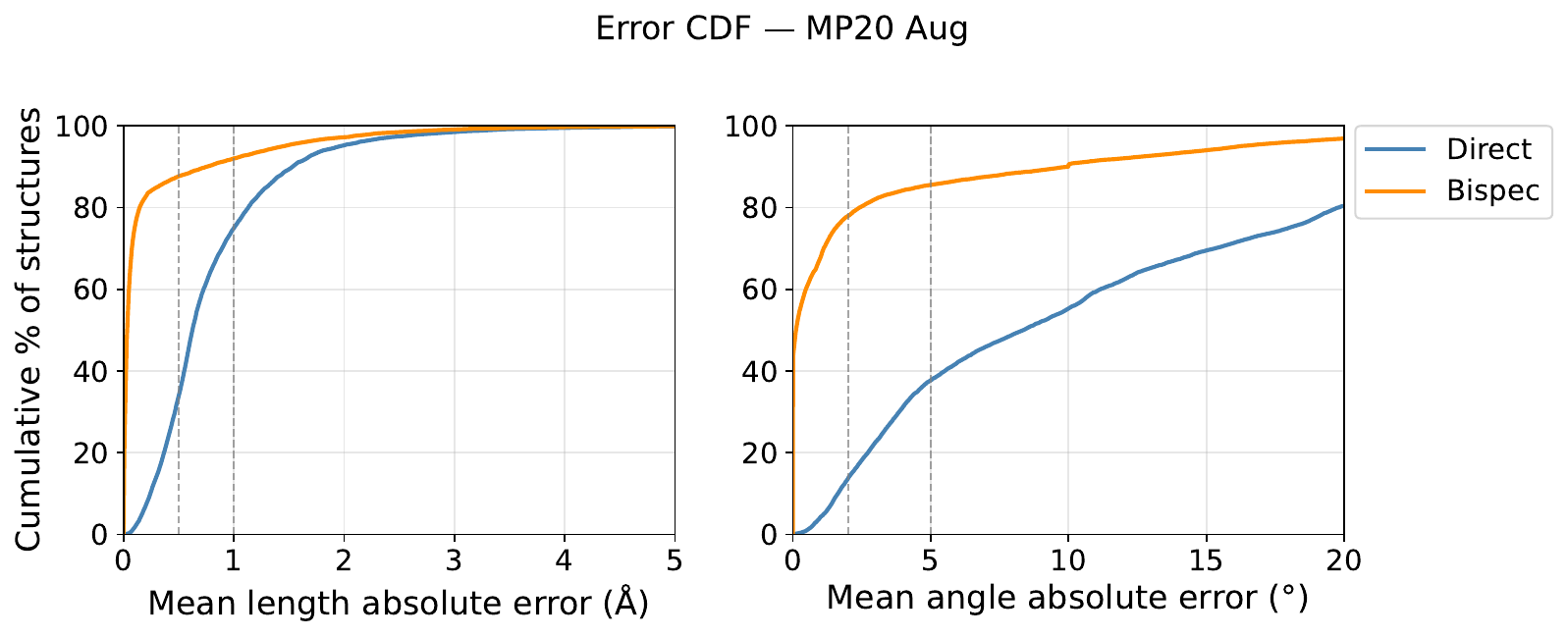}
    \caption{Cumulative distribution function for length and angle MAE MP20 augmented.
    }
    \label{fig:error_cdf_mp20_aug}
\end{figure}

\begin{figure}[htbp]
    \centering
    \begin{subfigure}[t]{0.7\textwidth}
        \centering
        \includegraphics[width=\textwidth]{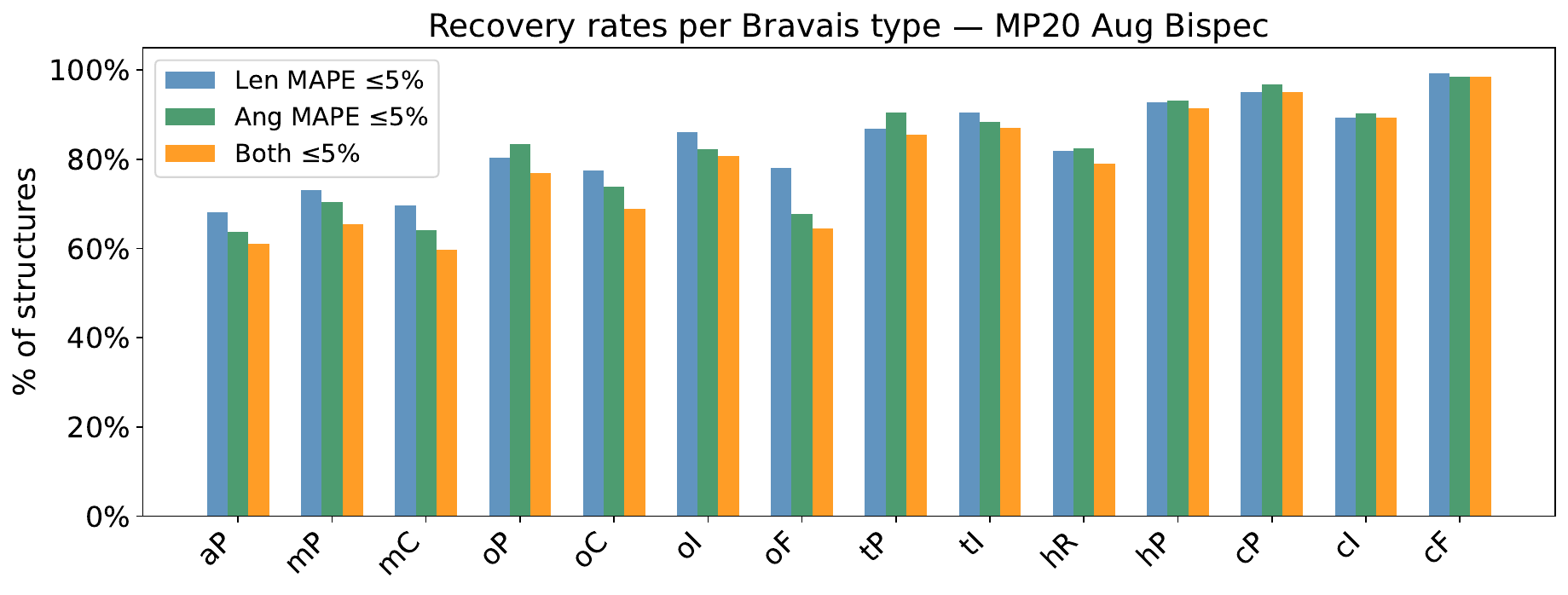}
    \end{subfigure}
    \hfill
    \begin{subfigure}[t]{0.7\textwidth}
        \centering
        \includegraphics[width=\textwidth]{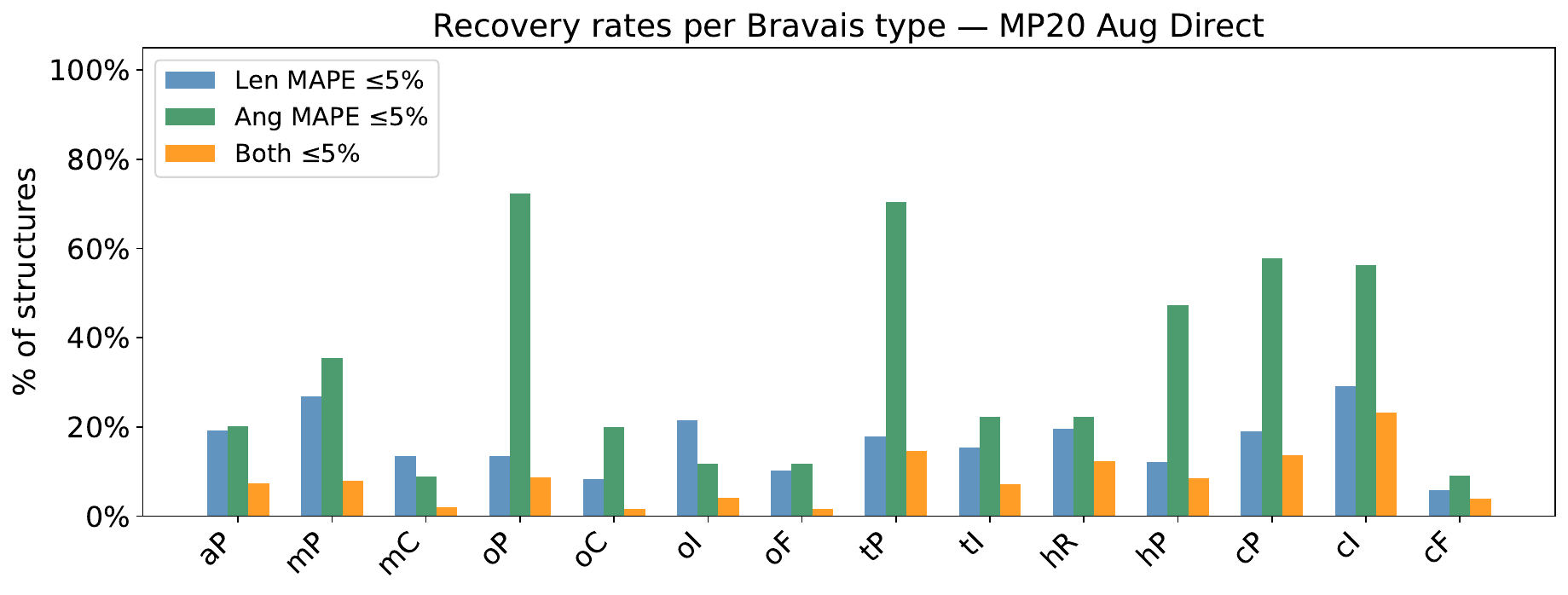}
    \end{subfigure}
    \caption{Recovery rates per Bravais lattice for the MP20 dataset augmented.}
    \label{fig:mp20aug_threshold}
\end{figure}

\begin{figure}[htbp]
    \centering
    \begin{subfigure}[t]{0.7\textwidth}
        \centering
        \includegraphics[width=\textwidth]{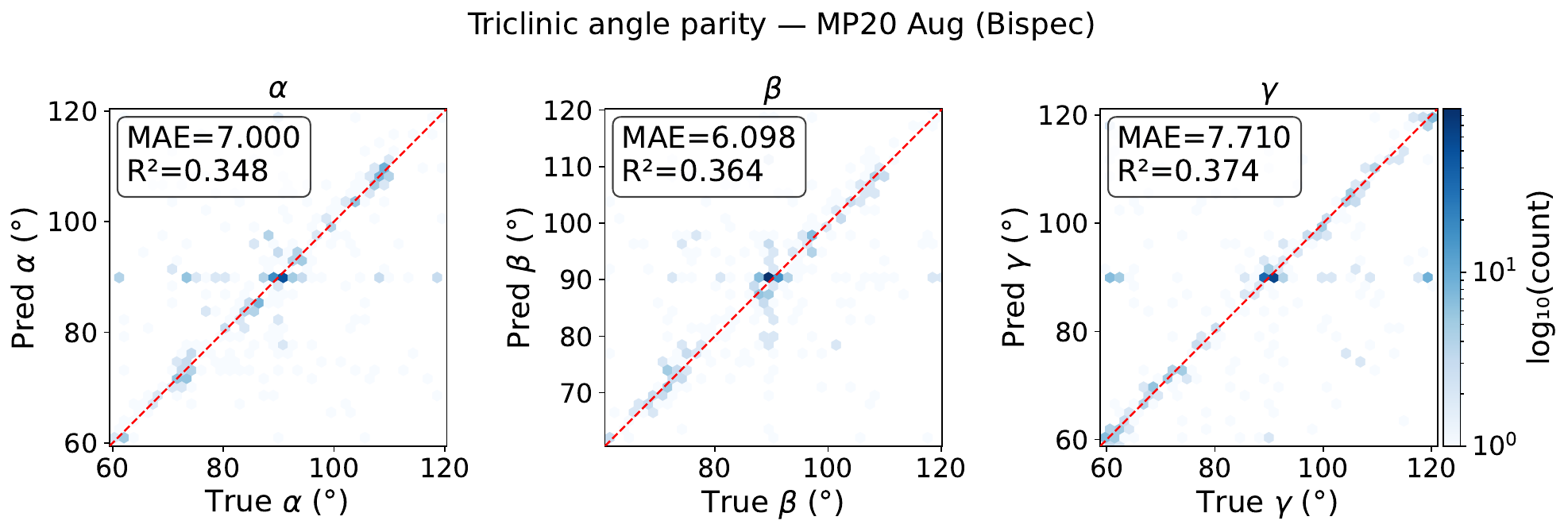}
    \end{subfigure}
    \hfill
    \begin{subfigure}[t]{0.7\textwidth}
        \centering
        \includegraphics[width=\textwidth]{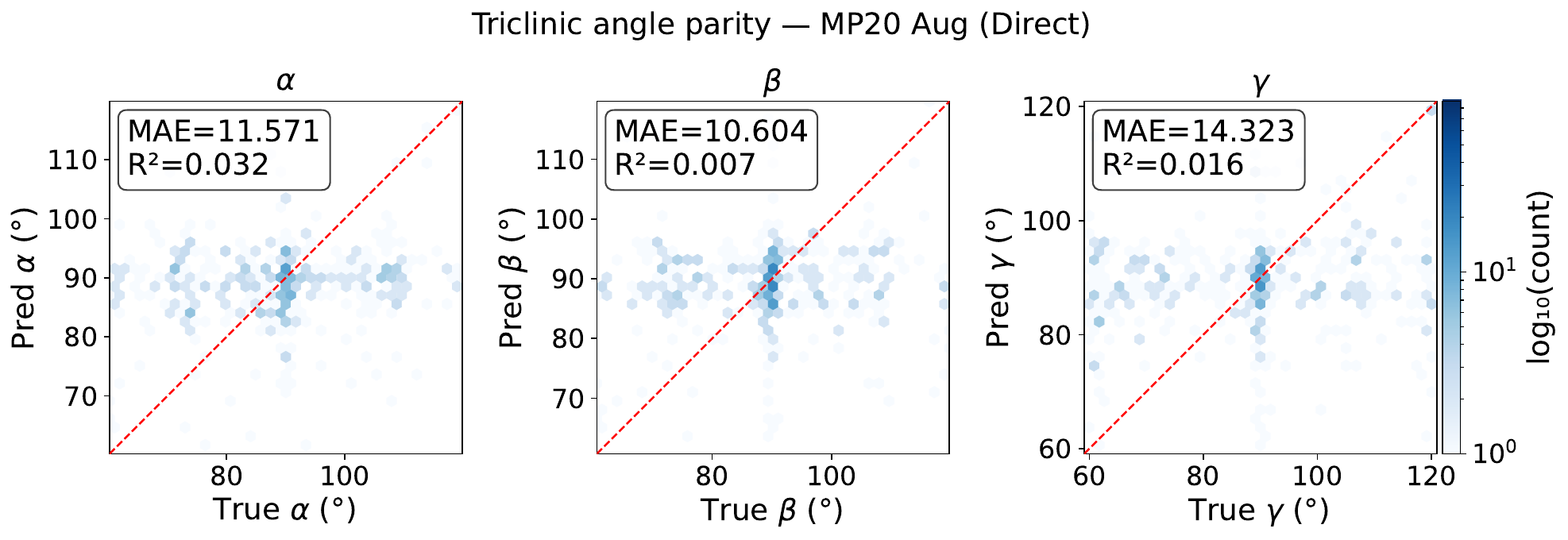}
    \end{subfigure}
    \caption{Parity plots for triclinic systems for the MP20 test set trained on MP20 Aug.}
    \label{fig:mp20aug_triclinicparity}
\end{figure}

\clearpage
\subsubsection{MP Full}\label{supp_info:mpfull_res}
We report length MAE/angle MAE per Bravais lattice training with the MPFull dataset (without augmentations) and evaluated on the MPFull test set.
\begin{figure}[htbp]
    \centering
    \includegraphics[width=1.0\textwidth]{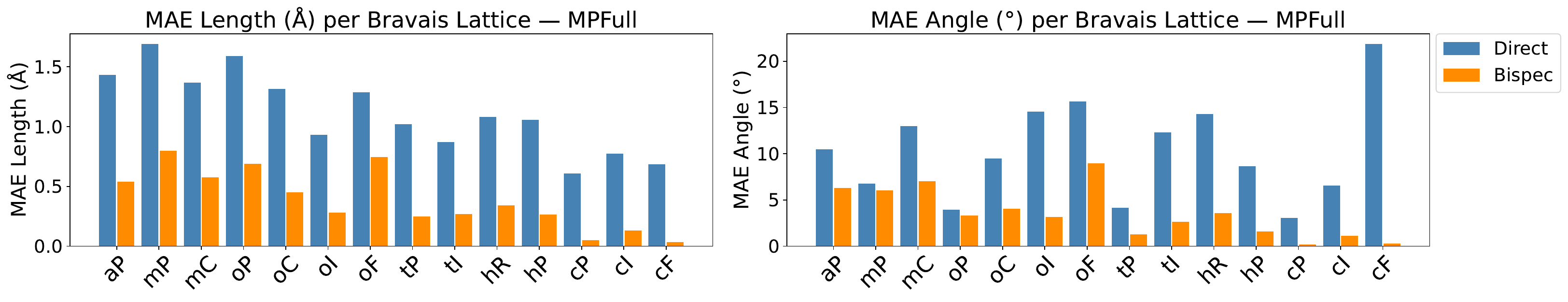}
    \caption{MAE for bispectrum + inversion compared to direct predictions per bravais lattice for MP full.
    }
    \label{fig:mpfull_res}
\end{figure}
\begin{figure}[htbp]
    \centering
    \begin{subfigure}[t]{0.7\textwidth}
        \centering
        \includegraphics[width=\textwidth]{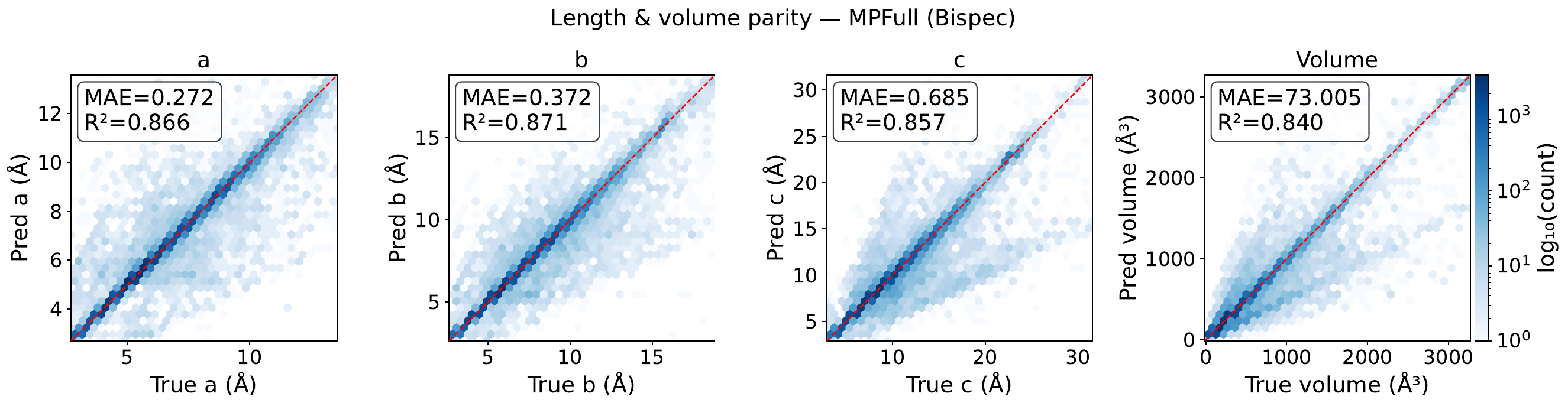}
    \end{subfigure}
    \hfill
    \begin{subfigure}[t]{0.7\textwidth}
        \centering
        \includegraphics[width=\textwidth]{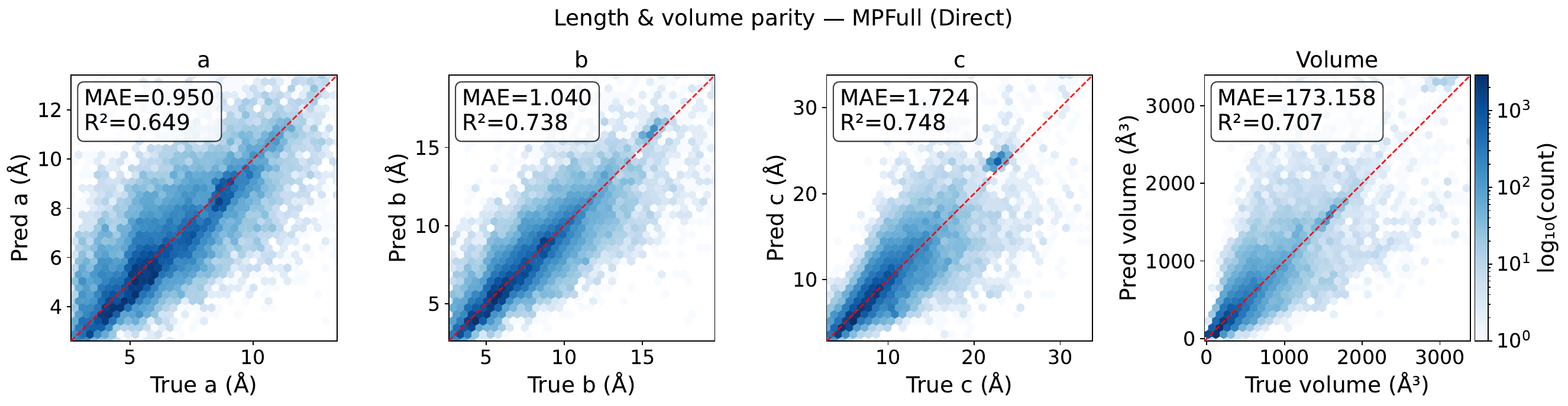}
    \end{subfigure}
    \caption{Predicted vs. true for $a,b,c$ and volume for bispec + inversion and directly predicting parameters using the MPFull dataset (no augmentations).}
    \label{fig:pred_true_parity_mpfull}
\end{figure}

\begin{figure}[htbp]
    \centering
    \begin{subfigure}[t]{0.7\textwidth}
        \centering
        \includegraphics[width=\textwidth]{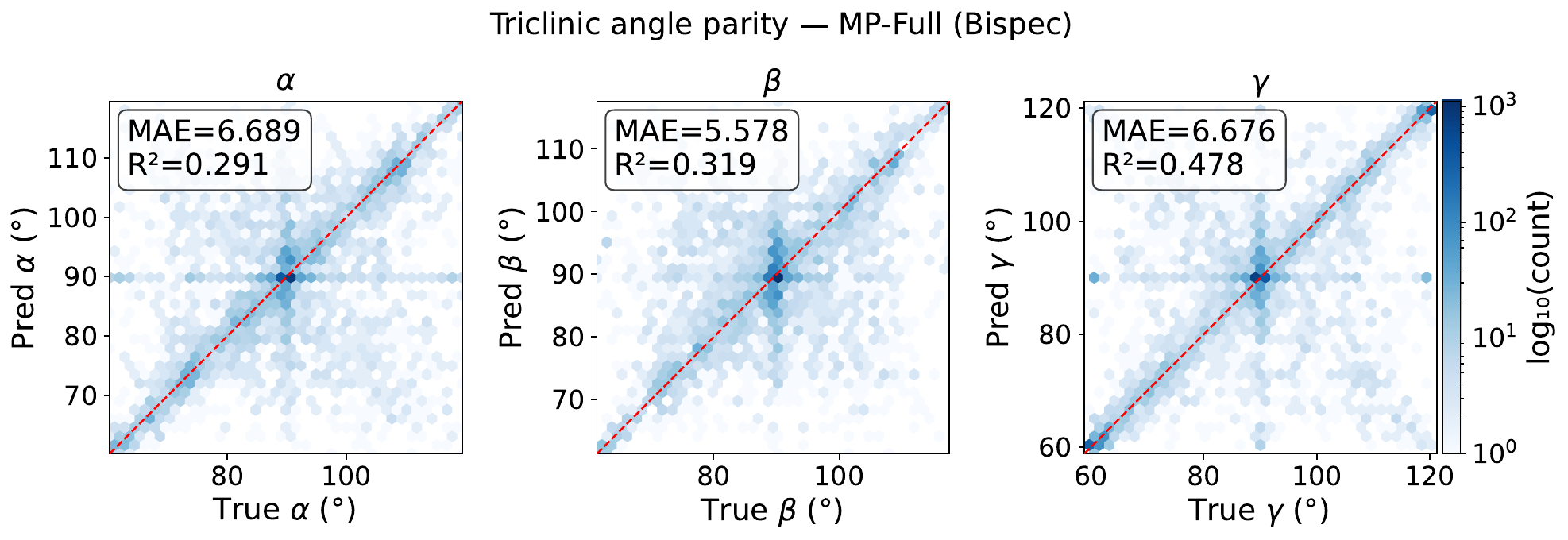}
    \end{subfigure}
    \hfill
    \begin{subfigure}[t]{0.7\textwidth}
        \centering
        \includegraphics[width=\textwidth]{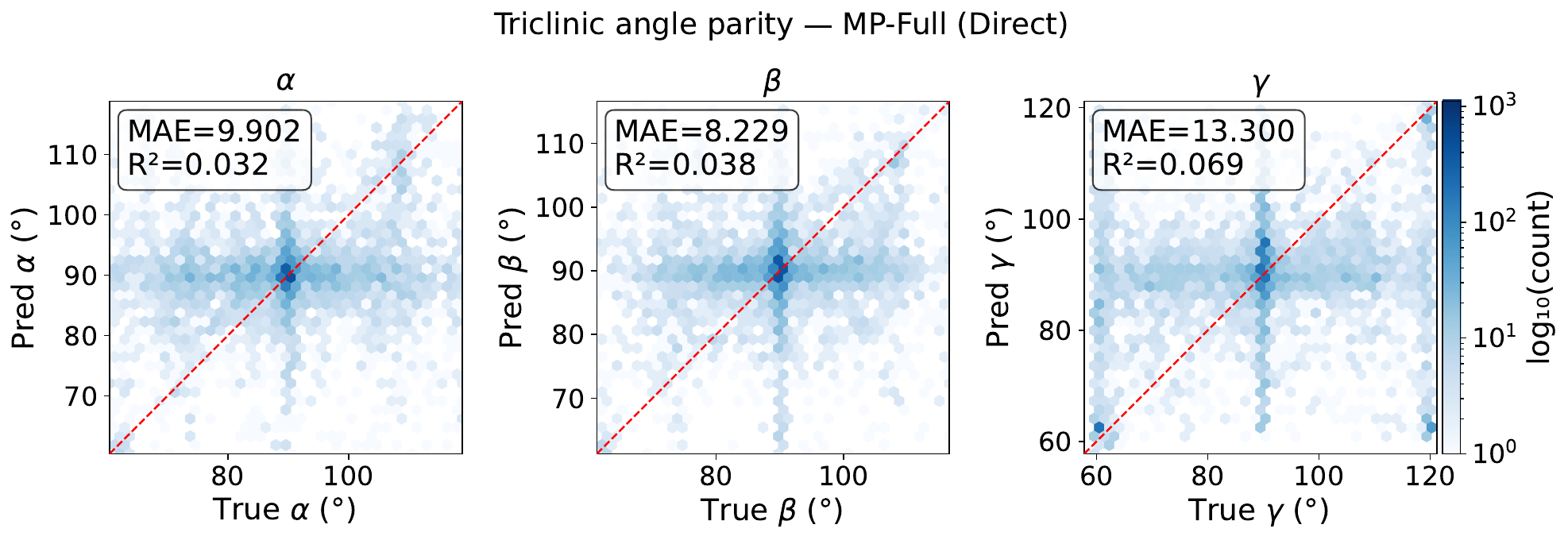}
    \end{subfigure}
    \caption{Parity plots for triclinic systems for the MP Full test set trained on MP Full (no augmentations).}
    \label{fig:pred_true_trinclinic_mpfull}
\end{figure}

\begin{figure}[htbp]
    \centering
    \includegraphics[width=0.7\textwidth]{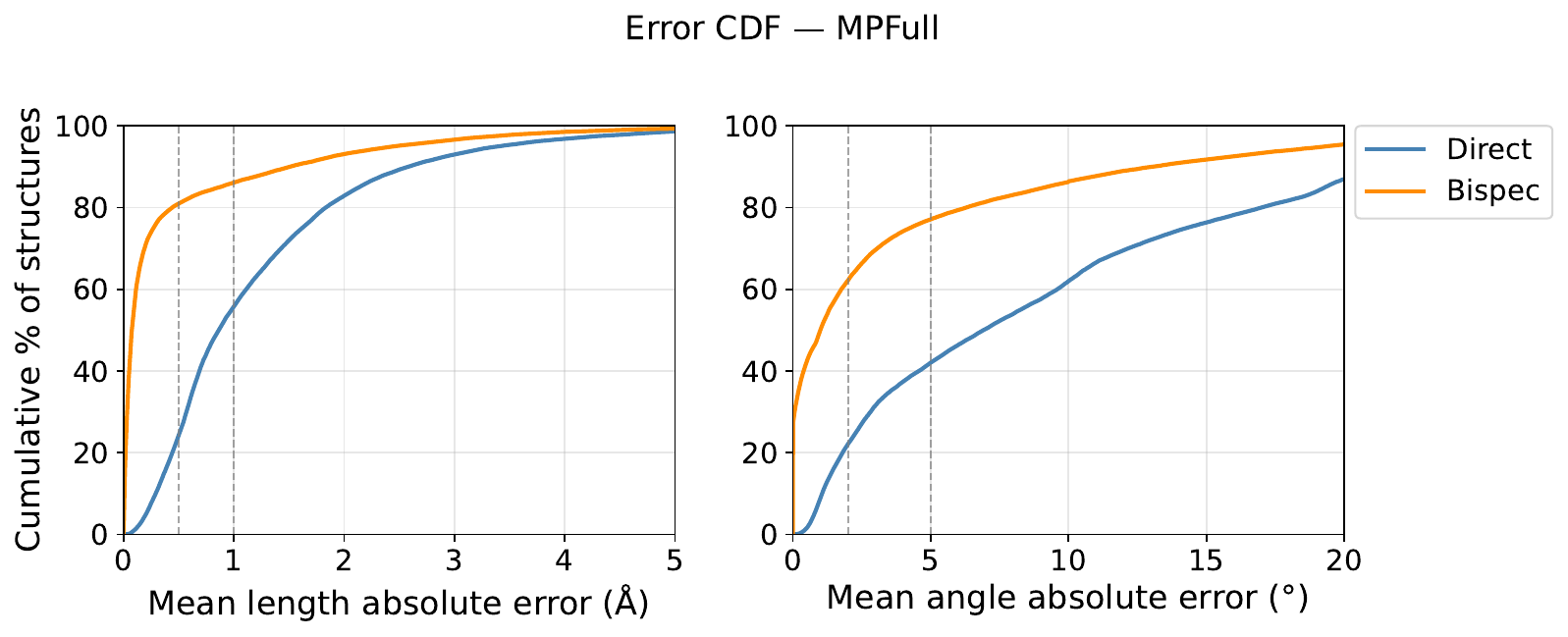}
    \caption{Cumulative distribution function for length and angle MAE MPFull.
    }
    \label{fig:error_cdf_mpfull}
\end{figure}

\begin{figure}[htbp]
    \centering
    \begin{subfigure}[t]{0.7\textwidth}
        \centering
        \includegraphics[width=\textwidth]{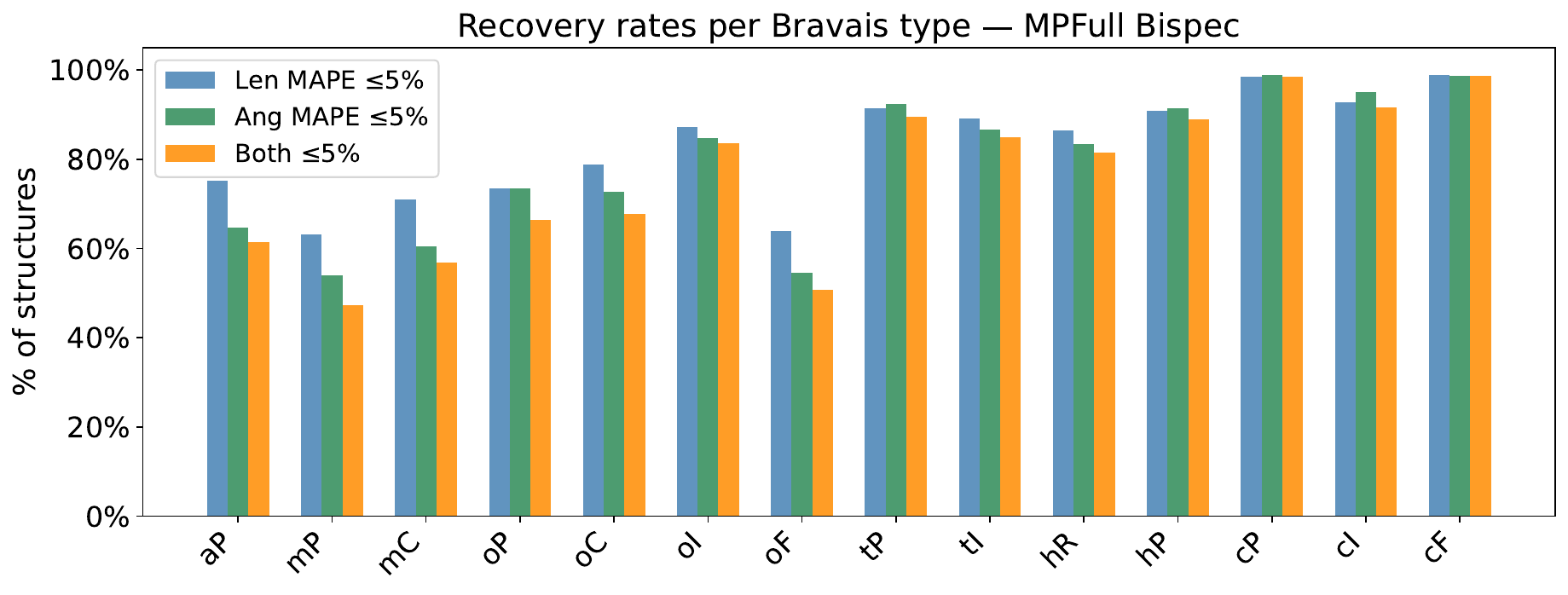}
    \end{subfigure}
    \hfill
    \begin{subfigure}[t]{0.7\textwidth}
        \centering
        \includegraphics[width=\textwidth]{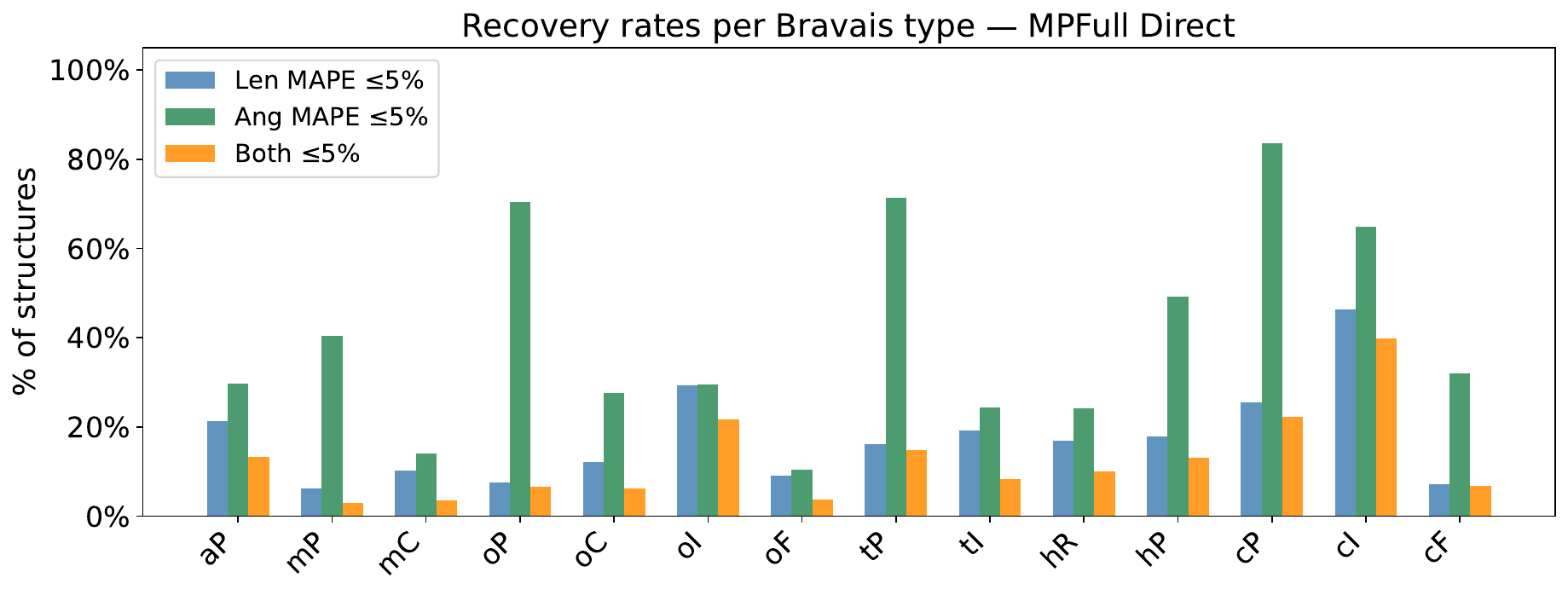}
    \end{subfigure}
    \caption{Recovery rates per Bravais lattice for the MPFull dataset (no augmentations).}
    \label{fig:mpfull_threshold}
\end{figure}

\clearpage
\subsubsection{MP Full Augmented}\label{supp_info:mpfull_aug_res}

\begin{figure}[htbp]
    \centering
    \begin{subfigure}[t]{0.7\textwidth}
        \centering
        \includegraphics[width=\textwidth]{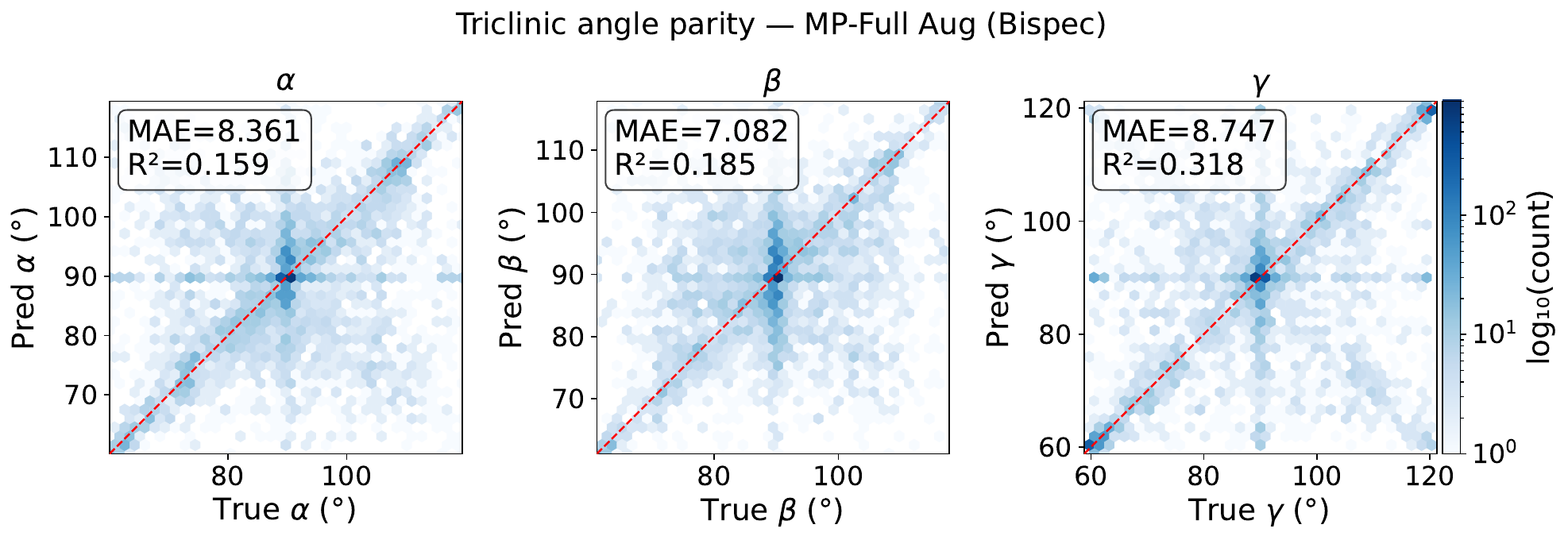}
    \end{subfigure}
    \hfill
    \begin{subfigure}[t]{0.7\textwidth}
        \centering
        \includegraphics[width=\textwidth]{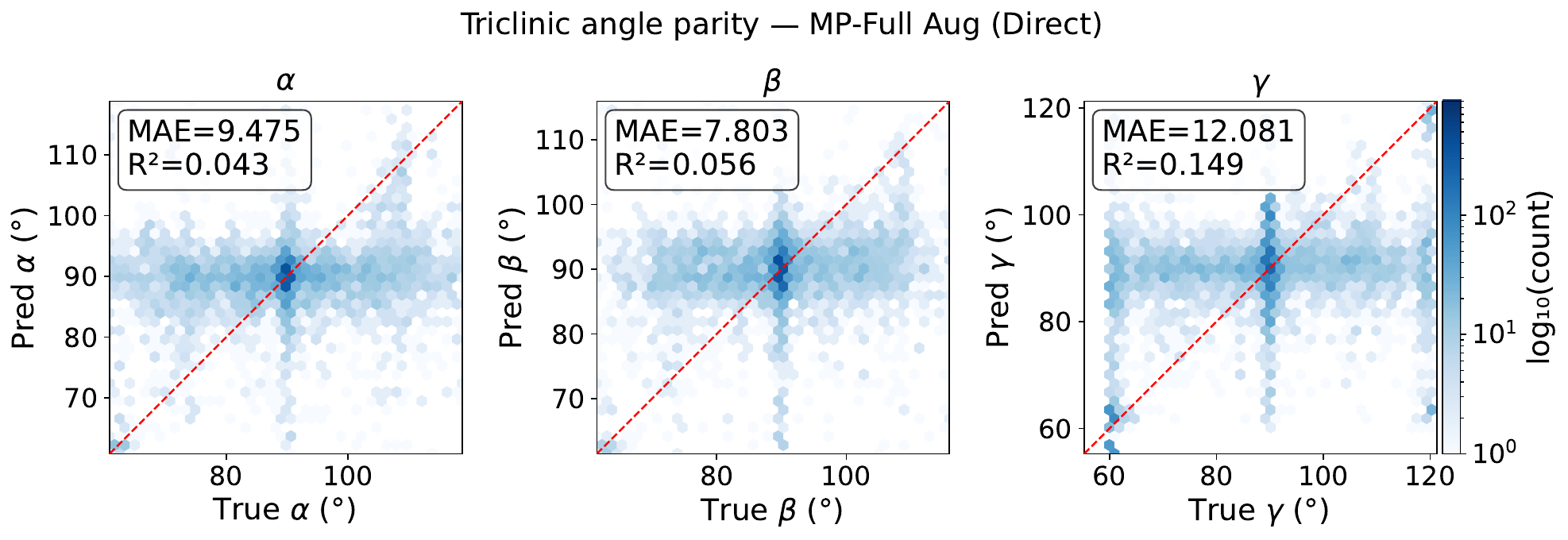}
    \end{subfigure}
    \caption{Parity plots for triclinic systems for the MP Full test set trained on MP Full Aug.}
    \label{fig:pred_true_trinclinic_mpfull_aug}
\end{figure}

\begin{figure}[htbp]
    \centering
    \includegraphics[width=0.7\textwidth]{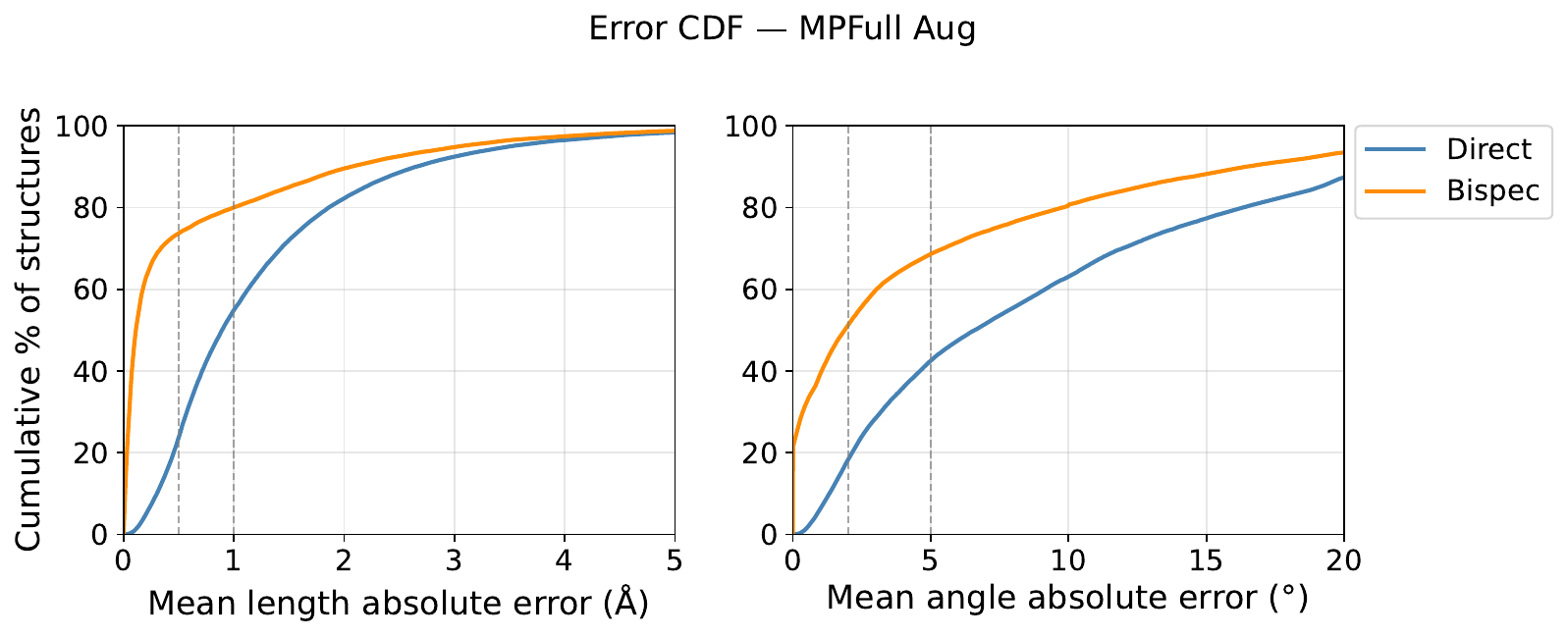}
    \caption{Cumulative distribution function for length and angle MAE MPFull augmented.
    }
    \label{fig:error_cdf_mpfull}
\end{figure}

\begin{figure}[htbp]
    \centering
    \begin{subfigure}[t]{0.7\textwidth}
        \centering
        \includegraphics[width=\textwidth]{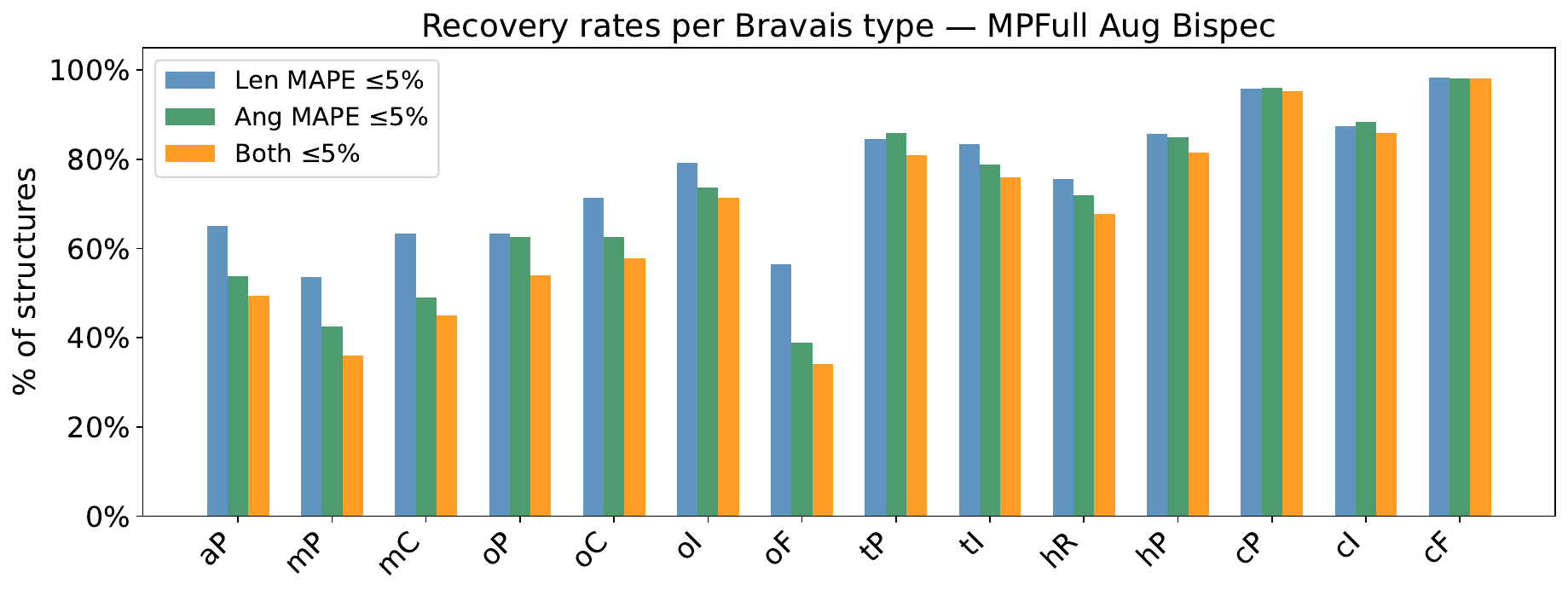}
    \end{subfigure}
    \hfill
    \begin{subfigure}[t]{0.7\textwidth}
        \centering
        \includegraphics[width=\textwidth]{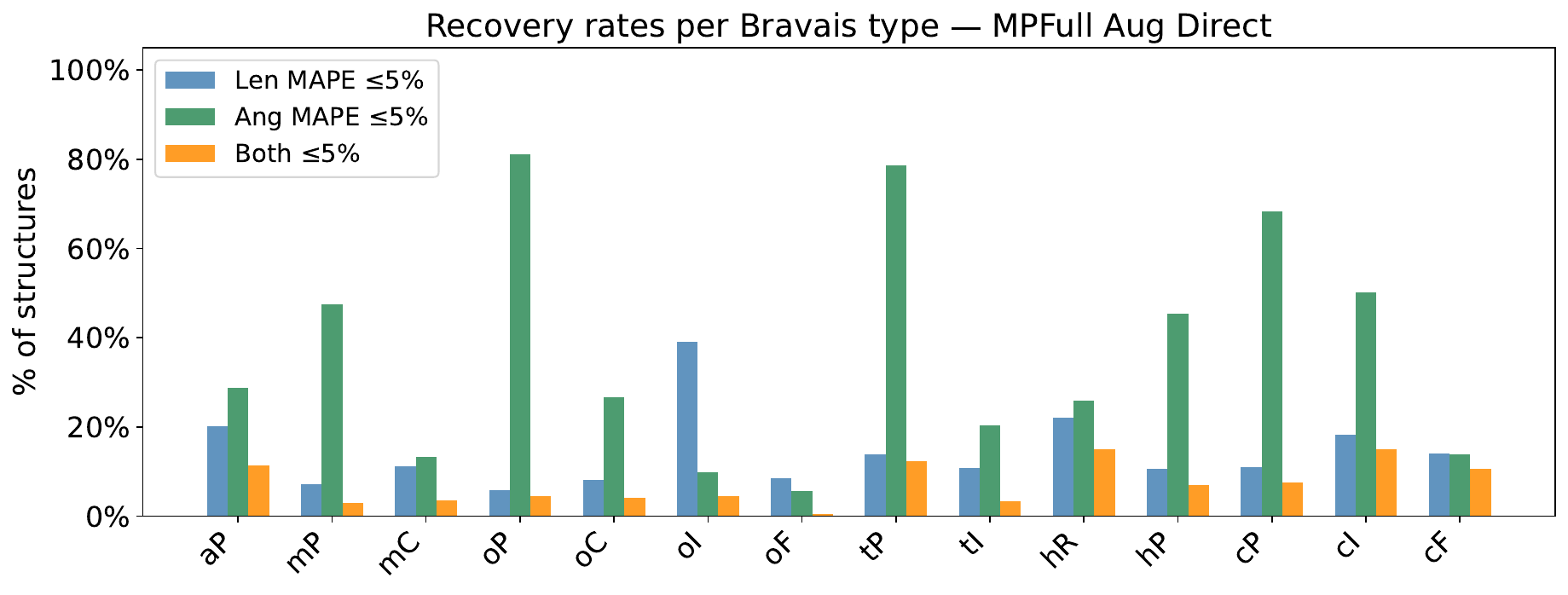}
    \end{subfigure}
    \caption{Recovery rates per Bravais lattice for the MPFull Augmented dataset.}
    \label{fig:mpfullaug_threshold}
\end{figure}

\clearpage
\subsubsection{RRUFF Datasets}\label{supp_info:rruff_res}
\begin{figure}[htbp]
    \centering
    \begin{subfigure}[t]{0.7\textwidth}
        \centering
        \includegraphics[width=\textwidth]{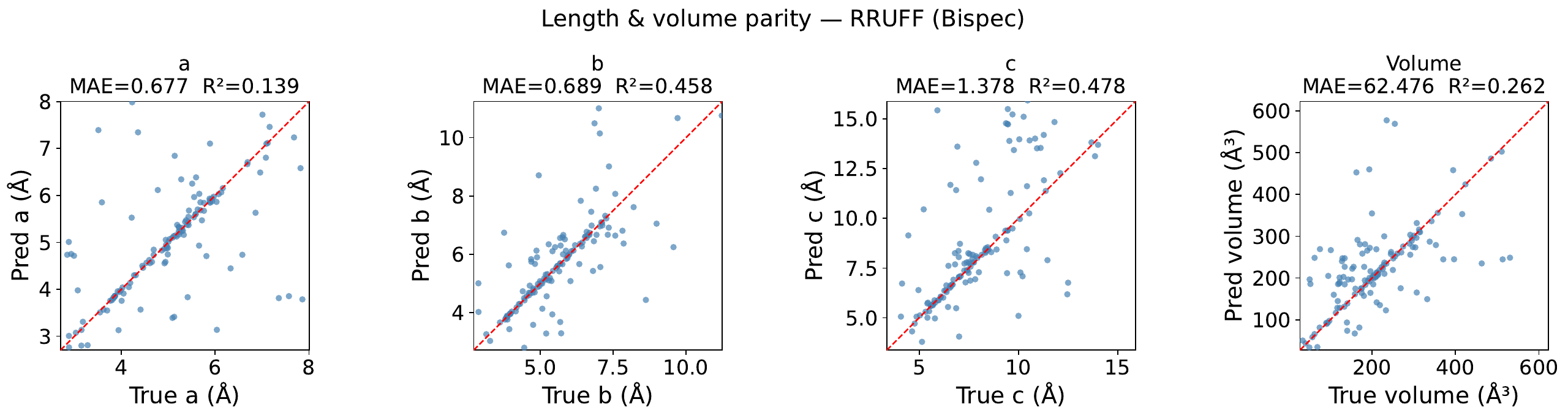}
    \end{subfigure}
    \hfill
    \begin{subfigure}[t]{0.7\textwidth}
        \centering
        \includegraphics[width=\textwidth]{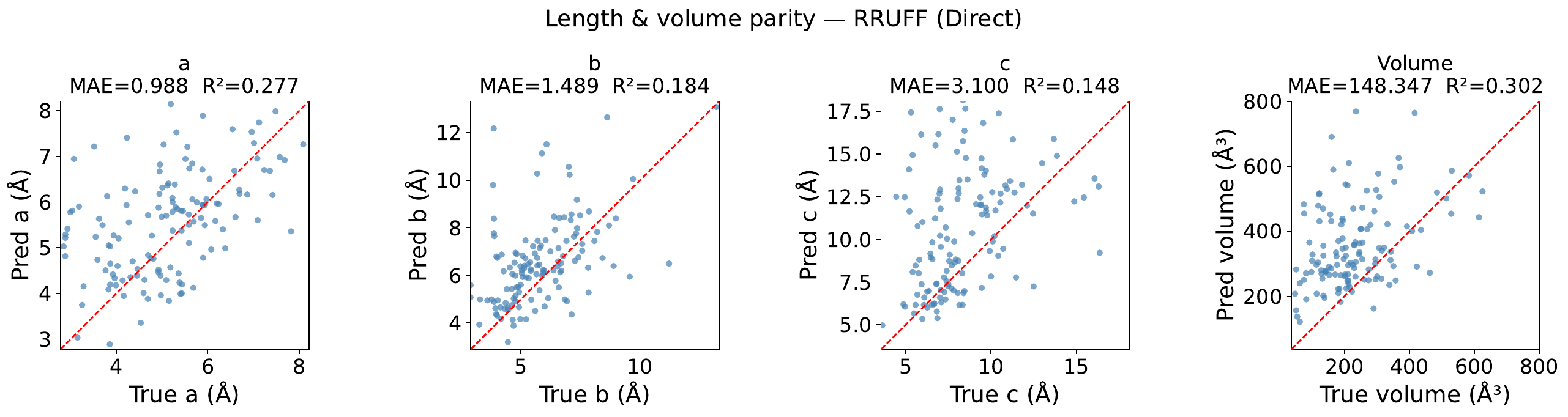}
    \end{subfigure}
    \begin{subfigure}[t]{0.7\textwidth}
        \centering
        \includegraphics[width=\textwidth]{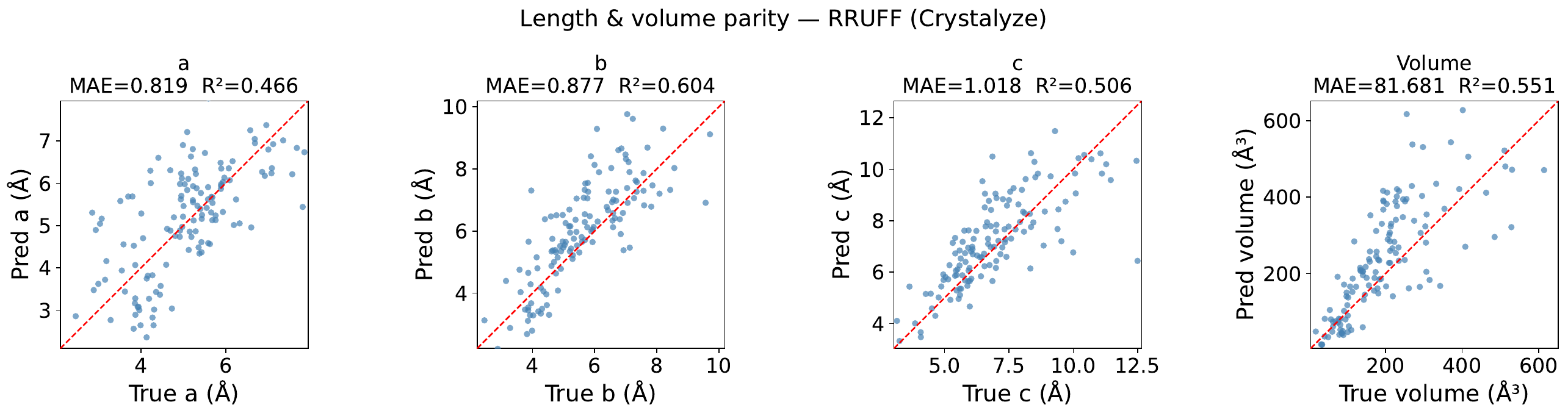}
    \end{subfigure}
    \caption{Predicted vs. true for $a,b,c$ and volume for bispec + inversion, directly predicting parameters, and Crystalyze comparison on the Crystalyze RRUFF test set using the MP20 augmented dataset.}
    \label{fig:crystalyze_rruff_scatter}
\end{figure}

We first include additional plots for RRUFF evaluated with the Crystalyze test split (for patterns with fewer than 20 atoms in the unit cell). Scatterplots are shown in \Cref{fig:crystalyze_rruff_scatter}. We observe clear improvement with respect to $a,b$ axes when predicting the bispectrum. $c$ has more failures, perhaps indicative of the need for better or more data augmentation, see \Cref{supp_info:augfuture}.

\begin{figure}[htbp]
    \centering
    \begin{subfigure}[t]{0.7\textwidth}
        \centering
        \includegraphics[width=\textwidth]{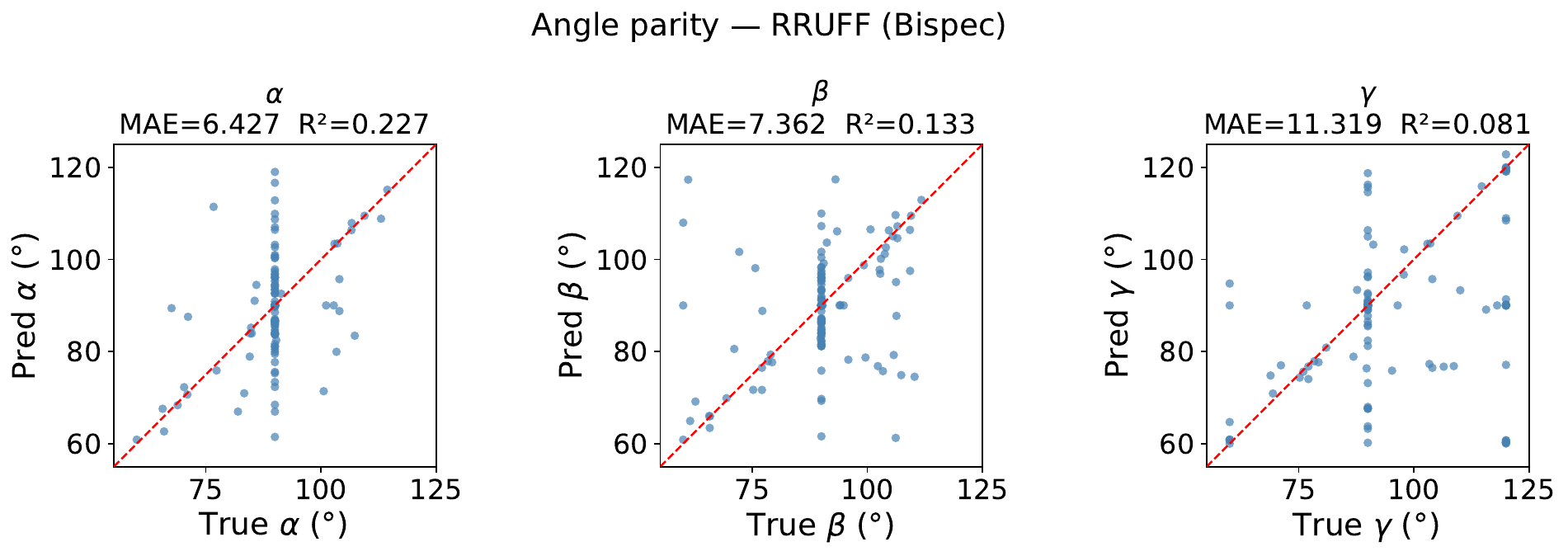}
    \end{subfigure}
    \hfill
    \begin{subfigure}[t]{0.7\textwidth}
        \centering
        \includegraphics[width=\textwidth]{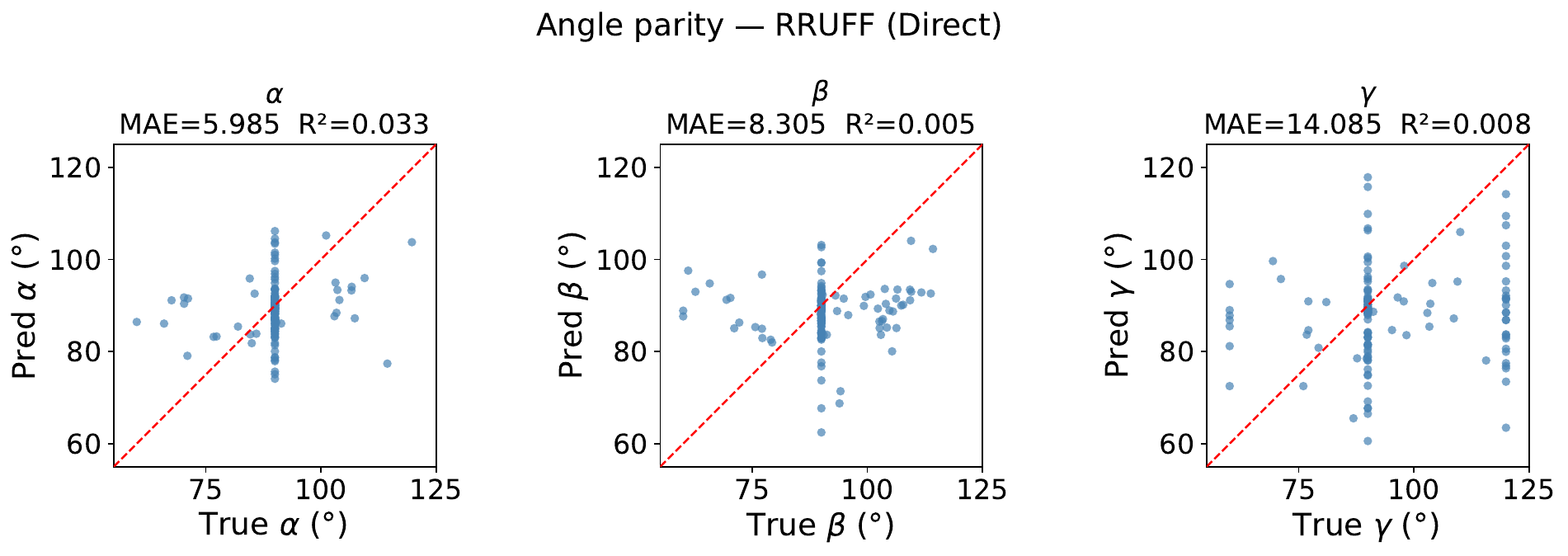}
    \end{subfigure}
    \begin{subfigure}[t]{0.7\textwidth}
        \centering
        \includegraphics[width=\textwidth]{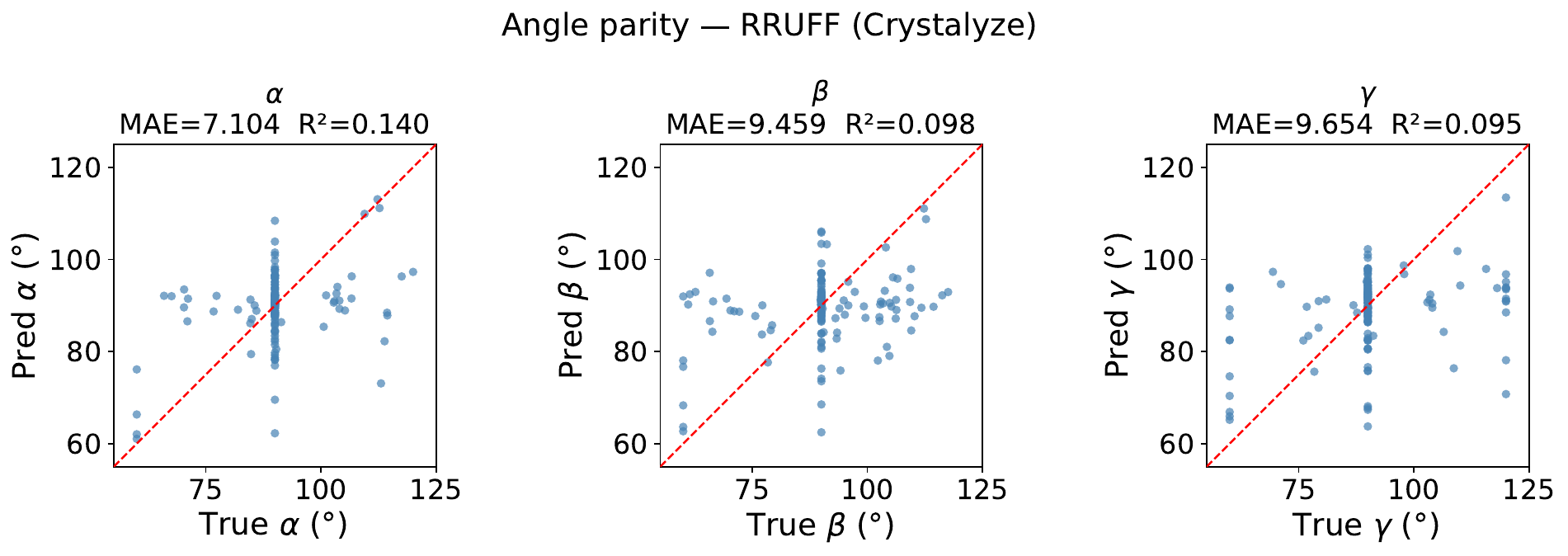}
    \end{subfigure}
    \caption{Predicted vs. true $\alpha,\beta,\gamma$ on the same 134 RRUFF patterns, for Crystalyze, direct prediction and bispec + inversion. The vertical columns at 90° and 120° are structures whose symmetry-fixed angles were predicted incorrectly.}
    \label{fig:crystalyze_rruff_scatter_angles}
\end{figure}

Nonetheless, our data augmentation methodology still clearly aids in model performance.

\begin{figure}[htbp]
    \centering
    \begin{subfigure}[t]{0.7\textwidth}
        \centering
        \includegraphics[width=\textwidth]{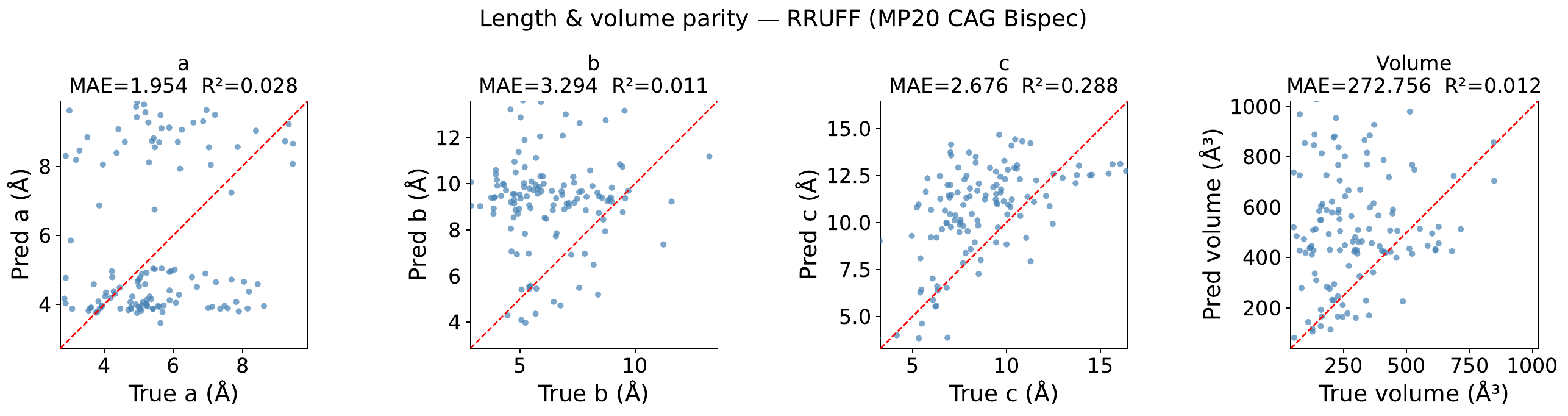}
    \end{subfigure}
    \hfill
    \begin{subfigure}[t]{0.7\textwidth}
        \centering
        \includegraphics[width=\textwidth]{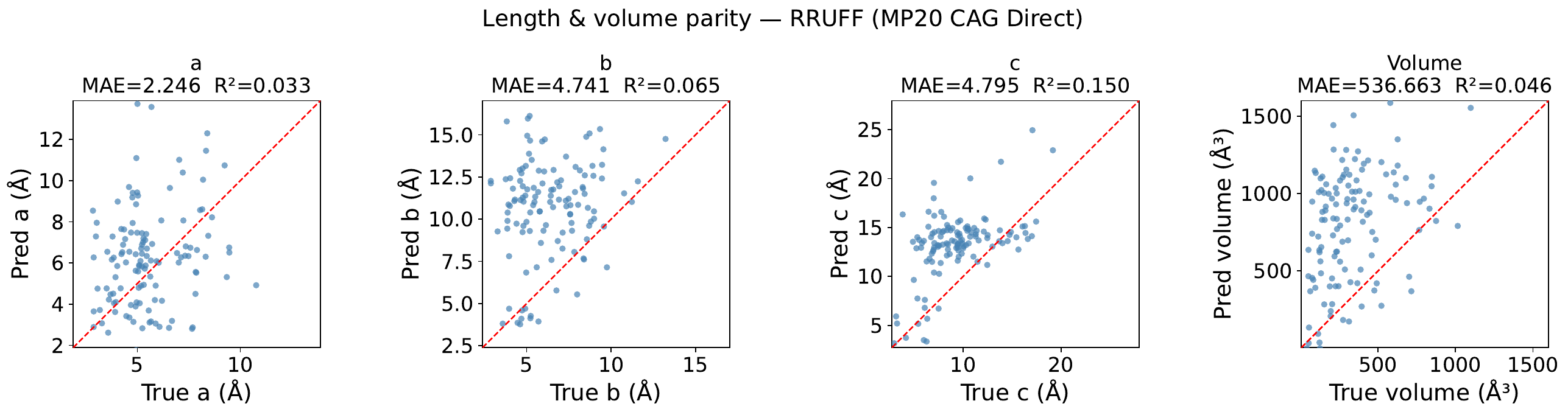}
    \end{subfigure}
    \caption{Predicted vs. true for $a,b,c$ and volume for bispec + inversion and directly predicting parameters using the MP20 dataset with no augmentations evaluated on the Crystalyze RRUFF test set.}
    \label{fig:crystalyze_rruff_scatter_noaug}
\end{figure}

\begin{figure}[htbp]
    \centering
    \begin{subfigure}[t]{0.7\textwidth}
        \centering
        \includegraphics[width=\textwidth]{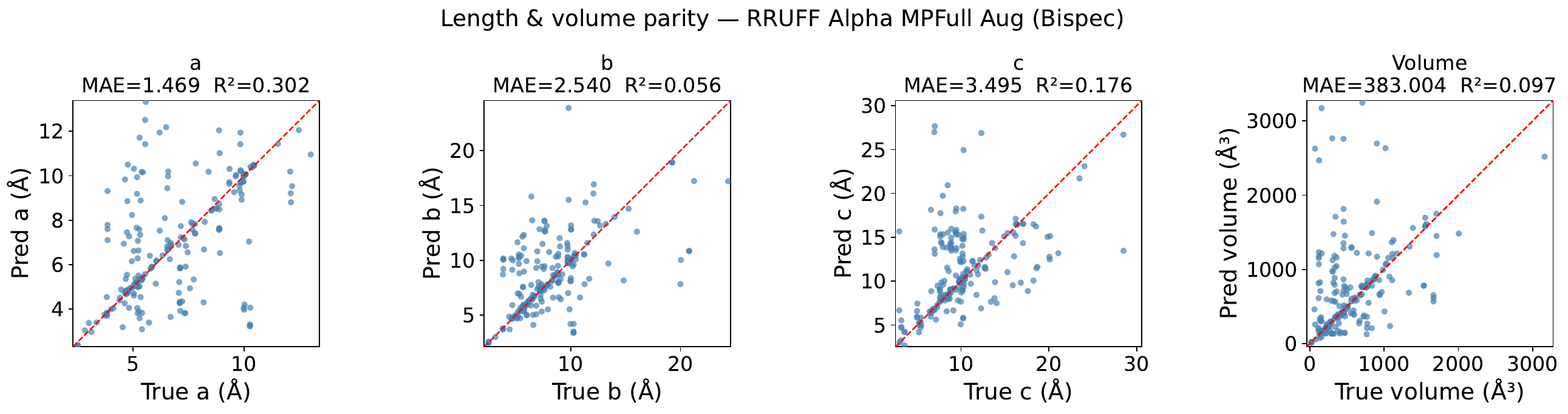}
    \end{subfigure}
    \hfill
    \begin{subfigure}[t]{0.7\textwidth}
        \centering
        \includegraphics[width=\textwidth]{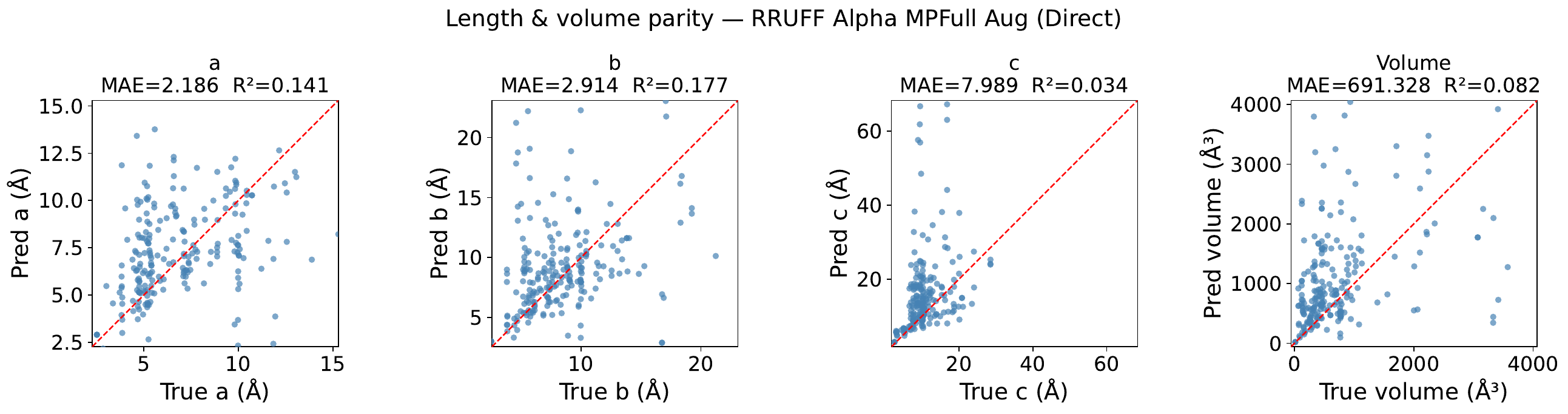}
    \end{subfigure}
    \caption{Predicted vs. true for $a,b,c$ and volume for bispec + inversion and directly predicting parameters using the MPFull dataset with augmentations evaluated on the RRUFF Alpha test set.}
    \label{fig:alpha_rruff_scatter_aug}
\end{figure}

We next include additional analysis for the more difficult RRUFF Alpha test set, which is not limited to 20 atoms per unit cell. We do still observe our data augmentation strategy aiding in model performance. Without data augmentation, the model is essentially unable to generalize to experimental data, see \Cref{fig:alpha_rruff_scatter_aug}.

\begin{figure}[htbp]
    \centering
    \begin{subfigure}[t]{0.7\textwidth}
        \centering
        \includegraphics[width=\textwidth]{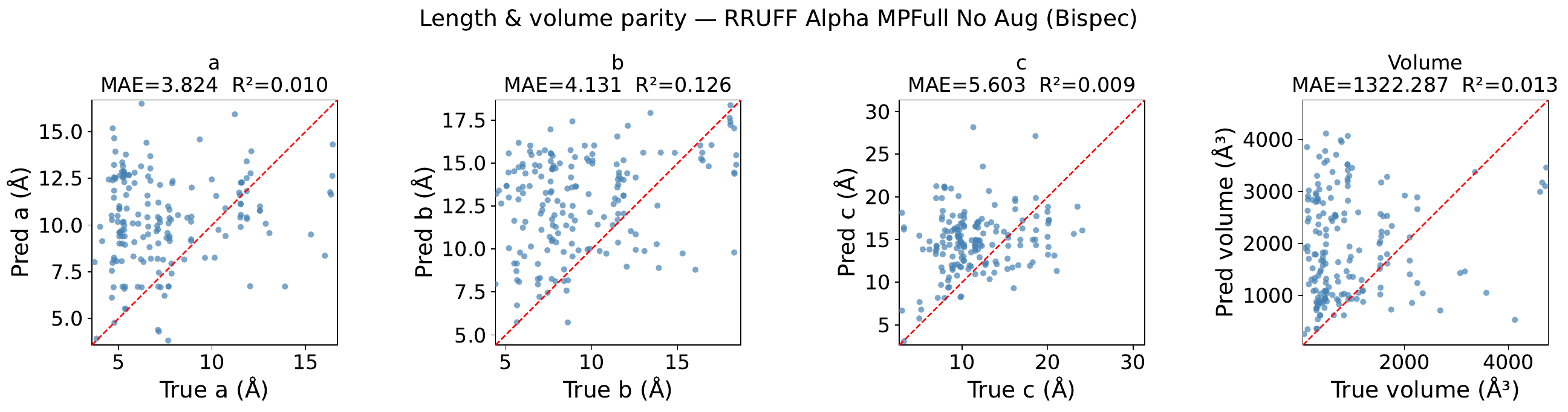}
    \end{subfigure}
    \hfill
    \begin{subfigure}[t]{0.7\textwidth}
        \centering
        \includegraphics[width=\textwidth]{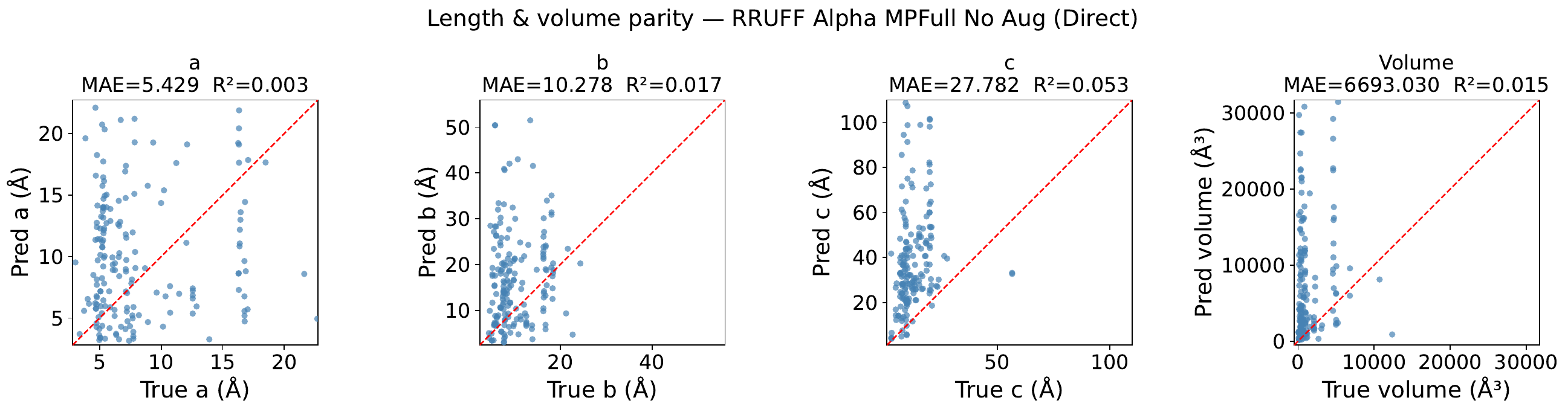}
    \end{subfigure}
    \caption{Predicted vs. true for $a,b,c$ and volume for bispec + inversion and directly predicting parameters using the MPFull dataset with no augmentations evaluated on the RRUFF Alpha test set.}
    \label{fig:alpha_rruff_scatter_no_aug}
\end{figure}

\clearpage
\subsubsection{\revised{Preliminary opXRD CNRS Analysis}}\label{supp_info:cnrs_res}

\revised{We evaluate on RRUFF to enable comparison to other models such as Crystalyze \cite{rieselCrystalStructureDetermination2024}, which are evaluated on this dataset. Our training augmentation follows Crystalyze with a Gamma-distributed intensity perturbation and Gamma background noise with parameters fit to RRUFF to enable a fair comparison. Fitting noise parameters to a reference dataset is standard practice, but it may limit generalization to other experimental datasets with different noise and background characteristics.}

\revised{To evaluate this, we use the opXRD CNRS/COD contribution. Our primary evaluation set is a high-quality, single-phase, curated subset of 599 patterns from the CNRS/COD opXRD dataset. Within this evaluation set, we also report a 116 pattern small-cell subset to compare to the RRUFF Crystalyze results, which are restricted to at most 20 atoms in the unit cell. We indeed find that performance for both the direct and bispectrum models degrade on the CNRS set (see \Cref{tab:cnrs_rruff}), which we hypothesize is due to the background noise augmentation specifically being fit to the RRUFF dataset.}

\revised{To explore whether this degradation can be reduced, we retrained the MP-20 Aug and MP-Full Aug models with two changes: background noise informed by CNRS patterns mixed with RRUFF-fit noise, and a peak-list input representation. The peak list contains up to 30 peaks, found with \texttt{scipy.signal.find\_peaks} on the maximum-normalized pattern and ranked by intensity. Each peak is represented by its $2\theta$ position and its intensity, ordered by position and zero-padded. The position is embedded with a sinusoidal encoding and the intensity with a learned linear layer, and the same extraction is applied to the experimental patterns. The peak lists are extracted after the background noise is added to the simulated pattern. To separate the two changes, we also trained an MP-20 Aug model with CNRS-informed noise and the original full-pattern input.}

\revised{We view these ablations as exploratory. Crucially, both changes reduce error on the CNRS set, demonstrating that the noise model is crucial to performance on experimental data. For MP-20 Aug on the CNRS small-cell subset, training with CNRS-informed noise alone, with the input unchanged, lowers the length MAPE of the bispectrum model from $43.4\%$ to $26.6\%$ and the angle MAPE from $15.0\%$ to $11.7\%$. Adding the peak-list representation lowers them further, to $17.3\%$ and $10.0\%$. One possible explanation for the additional gain is that peak extraction removes most of the background, making the model less sensitive to background differences between simulated and experimental patterns. At MP-Full scale, the peak-list model with CNRS-informed noise lowers the length MAPE on all verified CNRS patterns from $68.8\%$ to $25.0\%$. The bispectrum route gives lower length error than direct prediction with the new recipe ($17.3\%$ vs.\ $26.4\%$ for MP-20 Aug, $25.0\%$ vs.\ $31.2\%$ for MP-Full Aug), while direct prediction gives lower angle error for MP-Full Aug ($8.2\%$ vs.\ $11.8\%$). Conversely, on RRUFF, whose noise statistics the original augmentation was fit to, the original model remains best ($14.4\%$ vs.\ $18.1\%$ length MAPE). Models therefore perform best on the experimental dataset whose noise characteristics they were trained with.}

\revised{In summary, these results suggest that models trained with noise fit to one experimental dataset may not generalize to others, which has implications beyond our work. The sim-to-real gap remains large, particularly for angles, and we do not claim to have closed it. The main contribution of our paper is the invariant bispectrum representation and exploring its effect on prediction quality. Identifying training procedures that generalize better to experimental data, such as more realistic augmentation, improved input representations, or training on experimental patterns, is an important direction for future work.}

\begin{figure}[htbp]
    \centering
    \includegraphics[width=0.7\textwidth]{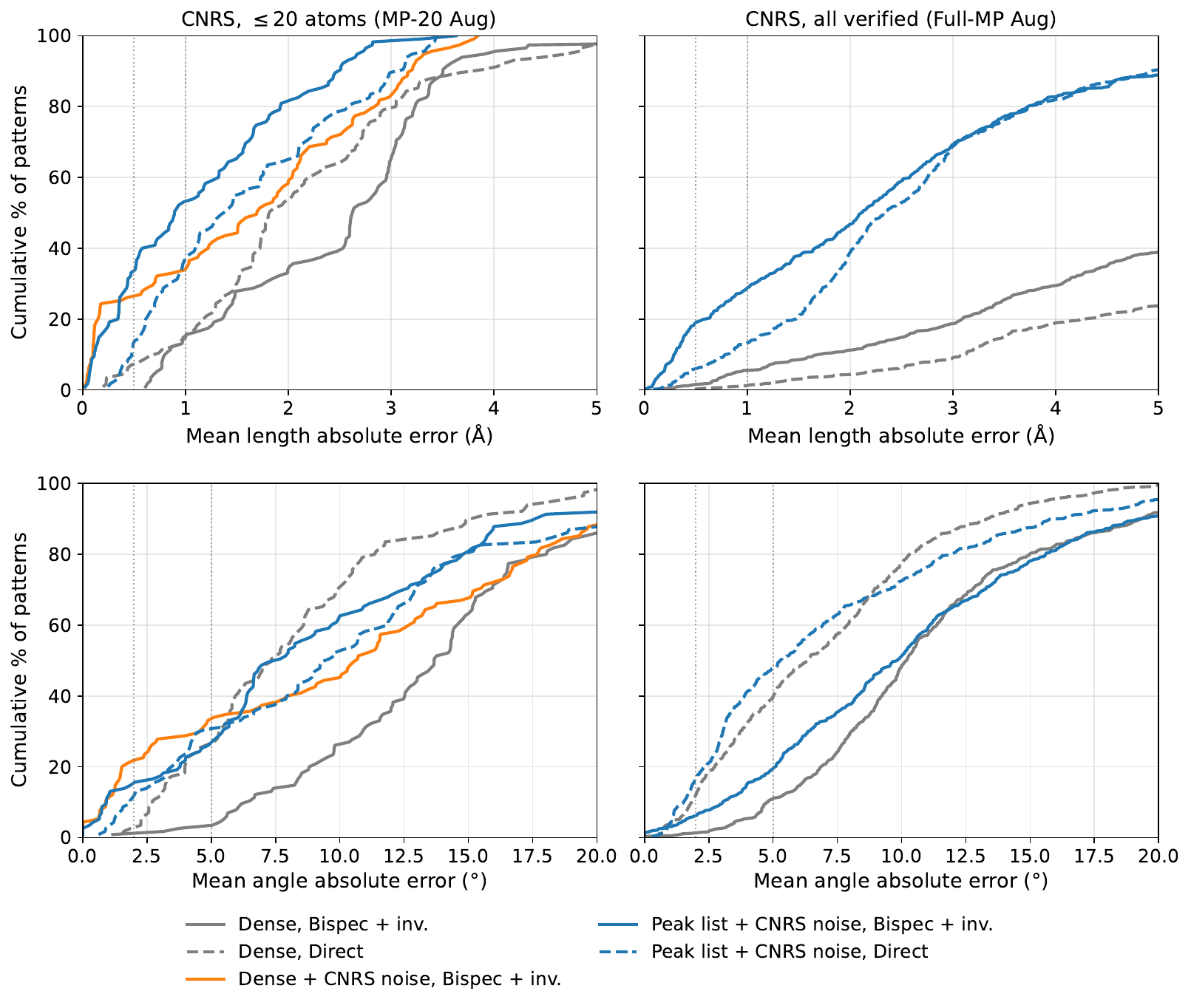}
    \caption{\revised{Cumulative distribution of mean length (top) and angle (bottom) absolute errors on CNRS opXRD patterns, for patterns with at most 20 atoms per primitive cell with MP-20 Aug models and all verified patterns with MP-Full Aug models. Each panel uses the patterns shared by all models shown, as in \Cref{tab:cnrs_small} and \Cref{tab:cnrs_rruff}. Color indicates the training recipe (dense input, dense input with CNRS-informed noise, or peak-list input with CNRS-informed noise) and line style the method (solid: bispec + inversion; dashed: direct prediction). Dotted vertical lines mark 0.5 and 1~\AA{} (length) and 2$^\circ$ and 5$^\circ$ (angle).}
    }
    \label{fig:error_cdf_cnrs}
\end{figure}

\begin{figure}[htbp]
    \centering
    \begin{subfigure}[t]{0.7\textwidth}
        \centering
        \includegraphics[width=\textwidth]{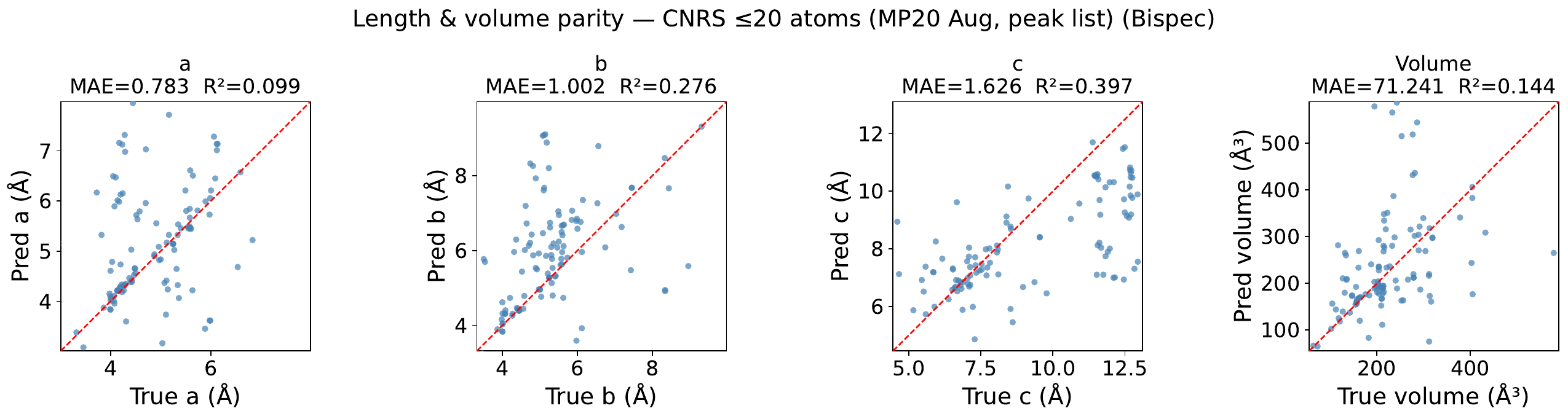}
    \end{subfigure}
    \hfill
    \begin{subfigure}[t]{0.7\textwidth}
        \centering
        \includegraphics[width=\textwidth]{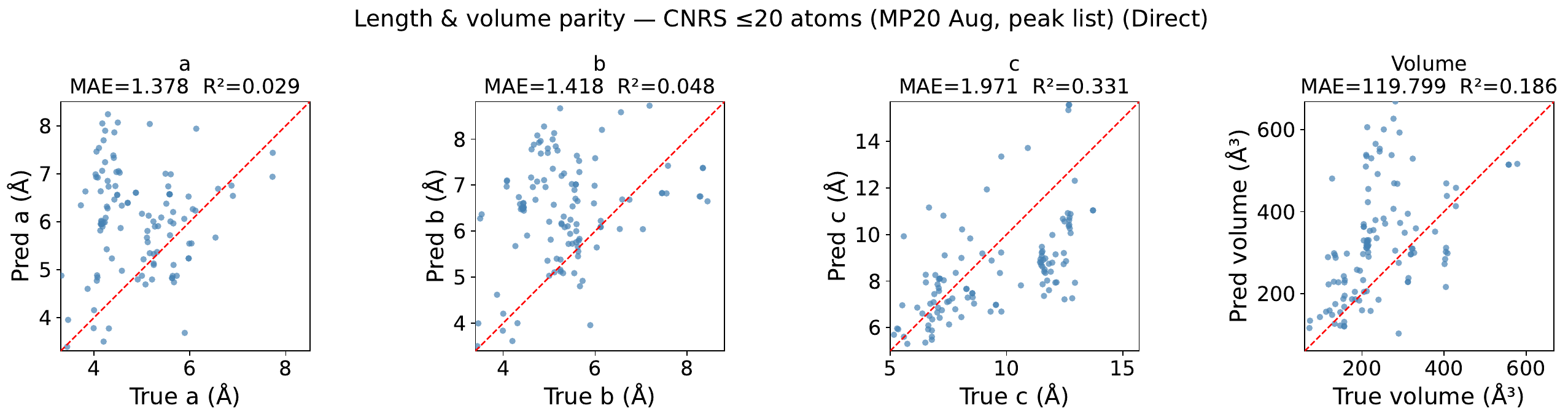}
    \end{subfigure}
    \caption{\revised{Predicted vs. true for $a,b,c$ and volume for bispec + inversion vs directly predicting parameters on the small CNRS set using the MP20 augmented dataset for the peak list representation.}}
    \label{fig:cnrs_small_scatter}
\end{figure}

\begin{figure}[htbp]
    \centering
    \begin{subfigure}[t]{0.7\textwidth}
        \centering
        \includegraphics[width=\textwidth]{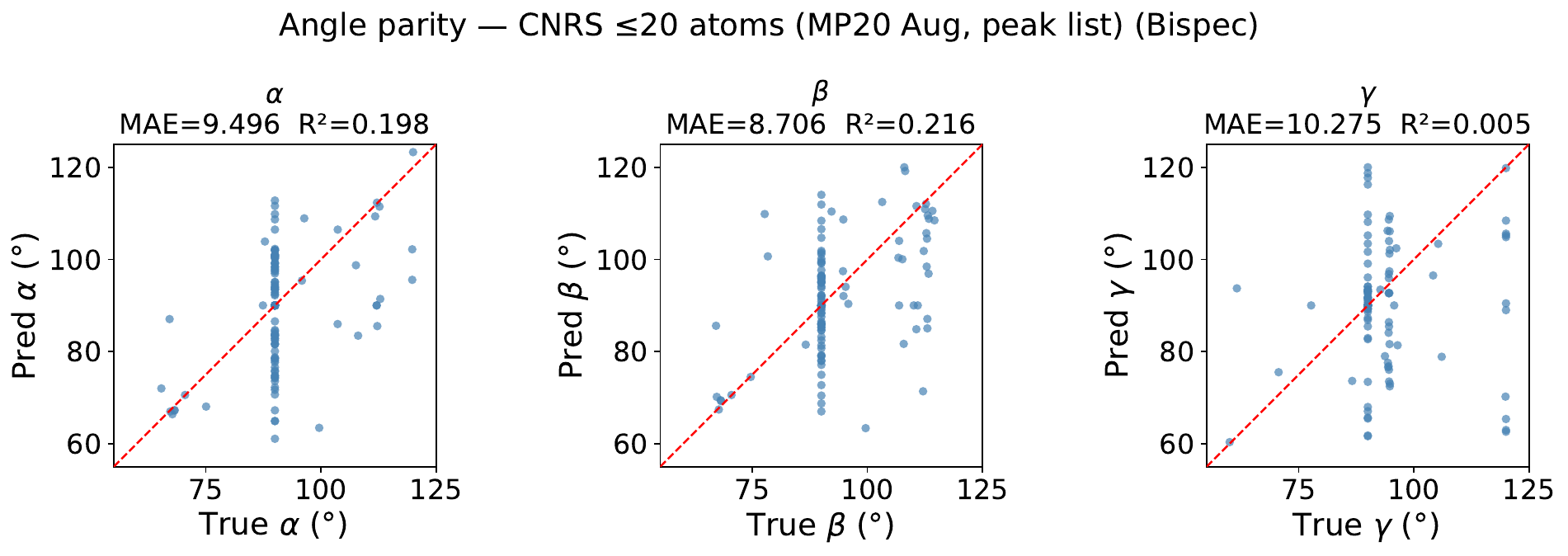}
    \end{subfigure}
    \hfill
    \begin{subfigure}[t]{0.7\textwidth}
        \centering
        \includegraphics[width=\textwidth]{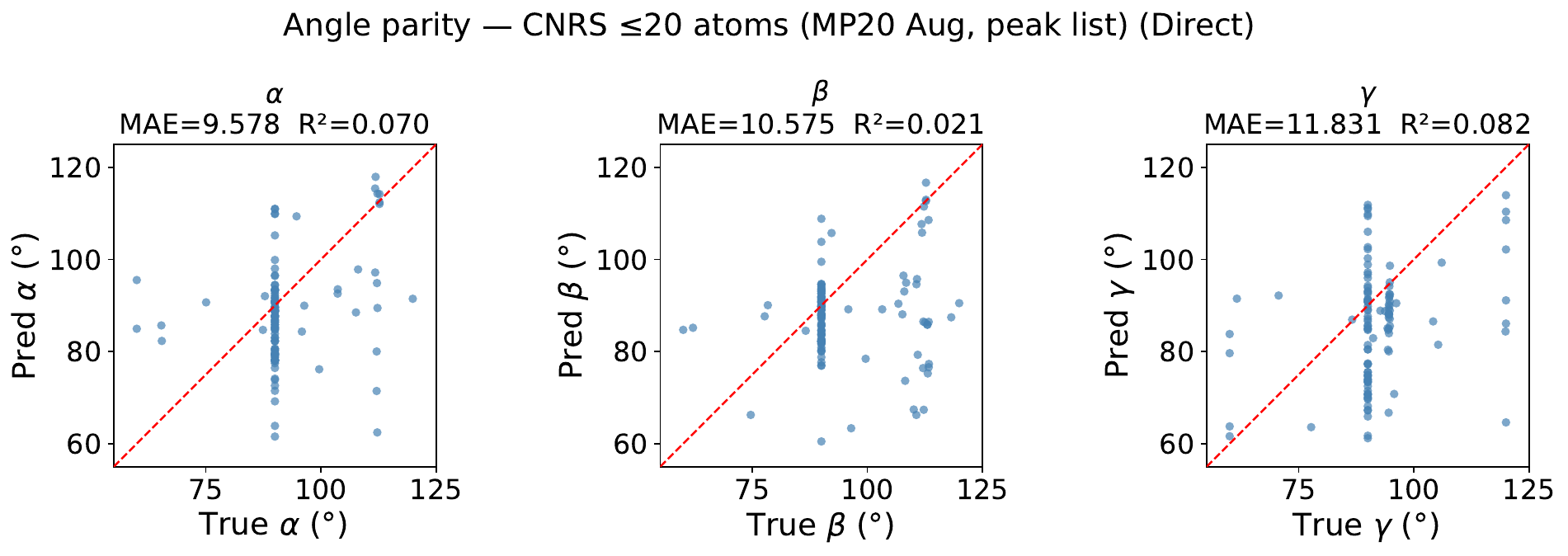}
    \end{subfigure}
    \caption{\revised{Predicted vs. true for $\alpha,\beta,\gamma$ for bispec + inversion vs directly predicting parameters on the small CNRS set using the MP20 augmented dataset for the peak list representation.}}
    \label{fig:cnrs_small_angle_scatter}
\end{figure}

\begin{table}[h]
\centering
\caption{\revised{MAE and MAPE on experimental data (CNRS opXRD and RRUFF). All predictions are post-processed using cctbx. ``Dense'' denotes the full-pattern input used for the models in the manuscript; ``+ CNRS noise'' adds background noise informed by CNRS patterns during training; ``Peak list'' uses up to 30 extracted peaks as input instead of the full pattern. Models other than the Dense rows were trained with the modified recipes described above.}}
\footnotesize
\resizebox{\columnwidth}{!}{
\begin{tabular}{llcccc}
\toprule
Training data & Method & Length MAE (\AA) & Length MAPE (\%) & Angle MAE ($^\circ$) & Angle MAPE (\%) \\
\midrule
\multicolumn{6}{l}{\emph{CNRS, verified patterns with $\leq$20 atoms per primitive cell}} \\
MP-20 Aug Dense & Bispec + inv. & 2.54 & 43.36 & 13.85 & 14.95 \\
MP-20 Aug Dense & Direct & 2.15 & 39.17 & 8.26 & 8.88 \\
MP-20 Aug Dense + CNRS noise & Bispec + inv. & 1.62 & 26.60 & 10.87 & 11.67 \\
MP-20 Aug Peak list + CNRS noise & Bispec + inv. & 1.14 & 17.34 & 9.49 & 10.01 \\
MP-20 Aug Peak list + CNRS noise & Direct & 1.59 & 26.37 & 10.66 & 11.40 \\
\midrule
\multicolumn{6}{l}{\emph{CNRS, all verified patterns}} \\
Full-MP Aug Dense & Bispec + inv. & 6.10 & 68.81 & 11.30 & 12.39 \\
Full-MP Aug Dense & Direct & 10.08 & 117.41 & 6.98 & 7.60 \\
Full-MP Aug Peak list + CNRS noise & Bispec + inv. & 2.55 & 24.98 & 10.68 & 11.83 \\
Full-MP Aug Peak list + CNRS noise & Direct & 2.79 & 31.24 & 7.45 & 8.21 \\
\midrule
\multicolumn{6}{l}{\emph{RRUFF (Crystalyze)}} \\
MP-20 Aug Dense & Bispec + inv. & 0.91 & 14.41 & 8.37 & 8.94 \\
MP-20 Aug Dense & Direct & 1.86 & 30.92 & 9.46 & 10.29 \\
MP-20 Aug Peak list + CNRS noise & Bispec + inv. & 1.08 & 18.12 & 10.08 & 11.03 \\
\bottomrule
\end{tabular}
}
\label{tab:cnrs_rruff}
\end{table}

\end{document}